\documentclass[longauth]{aa}

\usepackage{graphicx}
\usepackage{txfonts}
\usepackage{placeins}
\usepackage[colorlinks=true,linkcolor=blue,citecolor=blue]{hyperref}
\usepackage[dvipsnames]{xcolor}

\DeclareUnicodeCharacter{03B1}{$\alpha$}
\DeclareUnicodeCharacter{2013}{--}
\DeclareUnicodeCharacter{2014}{---}

\def\Msol{M_\odot}
\def\Mstar{M_\star}
\def\Mform{M_{\mathrm{form}}}

\def\T50{T$_{50\%}$}
\def\Rnorm{R_{\mathrm{norm}}}
\def\Rgroup{R_{\mathrm{group}}}
\def\Phihalf{\Phi_{50}}
\def\rhalf{r_{50}}

\def\CWeb{COSMOS-Web}

\def \cigale{{\texttt{CIGALE}}}
\def\photoz{\mbox{photo-\textit{z}}}

\def\lephare{\texttt{LePHARE}}

\def \amico{{\texttt{AMICO}}}
\newcommand{\tlb}{t_\mathrm{lb}}

\definecolor{Blue}{rgb}{0,0.25,0.9}
\definecolor{Red}{rgb}{0.85,0.08,0.05}
\definecolor{Green}{rgb}{0.35,0.45,0.25}
\definecolor{Orange}{rgb}{1.0,0.5,0.15}
\definecolor{Brown}{rgb}{0.7,0.25,0.0}

\begin{document}

\title{COSMOS-Web: From early star-formation enhancement to late suppression in galaxy groups}

\author{Arango-Toro, R. C. \inst{\ref{IAP}}\thanks{email: claudino.arango@gmail.com}{}
\and Laigle, C.\inst{\ref{IAP}}
\and Lewis, J.\inst{\ref{IAP},\ref{Saclay}}
\and Joly, M. \inst{\ref{IAP}}
\and Toni, G.\inst{\ref{Bologna},\ref{INAFBologna},\ref{Heidelberg}}
\and Casey, C. M.\inst{\ref{UCSB},\ref{DAWN}}
\and Kartaltepe, J. S.\inst{\ref{RIT}}
\and Ilbert, O.\inst{\ref{LAM}}
\and Shuntov, M.\inst{\ref{DAWN},\ref{NBI},\ref{Geneva}}
\and Akins, H. B.\inst{\ref{UTAustin}}
\and Paquereau, L.\inst{\ref{Chalmers}}
\and Dubois, Y.\inst{\ref{IAP}}
\and Franco, M.\inst{\ref{Saclay}}
\and Mamon, G. A.\inst{\ref{IAP}}
\and Aufort, G.\inst{\ref{Berkeley}}
\and Faisst, A. L.\inst{\ref{Caltech}}
\and Ghaffari, Z.\inst{\ref{INAFBologna},\ref{IFPU}}
\and Gozaliasl, G.\inst{\ref{Aalto},\ref{Helsinki}}
\and Hatamnia, H.\inst{\ref{UCRiverside},\ref{Carnegie}}
\and Hirschmann, M.\inst{\ref{EPFL}}
\and Huertas-Company, M.\inst{\ref{IAC},\ref{ULaguna},\ref{LERMA}}
\and Koekemoer, A. M.\inst{\ref{STScI}}
\and Laishram, R.\inst{\ref{NAOJ}}
\and Le Borgne, D.\inst{\ref{IAP}}
\and Leroy, G.\inst{\ref{Durham}}
\and McCracken, H. J.\inst{\ref{IAP}}
\and Moscardini, L.\inst{\ref{Bologna},\ref{INAFBologna},\ref{INFNBologna}}
\and Nuñez-Castiñeyra, A.\inst{\ref{IAP}}
\and Rhee, J.\inst{\ref{IAP}}
\and Samir, R. M.\inst{\ref{NRIAG}}
\and Trebitsch, M.\inst{\ref{ParisObs}}
\and Tresse, L.\inst{\ref{LAM}}
}
\institute{
Institut d'Astrophysique de Paris, UMR 7095, CNRS, and Sorbonne Université, 98bis boulevard Arago, 75014 Paris, France \label{IAP}
\and Université Paris-Saclay, Université Paris Cité, CEA, CNRS, AIM, 91191 Gif-sur-Yvette, France \label{Saclay}
\and University of Bologna - Department of Physics and Astronomy ``Augusto Righi'' (DIFA), Via Gobetti 93/2, I-40129 Bologna, Italy \label{Bologna}
\and INAF-Osservatorio di Astrofisica e Scienza dello Spazio, Via Gobetti 93/3, I-40129 Bologna, Italy \label{INAFBologna}
\and Zentrum für Astronomie, Universität Heidelberg, Philosophenweg 12, D-69120, Heidelberg, Germany \label{Heidelberg}
\and Department of Physics, University of California, Santa Barbara, Santa Barbara, CA 93106, USA \label{UCSB}
\and Cosmic Dawn Center (DAWN), Denmark \label{DAWN}
\and Laboratory for Multiwavelength Astrophysics, School of Physics and Astronomy, Rochester Institute of Technology, 84 Lomb Memorial Drive, Rochester, NY 14623, USA \label{RIT}
\and Aix Marseille Univ, CNRS, CNES, LAM, Marseille, France \label{LAM}
\and Niels Bohr Institute, University of Copenhagen, Jagtvej 128, DK-2200, Copenhagen, Denmark \label{NBI}
\and Department of Astronomy, University of Geneva, Chemin Pegasi 51, 1290 Versoix, Switzerland \label{Geneva}
\and The University of Texas at Austin, 2515 Speedway Blvd Stop C1400, Austin, TX 78712, USA \label{UTAustin}
\and Department of Physics and Astronomy, Chalmers University of Technology, SE-412 96 Gothenburg, Sweden \label{Chalmers}
\and Department of Astronomy, University of California, Berkeley, CA 94720, USA \label{Berkeley}
\and Caltech/IPAC, MS 314-6, 1200 E. California Blvd. Pasadena, CA 91125, USA \label{Caltech}
\and IFPU: Institute for Fundamental Physics of the Universe, Via Beirut 2, 34151 Trieste, Italy \label{IFPU}
\and Department of Computer Science, Aalto University, PO Box 15400, Espoo, FI-00076, Finland \label{Aalto}
\and Department of Physics, University of Helsinki, P. O. Box 64, FI-00014 Helsinki, Finland \label{Helsinki}
\and Department of Physics and Astronomy, University of California, Riverside, 900 University Avenue, Riverside, CA 92521, USA \label{UCRiverside}
\and Observatories of the Carnegie Institution for Science, 813 Santa Barbara St., Pasadena, CA 91101, USA \label{Carnegie}
\and Institute of Physics, GalSpec, Ecole Polytechnique Federale de Lausanne, Observatoire de Sauverny, Chemin Pegasi 51, 1290 Versoix, Switzerland \label{EPFL}
\and Instituto de Astrofísica de Canarias, La Laguna, Tenerife, Spain \label{IAC}
\and Universidad de La Laguna, La Laguna, Tenerife, Spain \label{ULaguna}
\and Université de Paris, LERMA – Observatoire de Paris, PSL, Paris, France \label{LERMA}
\and Space Telescope Science Institute, 3700 San Martin Dr., Baltimore, MD 21218, USA \label{STScI}
\and National Astronomical Observatory of Japan, 2-21-1 Osawa, Mitaka, Tokyo 181-8588, Japan \label{NAOJ}
\and Institute for Computational Cosmology, Department of Physics, Durham University, South Road, Durham DH1 3LE, United Kingdom \label{Durham}
\and INFN-Sezione di Bologna, Viale Berti Pichat 6/2, I-40127 Bologna, Italy \label{INFNBologna}
\and Department of Astronomy, National Research Institute of Astronomy and Geophysics (NRIAG), Cairo, 11421, Egypt \label{NRIAG}
\and Observatoire de Paris, PSL University, Sorbonne Université, CNRS, 75014 Paris, France \label{ParisObs}
}
 
\date{Received --; accepted --}

\abstract
{Galaxy groups trace dense environments where interactions, gas removal, and reduced accretion may drive quenching. Common diagnostics trace star formation only over short timescales ($\lesssim100$ Myr), making time-resolved star formation histories (SFHs) necessary to separate recent changes from longer-term evolution at fixed mass and redshift.}
{Using current \CWeb\ data products, we test how group environment correlates with star-formation activity, how this evolves with cosmic time and group-centric distance, and how high-richness groups contrast with field galaxies (those excluded from the group sample).}
{We combine COSMOS2025/\CWeb\ stellar masses and non-parametric SFHs with group detections and probabilistic memberships from \amico\ (Adaptive Matched Identifier of Clustered Objects). We compare stacked SFHs of group and field galaxies matched in mass and redshift, using the 30 richest groups in each redshift bin to measure how the SFH offset evolves with group-centric distance.}
{Massive galaxies ($\log(\Mstar/\Msol)>10.5$) show the largest offsets but are already quenched in both environments. For lower masses ($8.1<\log(\Mstar/\Msol)<10.5$), suppression is clearest at $z<1$ (recent deficits of up to $0.7$ dex), while at $z>1$ SFHs show weaker suppression or occasional enhancement. The radial signal evolves similarly: quenching is broad across radius at low $z$, the inner--outer contrast appears at $z>0.5$, and any ordering at $z\gtrsim2$ remains tentative given growing uncertainties in the \amico\ centroids.}
{These results suggest an evolving environmental picture: groups are more mixed at early epochs, with both suppressed and occasionally elevated SFHs (partly reflecting high-$z$ uncertainties), while suppression dominates at $z<1$, most clearly for low-to-intermediate-mass galaxies. This fits a picture where inner-region galaxies spend more time in the group potential, undergo more passages through dense intra-group gas, and accrete less pristine cold gas, making quenching progressively clearer with cosmic time.}


\maketitle

\section{Introduction}

Understanding when, where, and how galaxies quench remains a central problem in galaxy evolution \citep{bluckGalaxyQuenchingHigh2024,deluciaTracingQuenchingJourney2024,deluciaCosmicQuenching2025,whitaker_quenching_2026}. ``Environment'' itself is a broad concept (cluster or group membership, local density, protocluster association, cosmic-web position), and these definitions do not always select the same physical regime. Rich clusters give the clearest example of environmentally driven quenching \citep[e.g., ][]{dressler_galaxy_1980,boselli_environmental_2006}, while groups and forming structures probe lower-mass, often less dynamically evolved regimes \citep{paul_understanding_2017,lovisari_scaling_2021}. Classical diagnostics (color and quiescent fractions, star formation rate (SFR)--stellar mass ($\Mstar$) trends) establish that dense environments host a higher fraction of passive galaxies \citep{dressler_galaxy_1980,van_den_bosch_importance_2008,peng_mass_2010,wetzel_galaxy_2013}, but these methods resolve star formation in group and cluster galaxies only weakly in time, because they trace only the last $\sim100\,\rm Myr$. Star formation histories (SFHs) instead preserve the temporal sequence of star-formation changes, distinguishing short-lived fluctuations from sustained suppression or enhancement, and allow recovery of recent evolutionary timescales ($\sim0.1\mbox{--}1\,\rm Gyr$) separated from longer-term change at fixed stellar mass and redshift \citep{leja_deriving_2019,arango-toro_cosmos-web_2025}.

At low redshift ($z\lesssim1.5$), the picture is relatively well established: satellite galaxies in dense environments tend to quench after infall, following a delayed-then-rapid pattern in which SFRs stay close to the field for a few Gyr and then decline on a sub-Gyr timescale. At $z=0$, this delay is typically 2--3 Gyr at fixed mass and morphology \citep[e.g., ][]{wetzel_galaxy_2013} followed by fading on $\lesssim1$ Gyr \citep{oman_homogeneous_2021}. In massive clusters, even massive satellites may only complete their quenching after their first passage, once they bounce out of the cluster \citep{Mahajan+11}. This delay has been linked to the dynamical (crossing) timescale of the host halo, which decreases with increasing redshift, potentially explaining why shorter quenching delays are inferred at higher redshift; for instance, \citet{Foltz+18} infer a quenching timescale of $\sim1.3$ Gyr at $z\sim1$. Other work, however, favors gas starvation or internally driven outflows (``overconsumption'') over orbit-based stripping as the dominant driver of this redshift evolution \citep{mcgee_overconsumption_2014}, so the physical origin of the shorter delays remains debated. These effects are strongest in lower-mass satellites whose shallower potential wells make them more vulnerable to stripping and starvation \citep{boselli_environmental_2006,wetzel_satellite_2015,fillingham_under_2016}.

At earlier cosmic times ($z\gtrsim1.5$), the literature is less settled, partly because studies do not measure the same kind of environment: some use projected or 3D galaxy density, probing the SFR--density relation on local or large-scale-structure scales \citep{elbaz_reversal_2007,cooper_deep2_2007}, while others select clusters, protoclusters, or overdense structures via halo-scale membership or extended emission \citep{brodwin_era_2013,alberts_evolution_2014,hwang_evolution_2019,williams_alma_2022} -- choices that matter, since these regimes correspond to different stages of structure growth and processing \citep{alberts_clusters_2022}. This helps explain why some studies find a weakened or reversed SFR--density relation while others already detect heterogeneous suppression \citep{Jeon+22}. A recent James Webb Space Telescope (JWST) analysis based on COSMOS-Web data supports this view: the stellar-mass--density relation persists across redshift, but SFR--density trends change sign with epoch, and environmental quenching becomes clearest at $z\lesssim0.8$ for low-mass galaxies \citep[$\Mstar\lesssim10^{10}\,\Msol$;][]{hatamnia_large-scale_2025}.

Galaxy groups, gravitationally bound associations of up to $\sim50$ members sharing a common dark-matter halo \citep[$\log\,(M_\mathrm{halo}/\Msol)\sim13-14$;][]{yang_galaxy_2007,lovisari_scaling_2021}, occupy the intermediate regime between isolated galaxies and rich clusters. Their shallow potential wells make them particularly sensitive to non-gravitational processes \citep{oppenheimer_simulating_2021}: active galactic nuclei (AGN) and supernova feedback can inject energy comparable to the binding energy of the halo gas, redistributing or ejecting baryons \citep{popesso_hot_2026}. This makes groups a primary site of environmental preprocessing: satellites can lose $35$--$45\%$ of their mass through tidal stripping and begin quenching long before accretion into more massive clusters \citep{joshi_preprocessing_2017,reeves_gogreen_2021}.

This motivates extending the analysis across redshifts with homogeneous data and a consistent environmental reconstruction. The Cosmic Evolution Survey (COSMOS) field now offers this framework: survey and catalog legacies \citep{scoville_cosmic_2007,koekemoer_cosmos_2007,laigle_cosmos2015_2016,weaver_cosmos2020_2022}, density-field and cosmic-web mapping \citep{darvish_comparative_2015,darvish_cosmic_2017,cucciati_progeny_2018,hatamnia_large-scale_2025,jego_star-forming_2026}, and calibrated group catalogs \citep{gozaliasl_chandra_2019,toni_amico-cosmos_2024} together provide a robust basis for comparing group and field galaxies across a broad redshift range \citep[e.g., ][]{taamoli_cosmos2020_2024}.

The recent \CWeb\ survey \citep{casey_cosmos-web_2023}, covering $\sim0.5\,\mathrm{deg}^2$ of the COSMOS field with JWST/Near Infrared Camera (NIRCam) and Mid-Infrared Instrument (MIRI) imaging \citep{harish_cosmos-web_2025,franco_cosmos-web_2026}, adds the ingredients needed for time-resolved environmental tests: non-parametric SFHs and migration diagnostics in the mass--SFR plane \citep{shuntov_cosmos2025_2025,arango-toro_cosmos-web_2025}, and deep group catalogs built with \amico\ (Adaptive Matched Identifier of Clustered Objects), a matched-filter algorithm detecting spatial overdensities and assigning probabilistic memberships \citep{bellagamba_amico_2019, maturi_span_2019}. These datasets have already provided evidence that quenching can precede full morphological transformation \citep{ghaffari_star_2026} and measurements of quenched-fraction evolution in groups \citep{toni_cosmos-web_2026}. Still missing, however, is precisely the test these products now make possible: a time-resolved study over this full redshift range linking group environment to star-formation activity and its dependence on position within groups.

The goal of this paper is to quantify how strongly the group environment affects SFHs as a function of cosmic epoch and stellar mass, and to test whether this effect is modulated by projected distance from the group center. We adopt a matched group--field comparison with limited assumptions, combining \amico\ probabilistic memberships, the top-30 richest groups in each redshift bin, and normalized group-centric radius with non-parametric SFHs, and compare group galaxies with field controls matched in ($\Mstar$, $z$) bins. We use two complementary diagnostics: stacked SFHs and migration-vector statistics \citep{ciesla_goods-alma_2023,arango-toro_probing_2023,arango-toro_cosmos-web_2025}.

The remainder of this paper is organized as follows. The data and sample construction are described in Sects.~\ref{sec:data} and \ref{sec:sample}, respectively. The methodological framework is presented in Sect.~\ref{sec:methods}, followed by the results in Sect.~\ref{sec:results}. We discuss our findings in Sect.~\ref{sec:discussion} and summarize our conclusions in Sect.~\ref{sec:conclusions}. For consistency with the COSMOS2025/\CWeb\ SFH framework adopted in \citet{arango-toro_cosmos-web_2025}, we assume a flat $\Lambda$CDM cosmology with $\Omega_\mathrm{m}=0.3$, $\Omega_\Lambda=0.7$, and $H_0=70\,\mathrm{km\,s^{-1}\,Mpc^{-1}}$, a Chabrier initial mass function, and AB magnitudes.

\section{Data}\label{sec:data}
\subsection{COSMOS2025 / \CWeb\ catalog}
We use the COSMOS2025 catalog \citep{shuntov_cosmos2025_2025} built from the 255\,h JWST \CWeb\ survey \citep{casey_cosmos-web_2023}. \CWeb\ provides deep NIRCam imaging over $\sim0.54\,\mathrm{deg}^2$ in F115W, F150W, F277W, and F444W bands \citep{franco_cosmos-web_2026}, with parallel MIRI imaging over $\sim0.2\,\mathrm{deg}^2$ in F770W \citep{harish_cosmos-web_2025}, complemented by the legacy COSMOS multiwavelength dataset.
The COSMOS2025 catalog delivers homogeneous photometry from the optical to mid-infrared ($\sim0.3\mbox{--}8\,\mu$m), combining the four NIRCam bands with extensive ground-based, Hubble Space Telescope (HST), and Spitzer data. Source detection is performed on the stacked NIRCam images, while fluxes are measured using a profile-fitting approach designed to convolve surface-brightness profiles with the wide range of point spread functions (PSFs) across facilities and to handle deblending in the deep JWST data \citep{bertin_sourcextractor_2022}.
A set of physical parameters, including photometric redshifts ($\photoz$), is obtained from spectral energy distribution (SED) fitting via the \lephare\ (PHotometric Analysis for Redshift Estimate) code \citep{arnouts_measuring_2002,ilbert_accurate_2006}. The $\photoz$ setup is calibrated against a compilation of more than 11\,000 confirmed spectroscopic redshifts; the resulting precision reaches $\sigma_{\mathrm{NMAD}}\lesssim0.003$, where the normalized median absolute deviation is $\sigma_\mathrm{NMAD}=1.48\times\mathrm{median}(|\Delta z|/(1+z_\mathrm{spec}))$, with $\Delta z=z_\mathrm{phot}-z_\mathrm{spec}$; the catastrophic-outlier fraction is below 10\% at F444W $<28$ \citep{khostovan_cosmos_2026}. This accuracy is important for our analysis because \amico\ uses each galaxy's redshift probability distribution directly both to detect groups and to assign membership probabilities, rather than relying only on point estimates.

For this work, we adopt the COSMOS2025 products based on \cigale\ (Code Investigating GALaxy Emission), specifically stellar mass and the non-parametric SFH reconstructions \citep{arango-toro_cosmos-web_2025,shuntov_cosmos2025_2025}.
Throughout this paper, the SFH of a galaxy is its SFR as a function of lookback time, $\mathrm{SFR}(\tlb)$. The COSMOS2025 SFHs are delivered as piecewise-constant SFRs in bins of lookback time (hereafter ``SFH bins''), which we use for stacking and for constructing the formation-time metrics (Sect.~\ref{sec:methods}). We distinguish two stellar masses: $\Mstar$ is the current stellar mass of the catalog (i.e., after mass loss from stellar evolution), used for mass binning and completeness; $\Mform$ is the total mass formed, obtained by integrating the SFH without correcting for the returned fraction (Eq.~\ref{eq:Mform}).

\subsection{\amico\, group catalog and memberships}
Galaxy groups are identified with \amico{} \citep{bellagamba_amico_2019,maturi_span_2019}, following its application over $\sim1.7\,\mathrm{deg}^2$ of the COSMOS field up to $z\sim2$ \citep{toni_amico-cosmos_2024} and its \CWeb\ extension over $\sim0.5\,\mathrm{deg}^2$ to $z=3.7$ \citep{toni_cosmos-web_2025}. \amico\, is a linear matched-filter algorithm that combines galaxy sky positions, magnitudes, and photometric redshift information to detect groups or galaxy clusters and assign membership probabilities. In this framework, each galaxy is represented by a Gaussian redshift probability distribution centered on its best-fit redshift and scaled by its 1$\sigma$ uncertainty. The group redshift ($z_\mathrm{group}$) is the redshift preferred by the full matched-filter model, not a simple average of the member redshifts. In the \CWeb\ \amico\, group catalog, mocks are used to calibrate the mapping between significance and sample purity/completeness, which enables controlled selections from very deep to high-purity subsamples \citep{toni_cosmos-web_2025,toni_cosmos-web_2026}.

We use the COSMOS2025 \amico\, products\footnote{\href{https://cosmos2025.iap.fr/groups.html}{https://cosmos2025.iap.fr/groups.html}} distributed as a group catalog and a companion membership table. At the group level, we use: the \amico\, centroid (sky position and redshift of the matched-filter peak); \texttt{SN\_NOCL}, the detection signal-to-noise ratio computed from the background noise alone, i.e., without the shot-noise contribution of group members; \texttt{MSKFRC}, the fraction of the group area around the detection that is excluded by the survey mask; and \texttt{LAMBDA\_STAR} ($\lambda_\star$), the intrinsic richness proxy. Specifically, $\lambda_\star$ is the sum of the membership probabilities of galaxies within the model $R_{200}$ and brighter than $m_\star+1.5$, where $m_\star$ is the knee of the model luminosity function. It can therefore be read as a probability-weighted effective number of bright group members, rather than as a raw galaxy count. We adopt quality cuts consistent with the \CWeb\ \amico\, calibration, in which higher \texttt{SN\_NOCL} corresponds to higher expected purity \citep{toni_cosmos-web_2025}. Applying the $\lambda_\star$--$M_{200}$ scaling relation of \citet{toni_amico-cosmos_2024}, calibrated on X-ray masses, the median richness of the quality-selected sample (Sect.~\ref{sec:membership}) translates into $\log(M_{200}/\Msol)\simeq13.5$ (13.2--13.5 across redshift bins).

At the galaxy level, the membership table provides the linked group identifier plus two complementary probabilities: \texttt{ASSOC\_PROB}, the probability that a galaxy belongs to the detected group, and \texttt{FIELD\_PROB},  the probability that it is better described as a field galaxy. The expected positions and magnitudes of group galaxies are described by a truncated Navarro--Frenk--White profile and an evolving Schechter luminosity function, respectively \citep{toni_cosmos-web_2025}. In practice, a galaxy is assigned a higher group probability when it lies close to the detected group center, has a photometric redshift compatible with the group redshift, and has a magnitude typical of modeled group members. This expected group signal is compared with the expected number of unrelated field galaxies with similar magnitude and redshift. Moreover, rather than using a local annulus around each group, the field probability is estimated from the full input catalog: the field is split into four sky quadrants, the galaxy counts are measured in each magnitude--redshift bin, and the median of the four values is adopted to reduce the impact of real overdensities \citep{toni_cosmos-web_2025}.

\section{Sample selection and definitions} \label{sec:sample}
\subsection{Clean galaxy sample and mass completeness}
From the COSMOS2025 catalog, we build a clean parent galaxy sample with simple quality cuts and star removal\footnote{See the dedicated COSMOS2025 tutorial: \emph{\href{https://cosmos2025.iap.fr/tutorials/catalog_usage_tutorial.html}{Working with the \CWeb\ catalog}}}. We keep only secure galaxy classifications, remove sources with unreliable or contaminated photometry (including bright-star halos), and mask regions with known stellar contamination \citep{shuntov_cosmos2025_2025}.
Moreover, we require \texttt{type}=0 from the \lephare\ cross-matched sources, which excludes catalog-identified AGN, in particular those with Chandra detections (\texttt{flag\_chandra}=1) and compact sources preferentially fitted by AGN templates.

Following recent \CWeb\ analyses \citep[e.g., ][]{paquereau_tracing_2025, arango-toro_cosmos-web_2025, shuntov_stellar_2026} we adopt a flux-limited parent selection in F444W and compute stellar-mass completeness with the \citet{pozzetti_zcosmos_2010} estimator.
For each galaxy, the limiting mass is evaluated as
\begin{equation}
  \log M_{\star,\mathrm{lim}} = \log\,(\Mstar/\Msol) + 0.4\,(m - m_{\mathrm{lim}}),
\end{equation}
where $m$ is the observed magnitude and $m_{\mathrm{lim}}$ is the adopted flux limit in the same band. We adopt $m_{\mathrm{lim}}^{\mathrm{F444W}}=27.5$ and derive $M_\mathrm{lim}(z)$ from the faint-end envelope using percentile-based thresholds that define 95\% completeness.
This yields 95\%-completeness limits that rise from $\log(\Mstar/\Msol)=7.19$ at $0<z<0.5$ to $8.15$ at $3.5<z<4$.
In this work, we enforce a common stellar-mass completeness floor in all redshift bins, defined as the 95\%-completeness mass limit reached in the highest redshift bin used in the analysis. This choice makes the stellar-mass and redshift bins directly comparable throughout the paper; for the highest analysis bin ($2<z<3.7$), it corresponds to $\log(M_{\star,\mathrm{lim}}/\Msol)\simeq8.1$.

Throughout the manuscript, ``low mass'' refers to $8.1<\log(\Mstar/\Msol)<9.5$, ``intermediate mass'' to $9.5<\log(\Mstar/\Msol)<10.5$, and ``high mass'' to $\log(\Mstar/\Msol)>10.5$; the combined $8.1<\log(\Mstar/\Msol)<10.5$ sample is consistently called the low- to intermediate-mass population. We do not explicitly exclude the brightest group galaxy (BGG) from the high-mass bin. As centrals typically receive high \texttt{ASSOC\_PROB} and tend to be massive, they contribute to the high-mass stacked signal; their inclusion should be kept in mind when interpreting the high-mass results, where internal quenching mechanisms are already expected to dominate.

\subsection{Group, membership, and field definitions}
\label{sec:membership}
We begin by defining a high-reliability group sample. The quality selection applies two filters: $\texttt{SN\_NOCL}\geq10$ and $\texttt{MSKFRC}<0.2$, the latter requiring that less than 20\% of the local group area is masked. Of the 1678 groups in the parent catalog, 1141 (68.0\%) meet the masking requirement. The $\texttt{SN\_NOCL}$ cut is met by 665 groups (39.6\%). Applying both cuts leaves a final sample of 460 groups, or 27.4\% of the parent catalog. Together, these filters select robust detections and limit mask-driven biases, improving the expected purity, completeness, and stability of the working sample \citep{toni_cosmos-web_2025,toni_cosmos-web_2026,ghaffari_star_2026}. The top-30 sample (Sect.~\ref{sec:methods_staked-SFHs}) is drawn from these 460 groups.

We then assign galaxy memberships. Candidate members must satisfy a $\photoz$ consistency window, $\Delta z = \pm 0.01(1+z_\mathrm{group})$, so that assignments mitigate line-of-sight projection effects, remain consistent with the expected $\photoz$ uncertainty, and remove galaxies whose photometric redshifts are inconsistent with the group and therefore likely interlopers. We keep galaxies with $\texttt{ASSOC\_PROB}>0$ and propagate \texttt{ASSOC\_PROB} as a statistical weight in all aggregate measurements: every plausible member contributes, but higher-probability members contribute more.

Finally, we define the field control sample consistently with the same \amico\, framework. In \amico, the field probability corresponds to the fraction of ``unassigned'' probability remaining after iterative membership assignment, i.e., the component compatible with the background model. Accordingly, our control sample (hereafter the field sample) excludes all galaxies appearing as \amico\, membership candidates, removing potential satellites by construction. Once the field sample is defined, we match it to the group sample in stellar mass within each redshift bin so that the field stellar-mass distribution follows that of the groups, making any group versus field comparison easier to interpret. We note that galaxies excluded from the \amico\ membership catalog may still reside in cosmic filaments or other intermediate-density environments, which are themselves associated with some degree of environmental preprocessing \citep{Song+21}. Our ``field'' sample is therefore more accurately described as a non-group comparison sample rather than a purely isolated field.

\subsection{Normalized radius $R_\mathrm{norm}$ \label{sec:sample-radial}}
 
We employ an inertia tensor approach to characterize the spatial extent of galaxy groups. Second-moment (inertia) tensors of the member-galaxy distribution have long been the standard tool for quantifying the projected size, elongation and orientation of clusters and rich groups of galaxies \citep{carter1980, Basilakos2000, paz2006, wojtak2013}. This model-independent method does not assume any specific density profile \citep[e.g., ][]{plummer_problem_1911, king1962, navarro1997}, making it applicable to galaxy groups with arbitrary morphologies and relaxation states. The approach is based on the moment of inertia tensor computed from the positions and masses of member galaxies, weighted by their membership probabilities. The weighted barycenter is calculated as:
\begin{equation}
    \vec{r}_{\mathrm{bary}} = \frac{\sum_i m_i \, p_i \, \vec{r}_i}{\sum_i m_i \, p_i},
\label{eq:barycenter}
\end{equation}
where $m_i$ is the stellar mass of galaxy $i$, $p_i$ is its group membership probability (\texttt{ASSOC\_PROB}), and $\vec{r}_i$ is its position vector. The inertia tensor is then constructed as:
\begin{equation}
    I_{jk} = \frac{\sum_i m_i \, p_i \, (r_{i,j} - r_{\mathrm{bary},j})(r_{i,k} - r_{\mathrm{bary},k})}{\sum_i m_i\,p_i}.
\label{eq:inertia_tensor}
\end{equation}
Diagonalizing this matrix yields the eigenvalues $\lambda_1 \geq \lambda_2$, whose square roots, $a=\sqrt{\lambda_1}$ and $b=\sqrt{\lambda_2}$, correspond respectively to the semi-major and semi-minor axes of the ellipse representing the spatial distribution of the group, centered on the barycenter \citep{paz2006, wojtak2013}. We do not consider the orientation of the principal axes (i.e., the eigenvectors) in this work, retaining only the axis lengths. The characteristic scale radius $\Rgroup$ is obtained by averaging these two semi-axes, to finally define the normalized projected radius as
\begin{equation}
    R_\mathrm{norm}=\frac{R_\mathrm{proj}}{\Rgroup}, \label{eqn:Rnorm}
\end{equation}
with $R_\mathrm{proj}$ as the physical distance (in kpc) of a galaxy from the group barycenter (Eq.~\ref{eq:barycenter}) measured in the plane of the sky, computed from the angular separation and the angular diameter distance at the group's redshift. Normalizing radial distances by an empirically defined, projected group radius rather than by a dynamically inferred one follows the practice adopted in wide-field group catalogs, where such radii are constructed directly from the observed member distribution \citep[e.g., ][]{robotham2011}.
 
Additionally, uncertainties on these quantities are estimated via Bernoulli draws, i.e., a probabilistic resampling of member galaxies according to their membership probability $p_i$, allowing the uncertainty associated with ambiguous group membership to be propagated through the calculation. This is analogous to the bootstrap resampling of member galaxies commonly used to assign errors to tensor-based shape and size estimates \citep{wojtak2013}, but weighted by the membership probability rather than drawn uniformly. For each group, 200 realizations are generated via Bernoulli draws on the member galaxies, and the uncertainty on each quantity is estimated from the standard deviation of the resulting distribution over these 200 realizations.
 
We stress that $R_{\mathrm{group}}$ remains a descriptive scale of the projected galaxy distribution, not a virial radius or an indicator of dynamical relaxation \citep[cf.][]{robotham2011}: substructure, interlopers, and sparsely sampled memberships can broaden the inertia tensor and bias both the inferred center and radius \citep{plionis2004, paz2006, zemp2011}. We return to these caveats, and their impact on our radial results, in Sect.~\ref{sec:disc-radial}.

\section{Methods}
\label{sec:methods}
\subsection{SFH stacking\label{sec:methods_staked-SFHs}}
We analyze the SFHs in bins of redshift and stellar mass using the top-30 sample: in each redshift bin, we rank the quality-selected groups (Sect.~\ref{sec:membership}) by $\lambda_\star$ and retain up to the 30 richest systems with $\lambda_\star>10$. We first compare their members with matched field controls to establish a high-contrast reference for the environmental signal, and then examine how this signal varies with normalized group-centric radius, $\Rnorm$. The value 30 is not a physically motivated threshold but a pragmatic fixed-$N$ choice that keeps the sample simple and high contrast, following recent \CWeb\ work \citep[e.g., the top-25 sample in][]{ghaffari_star_2026}; the dependence of our results on $N$ is not explored here.

\begin{figure}[h]
  \centering
  \includegraphics[width=\columnwidth]{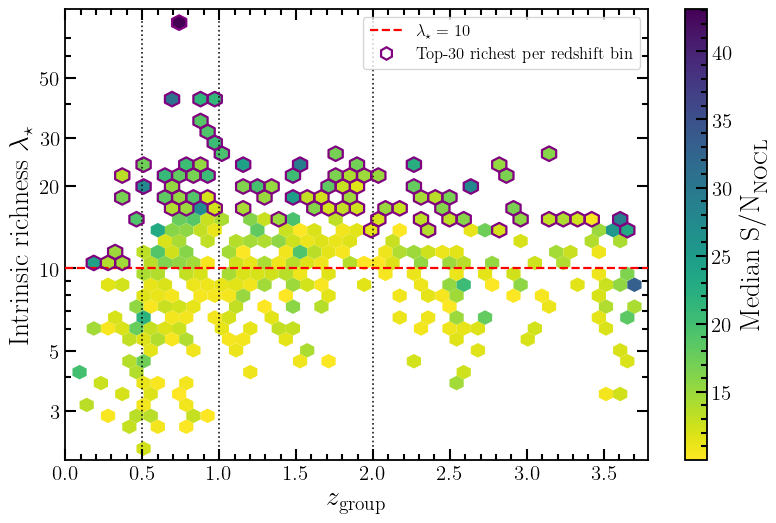}
  \caption{Intrinsic richness $\lambda_\star$ as a function of group redshift for the \CWeb\ \amico\ detections. Hexagonal bins are colored by the median \texttt{SN\_NOCL}. The dashed red line marks the richness limit $\lambda_\star=10$, and purple hexagonal outlines highlight the bins containing the top-30 richest groups retained in each redshift bin for the main analysis. Dotted vertical lines mark the redshift-bin edges ($z=0.5$, 1, and 2). The top-30 selection is applied only above $\lambda_\star>10$.}
  \label{fig:top30_selection}
\end{figure}

To visualize this selection, Fig.~\ref{fig:top30_selection} shows the parent \amico\ group population in the intrinsic-richness--redshift plane. Hexagonal bins are colored by the median \texttt{SN\_NOCL}, and the selected high-richness systems are highlighted. The selection yields 30 groups in each bin at $z>0.5$, with minimum selected richness $\lambda_\star=13.6$--$16.6$ and median richness $\lambda_\star=15.6$--$22.3$. At $0<z<0.5$, only eight systems satisfy $\lambda_\star>10$, so this bin contains eight groups rather than 30. This sample-size asymmetry could bias the comparison if richness or mass distributions differed systematically across redshift bins. The eight low-redshift systems are somewhat less rich ($\lambda_\star=10.4$--$21.5$; median $\simeq13.2$) than the higher-redshift selections (pooled median $\simeq18.8$), corresponding to a median $\log(M_{200}/\Msol)\simeq13.6$ at $z<0.5$ versus $13.7$--$13.8$ at $z>0.5$. This difference is small, and all analyses are performed separately in bins of stellar mass and redshift, decoupling mass-driven evolution from the redshift-dependent environmental signal. In total, the top-30 sample comprises 98 groups and 14\,312 member galaxies, with a median $\log(M_{200}/\Msol)\simeq13.7$ and a range of $13.5$--$14.4$ from the $\lambda_\star$--$M_{200}$ relation of Sect.~\ref{sec:data}. The upper end of this range reaches the poor-cluster regime ($\log(M_{200}/\Msol)>14$); for simplicity, we refer to all these systems as groups.

Before stacking, and after defining the top-30 sample and its matched field controls, we normalize each SFH by its total formed stellar mass\footnote{$\mathrm{SFR}_\mathrm{norm}(\tlb)=\mathrm{SFR}(\tlb)/\Mform$, so that $\int_0^{t_\mathrm{lb,age}}\mathrm{SFR}_\mathrm{norm}(t^\prime)\,\mathrm{d}t^\prime=1$.} to compare their relative shapes independently of galaxy mass. $\mathrm{SFR}_\mathrm{norm}$ is thus the probability density of the formation times of the stellar mass, which we express in $\mathrm{Gyr}^{-1}$. It should not be confused with the specific SFR, which is normalized by the stellar mass at each epoch.

For each $(\Mstar,z)$ bin and environment (group or field), we restrict the stack to the lookback-time interval that is actually sampled by a sufficient number of galaxies. This is not the removal of a fixed SFH bin, nor does it simply discard the first time step. The non-parametric SFHs are piecewise-constant SFRs in a small number of lookback-time bins (10 bins, logarithmically spaced; \citealt{arango-toro_cosmos-web_2025}), which set their actual time resolution; the 1 Myr grid is only a common sampling used to stack galaxies whose SFH bin edges differ. Because galaxies in the same redshift bin can have slightly different ages (i.e., maximum accessible lookback times), the oldest part of the grid can be populated by only a few objects. We therefore retain only time steps at which at least 10 galaxies have a defined SFH (i.e., within their maximum accessible lookback time), thereby defining a common, well-sampled lookback-time range before computing the stack.
Stacked trends are quantified with $n_\mathrm{boot}=1000$ realizations. In each realization, the group-stacked SFH is computed as the \texttt{ASSOC\_PROB}-weighted mean SFH at each lookback-time step, while the matched field stack, built from the field sample of Sect.~\ref{sec:membership} (galaxies excluded from the \amico\ group members), is computed as an unweighted mean. For the error budget, we add in quadrature the bootstrap uncertainty of the stacked SFH and the \cigale\, Bayesian SFH bin uncertainty propagated to the same weighted (groups) or unweighted (field) stack.
Uncertainty envelopes represent the resulting 16th--84th percentile range. They quantify sampling uncertainty and posterior uncertainty within the adopted \cigale\ model grid but do not account for systematic uncertainties associated with the different SFH priors and model parameterizations or with alternative SED-fitting assumptions, which are discussed in detail in Sect.~\ref{sec:discussion}. Stellar mass bin selection also follows the common mass-completeness criterion defined in Sect.~\ref{sec:sample} before applying the adopted stellar-mass-bin boundaries.

\subsection{Formation times $t_{10}$--$t_{90}$ from reconstructed SFHs}
To compare galaxies at equivalent stages of their SFH evolution, we derive formation-time percentiles directly from the reconstructed non-parametric SFHs.
Because the SFHs are expressed in lookback time, we accumulate formed mass from the oldest lookback time to the epoch of interest,
\begin{equation}
  \Mform(\tlb) = \int_{t_\mathrm{lb}}^{t_{\mathrm{lb,age}}} \mathrm{SFR}(t')\,\mathrm{d}t', \label{eq:Mform}
\end{equation}
where $t_\mathrm{lb}\in[0,t_{\mathrm{lb,age}}]$ is lookback time; $\Mform(\tlb)$ is the mass formed before lookback time $\tlb$, and $\Mform\equiv\Mform(0)$ is the total formed mass.
Here $t_{\mathrm{lb,age}}$ is the maximum lookback time available for each galaxy and operationally marks the onset of the first resolved SFH episode (i.e., the earliest star formation traced by the reconstruction).
We then define $t_{\rm f}$ as the lookback time satisfying
\begin{equation}
  \frac{\Mform(t_{\rm f})}{\Mform(0)}=f,
\end{equation}
where $f$ is the formed mass fraction, set to $f=0.1,\,0.5,\,0.9$ for $t_{10}$, $t_{50}$, and $t_{90}$, respectively.
Thus, $t_{10}$ traces the earliest star formation, $t_{50}$ is the median age of the stellar mass formed by $t_\mathrm{now}$, and $t_{90}$ traces the late tail of mass growth. Throughout this paper, ``assembly'' refers to the formation of stellar mass as traced by the SFH, not to its accretion through mergers.

This parametrization has two practical advantages: it compresses each SFH into physically interpretable reference times, directly comparable across redshift, mass, and environment; and it lets us compare recent evolution consistently across galaxies, since the same $f$ reference is used for all of them. Formation times therefore bridge the full SFH information and the migration-based summary statistics introduced below.

\subsection{Migration vector anchored to $t_{50}$ \label{sec:mig-vec}}
While the formation times capture when a galaxy assembled its stellar mass, they do not show whether its star formation has since been sustained, enhanced, or suppressed. We quantify this with the migration vector of \citet{arango-toro_cosmos-web_2025}, here anchored at $t_{50}$, which measures the displacement of a galaxy in the $\Mform\mbox{--}\mathrm{SFR}$ plane between $t_{50}$ and $t_\mathrm{now}$ and provides a common, galaxy-specific reference epoch.

\begin{figure}[h]
  \centering
  \includegraphics[width=\columnwidth]{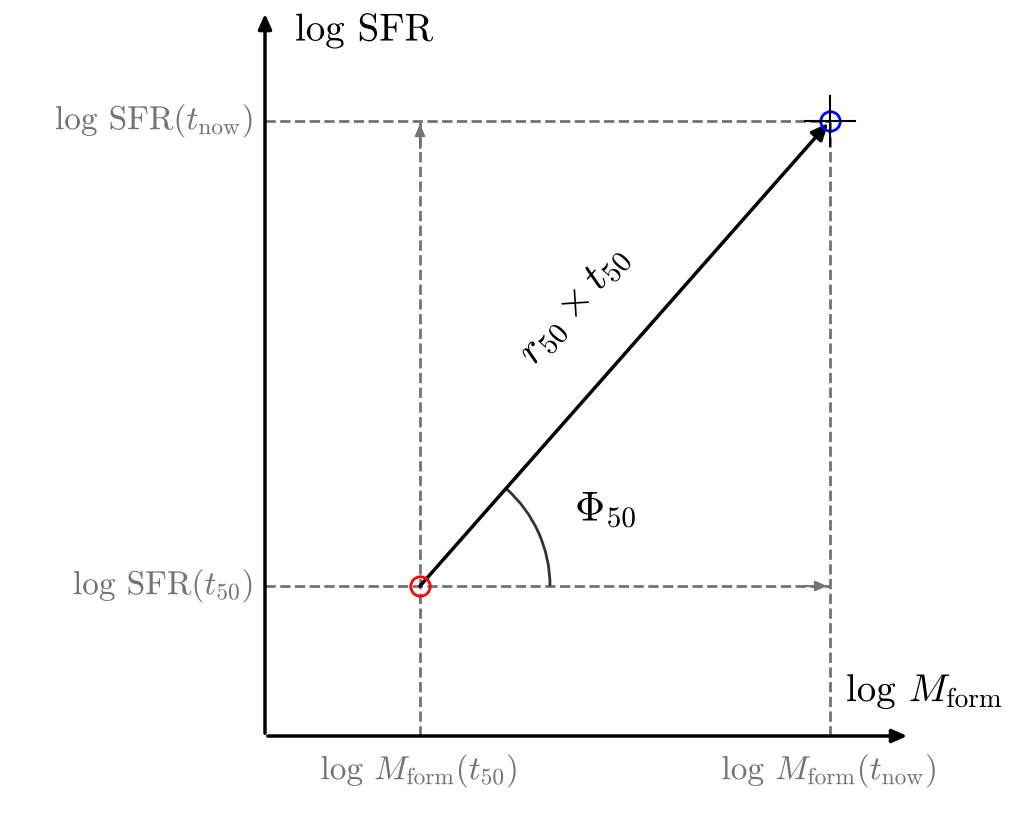}
  \caption{Schematic illustration of the migration-vector formalism anchored to $t_{50}$. The vector is defined in the log--log $\Mform\mbox{--}\mathrm{SFR}$ plane between the galaxy state at $t_{50}$, when 50\% of the final formed stellar mass is in place, and the observed state $t_\mathrm{now}$, corresponding to $t_\mathrm{lb}=0$. Its Cartesian components are given by $\Delta\log \Mform$ and $\Delta\log \mathrm{SFR}$, its orientation is quantified by $\Phi_{50}=\arctan(\Delta\log \mathrm{SFR}/\Delta\log \Mform)$, and its amplitude by $r_{50}\times t_{50}=\sqrt{(\Delta \log \Mform)^2 + (\Delta \log \mathrm{SFR})^2}$. Since $\Delta\log\Mform=\log2$ by construction, positive (negative) $\Phi_{50}$ corresponds to a higher (lower) SFR at $t_\mathrm{now}$ than at $t_{50}$; $|\Phi_{50}|=45^\circ$ marks a change by a factor of two.}
  \label{fig:migration_vector_t50}
\end{figure}

We denote the observation time by $t_\mathrm{now}$, corresponding to $t_\mathrm{lb}=0$. For each galaxy, we measure the displacement from $t_{50}$ to $t_\mathrm{now}$ in the log--log $\Mform\mbox{--}\mathrm{SFR}$ plane (Fig.~\ref{fig:migration_vector_t50}):
\begin{align}
  & \Delta \log \Mform = \log \Mform(0) - \log \Mform(t_{50}),\\
  & \Delta \log \mathrm{SFR} = \log \mathrm{SFR}(t_\mathrm{now}) - \log \mathrm{SFR}(t_{50}).
\end{align}
Because $t_{50}$ is defined as the lookback time at which half of the stellar mass formed by $t_\mathrm{now}$ was already in place, $\Mform(t_{50})=\Mform(0)/2$ by construction, and $\Delta\log\Mform=\log2\simeq0.30$ dex for every galaxy. The migration vector therefore compares all galaxies over the same fractional stellar-mass growth (a factor of two). From these components, we define\footnotemark
\begin{equation}
  \Phi_{50}=\arctan\left(\frac{\Delta \log \mathrm{SFR}}{\Delta \log \Mform}\right),
\end{equation}
\begin{equation}
  r_{50}=\frac{\sqrt{(\Delta \log \Mform)^2 + (\Delta \log \mathrm{SFR})^2}}{t_{50}}. \label{eqn:amplitude}
\end{equation}
\footnotetext{Although $\Delta \log \Mform$ and $\Delta \log \mathrm{SFR}$ originate from quantities with different physical units, here they are combined only as coordinates in the $(\log \Mform,\log \mathrm{SFR})$ plane. Accordingly, $\Phihalf$ and $r_{50}\times t_{50}$ should not be interpreted as a physical quantity with direct mass--SFR units, but as the Euclidean norm and direction of a displacement in log--log space, expressed in dex.}

Unlike an integrated change in SFR or specific SFR (sSFR; SFR per unit stellar mass) over a fixed time interval, the migration vector sets the baseline by each galaxy's own assembly: all galaxies are compared over the same fractional stellar-mass growth, from $t_{50}$ to $t_\mathrm{now}$, regardless of how long it took. Since $\Delta\log\Mform=\log2$ for every galaxy, $\Phihalf$ is a monotonic, bounded transformation of the SFR change alone, $\Delta\log\mathrm{SFR}=\log2\,\tan\Phihalf$ (e.g., $|\Phihalf|=45^\circ$, $70^\circ$, and $80^\circ$ correspond to changes of $0.3$, $0.8$, and $1.7$ dex), so our directional results can equivalently be read in terms of $\Delta\log\mathrm{SFR}$. We retain the angle because (i) it is bounded, so that galaxies with very low and poorly constrained present-day SFRs saturate near $\pm90^\circ$ instead of dominating the tail of $\Delta\log\mathrm{SFR}$ (Sect.~\ref{sec:result-staked-SFHS}); (ii) it allows a direct comparison with the main-sequence calibration of \citet{arango-toro_cosmos-web_2025}; and (iii) it is complemented by $\rhalf$, which, normalized by $t_{50}$, measures the rate of change. Since the angle is insensitive to large changes ($80^\circ$--$90^\circ$ covers all changes above $1.7$ dex), we do not interpret differences between large angles quantitatively. Note that $\Phihalf$ does not by itself identify a unique physical driver: a galaxy can remain on the main sequence while having $\Phihalf<0$ if its SFR decreases along the evolving sequence.

We report directional results in terms of $\Phi_{50}$ and treat $r_{50}$ as the trajectory-amplitude metric. As a robustness check against small-amplitude SFH fluctuations producing apparently large positive or negative $\Phihalf$ values, we repeated the group--field analysis after imposing minimum-excursion cuts of $r_{50}>0$, $0.3$, $0.4$, and $0.6$ dex/Gyr on the group sample and on the matched field controls. These cuts do not change the mass, redshift, radial, or migration-vector trends discussed below; they only reduce the confidence intervals nearly proportionally. This is expected because low-$r_{50}$ systems are mainly main-sequence (MS) galaxies, which in the \CWeb\ migration-vector calibration have median angles close to zero ($-30^\circ\lesssim\Phi_{50}^{\rm MS}\lesssim+35^\circ$) and contribute mainly a broad, approximately centered dispersion component \citep{arango-toro_cosmos-web_2025}.

\begin{figure}[h]
  \centering
  \includegraphics[width=\columnwidth]{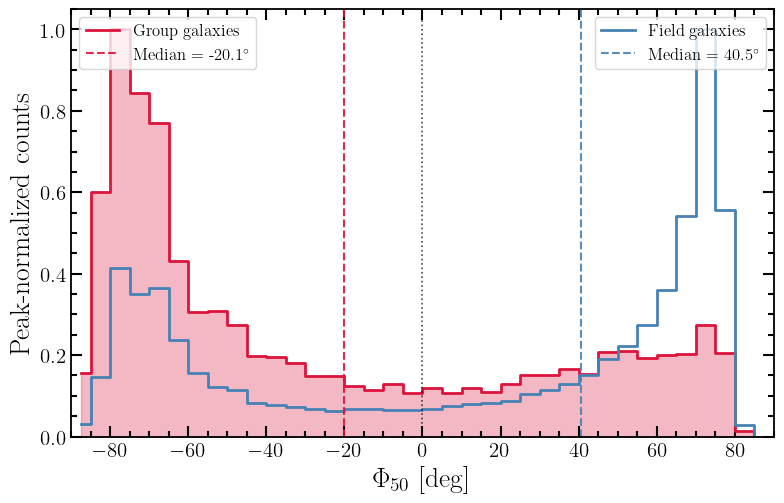}
  \caption{Peak-normalized $\Phi_{50}$ distributions for group (red) and field (blue) samples (matched in mass and redshift). Positive (negative) $\Phi_{50}$ values indicate rising (declining, quenching-oriented) recent SFHs.}
  \label{fig:bimodal_phi}
\end{figure}

\subsection{Bimodality framework in $\Phi_{50}$ for group versus field galaxies \label{sec:meth-bimodal}}
We define four redshift bins: $0.0<z\leq0.5$, $0.5<z\leq1.0$, $1.0<z\leq2.0$, and $2.0<z\leq3.7$.
In each of them we report normalized distributions of $\Phi_{50}$ separately for group and matched-field samples after the mass-completeness and membership-quality cuts described in Sects.~\ref{sec:data} and~\ref{sec:sample}.

\begin{figure*}
  \centering
  \includegraphics[width=\textwidth]{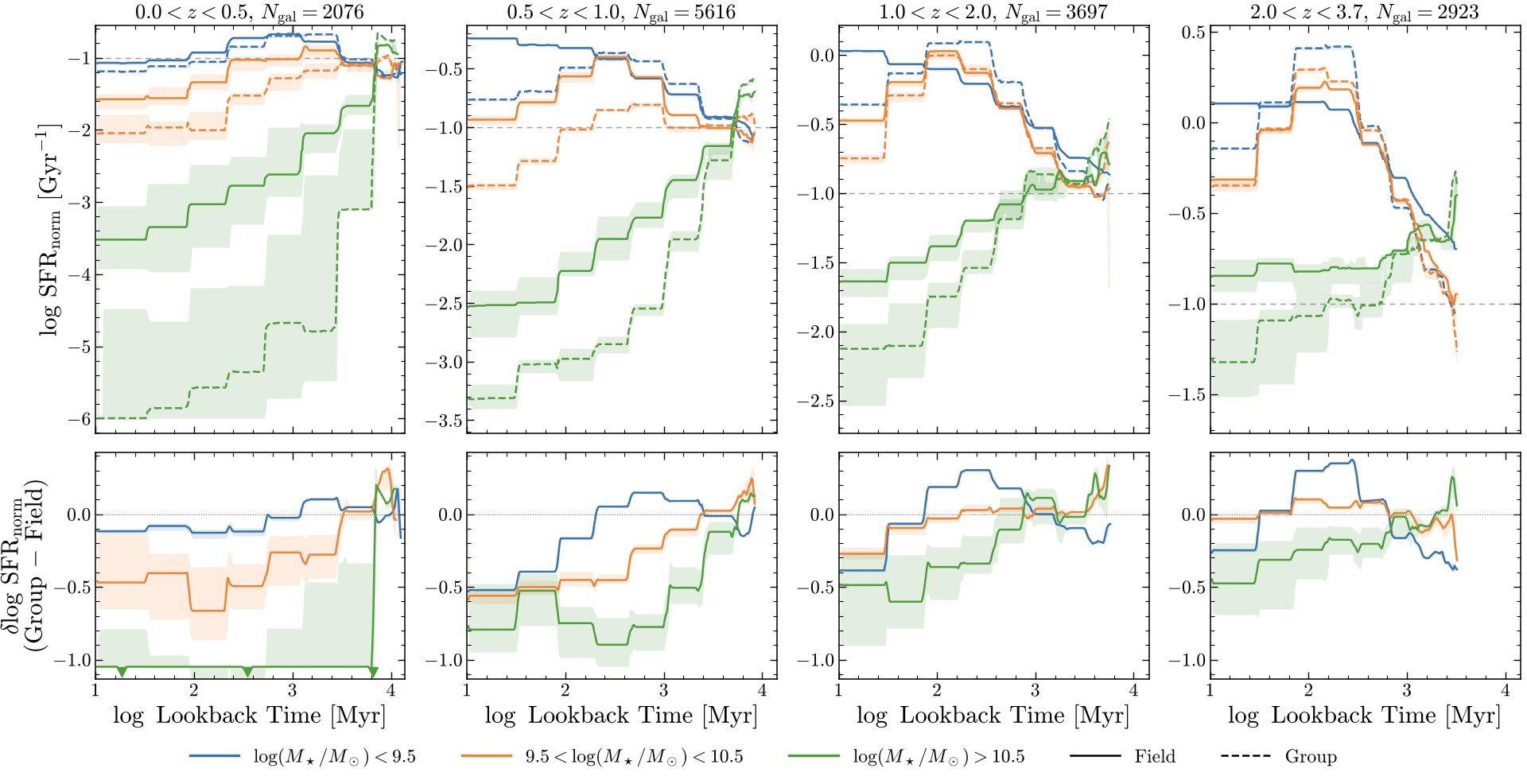}
  \caption{Stacked normalized SFHs ($\mathrm{SFR}_\mathrm{norm}(\tlb)=\mathrm{SFR}(\tlb)/\Mform$) (upper panels) for the top-30 richest groups in each redshift bin (dashed lines) compared to matched field controls (solid lines), split by stellar-mass bin (colors). $N_\mathrm{gal}$, given at the top of each column, is the number of group galaxies in each redshift bin, equal to the number of matched field galaxies. The gray dashed line in the upper panels marks $\log\mathrm{SFR}_\mathrm{norm}=-1$ to guide the eye. Shaded bands show the bootstrap 16th--84th percentile range. All curves are plotted on a common grid of lookback times; the $y$-axis scale differs among redshift bins. The lower panels show the group--field offset, $\delta\log\mathrm{SFR}_\mathrm{norm}\equiv\log\mathrm{SFR}_{\rm norm,\,group}-\log\mathrm{SFR}_{\rm norm,\,field}$, for each stellar-mass bin; negative (positive) values indicate lower (higher) $\mathrm{SFR}_\mathrm{norm}$ in groups than in the matched field sample, and the dotted horizontal line marks zero offset. Offsets below $-1$ dex are clipped at the lower plot limit (downward triangles). }
  \label{fig:topN_sfh}
\end{figure*}

Figure~\ref{fig:bimodal_phi} shows the global $\Phi_{50}$ distributions of the group and field samples (matched in $\Mstar$ and $z$, all stellar masses and redshifts combined), peak-normalized to unit maximum.
Both samples contain $N_{\rm group}=N_{\rm field}=14\,312$ galaxies, with medians $\widetilde{\Phi}_{50}^{\rm group}\simeq-20^\circ$ and $\widetilde{\Phi}_{50}^{\rm field}\simeq+40^\circ$.
Positive (negative) $\Phi_{50}$ values trace rising (declining) SFRs since $t_{50}$, the former including recent starburst-like episodes and the latter being more consistent with recent quenching. The strong peaks around $\pm80^\circ$ do not mean that most galaxies lie far from the main sequence: $\Phi_{50}$ describes the direction of evolution between $t_{50}$ and the present, not the instantaneous position in the $\log \Mform$--$\log\mathrm{SFR}$ plane, so these peaks trace galaxies whose SFR rose or fell by $\gtrsim1.7$ dex since $t_{50}$ (Sect.~\ref{sec:mig-vec}). The field distribution is mostly skewed toward the rising-SFH side while still retaining a non-negligible declining-SFH component; the group distribution shows the opposite balance. This population contrast motivates a bimodal analysis.

In each $(\Mstar,z)$ bin and environment, we smooth the $\Phi_{50}$ distribution with a moving average and split it at $\Phihalf=0^\circ$. We then compute the fractions of the distribution on the positive and negative sides, denoted $\rho^+$ and $\rho^-$.
From these, we define the bounded summary statistic:
\begin{equation}
  A_\rho\equiv\frac{\rho^+-\rho^-}{\rho^+ + \rho^-}. \label{eqn:A_rho}
\end{equation}
Here, $A_\rho$ traces the sign imbalance of the $\Phi_{50}$ distribution (the sign of $\Phi_{50}$ being that of $\Delta\log\mathrm{SFR}$ between $t_{50}$ and $t_\mathrm{now}$): $A_\rho>0$ indicates a population dominated by galaxies with rising recent SFRs, and $A_\rho<0$ one dominated by quenching-oriented galaxies, i.e., with a declining SFR over the $t_{50}\rightarrow t_\mathrm{now}$ interval. It is our primary metric of the balance between star-forming-oriented and quenching-oriented galaxies.

We report (i) the redshift evolution of $A_\rho$ for group and field samples in each ($\Mstar,z$) bin and (ii) the resulting contrast in star-formation activity between groups and the field, i.e., how much more or less quenching-dominated groups are than matched field galaxies ($\Delta A_{\rho}\equiv A_{\rho}^{\rm group}-A_{\rho}^{\rm field}$).
All side-specific quantities are reported only when that $\Phi_{50}$ side has a peak detected at $\geq 1\sigma$ above its bootstrap uncertainty; when a side is not significant, we interpret it as a non-detection.

Uncertainties are quantified with bootstrap resampling within each $(\Mstar,z,\mathrm{env})$ bin (preserving group membership weights), and we quote the median and 16th--84th percentile interval for all derived statistics (e.g., $A_{\rho,\,p16}$, $A_{\rho,\,p50}$, $A_{\rho,\,p84}$).
Group statistics are weighted by \texttt{ASSOC\_PROB} (Sect.~\ref{sec:membership}), so membership uncertainty is propagated at first order.

\begin{figure*}
  \centering
  \begin{minipage}[t]{1.0\textwidth}
    \centering
    \includegraphics[width=\textwidth]{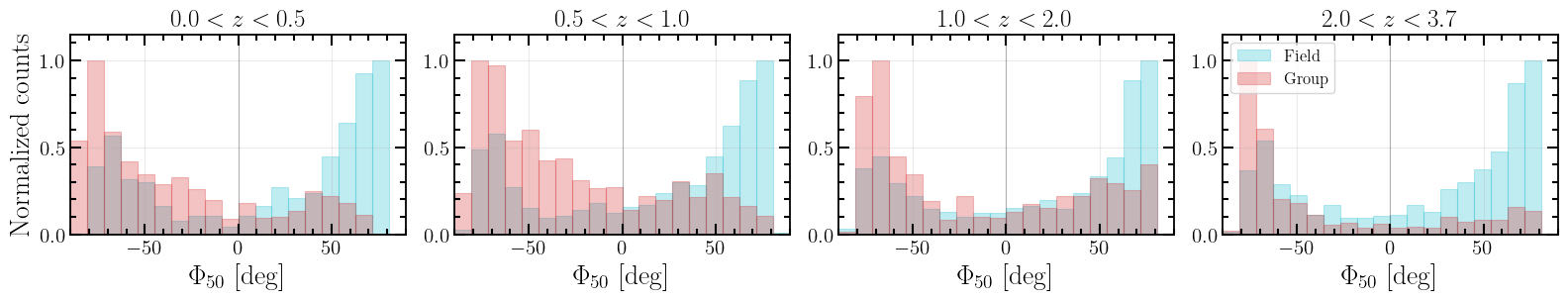}

    \small (a) $\Phihalf$ distribution for the low mass bin at $8.1<\log(\Mstar/\Msol)<9.5$
  \end{minipage}

  \vspace{2mm}
  \begin{minipage}[t]{1.0\textwidth}
    \centering
    \includegraphics[width=\textwidth]{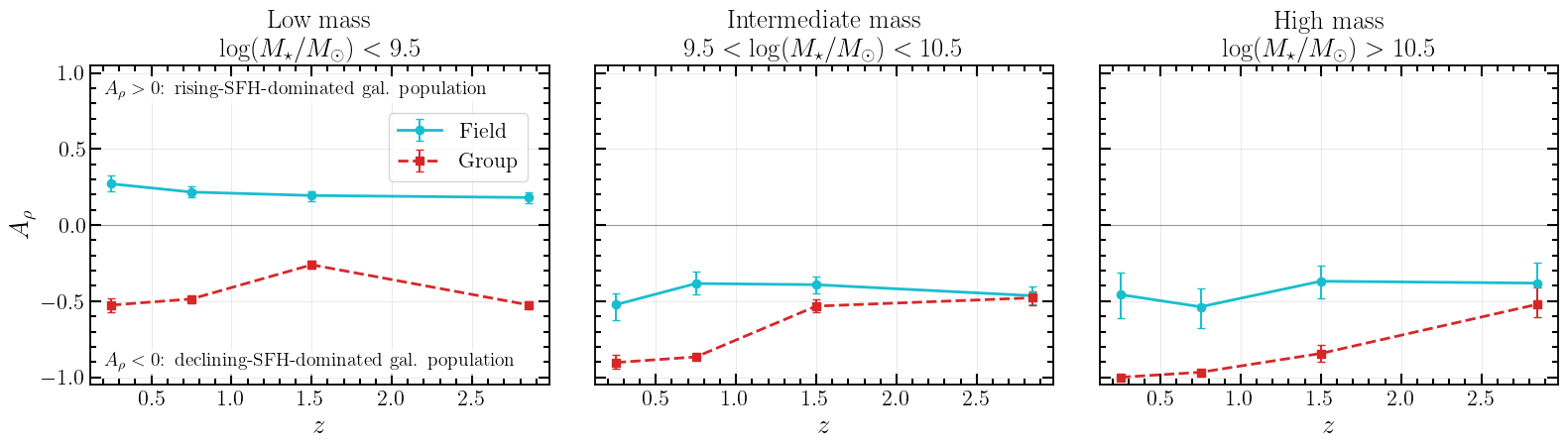}

    \small (b) Redshift evolution of $A_\rho$ in all mass bins
  \end{minipage}
  \caption{Top-30 richest-group migration-angle diagnostics. The upper panels (a) show the $\Phi_{50}$ peak-normalized distributions for the low-mass sample. The bottom panels (b) report the redshift and stellar-mass evolution of $A_\rho$ which quantifies the balance between rising and declining SFHs in samples of groups (red scatter dots and dashed lines) and field (blue scatter dots and solid lines) galaxies. The figure highlights the shift of group galaxies toward declining SFHs at low $z$.}
  \label{fig:topN_theta_triptych}
\end{figure*}

\section{Results \label{sec:results}}
All results are based on the members of the top-30 sample (Sect.~\ref{sec:methods_staked-SFHs}) and on their matched field controls (Sect.~\ref{sec:membership}), in bins of redshift and stellar mass.

\subsection{Stacked SFHs \label{sec:result-staked-SFHS}}
Figure \ref{fig:topN_sfh} compares the stacked SFHs of galaxies in the top-30 richest groups (dashed lines) with matched field controls (solid lines); the lower panels show $\delta \log \mathrm{SFR}_\mathrm{norm} \equiv \log \mathrm{SFR}_{\rm norm,\,group}-\log \mathrm{SFR}_{\rm norm,\,field}$, with negative values indicating lower $\mathrm{SFR}_\mathrm{norm}$ in groups at a given lookback time. Note that the ``recent'' part of an SFH refers to lookback times near zero at each galaxy's own redshift, not to $z=0$.

The most striking feature of Fig.~\ref{fig:topN_sfh} is that massive group galaxies ($\log\,(\Mstar/\Msol)>10.5$) have lower $\mathrm{SFR}_\mathrm{norm}$ than their field counterparts over most of the lookback-time range and in all redshift bins, with offsets reaching $\delta\log\mathrm{SFR}_\mathrm{norm}\lesssim-1$ dex at $z<0.5$. The two environments are similar at the oldest lookback times, and the divergence builds up during the subsequent evolution. In both environments, these SFHs decline early and monotonically, as expected from downsizing, whereby massive galaxies formed their stars earlier \citep[e.g., ][]{downsizing_Cowie_1996,Downsizing_Carnall_2019,Downsizing_Leung_2026}. Because these populations are already largely quenched, their recent normalized SFHs are very low, so large logarithmic offsets can arise from residual SFRs that are only weakly constrained by SED fitting; this bin also includes the BGGs (Sect.~\ref{sec:sample}) and has the largest stacked-SFH uncertainties. Low- to intermediate-mass galaxies ($\log\,(\Mstar/\Msol)<10.5$) show a smaller but better-constrained recent deficit ($\delta\log\,\mathrm{SFR}_\mathrm{norm}<0$ at $t_\mathrm{lb}\lesssim100$ Myr) at $z<2$, indicating environmental suppression; the signal weakens and becomes more heterogeneous at $2<z<3.7$, where occasional enhancement is also seen.

This deficit differs in detail between the two mass bins. In the low-mass bin ($\log\,(\Mstar/\Msol)<9.5$), the recent negative offset (modest, $\sim-0.1$ dex at $z<0.5$, strengthening to $\sim-0.25$ to $-0.5$ dex at $z>0.5$) is preceded, at earlier lookback times, by a positive offset that grows with redshift (from $\sim+0.1$ dex at $z<0.5$ to $\sim+0.4$ dex at $2<z<3.7$) and shifts to smaller lookback times ($\log(\tlb/{\rm Myr})\sim3$ at $z<0.5$ down to $\sim1.8$ at $2<z<3.7$). This is a structured signature combining late suppression with earlier excess, rather than a single monotonic quenching sequence. The intermediate-mass bin ($9.5<\log(\Mstar/\Msol)<10.5$) shows a cleaner, suppression-dominated signal, with the deficit reaching $\simeq-0.7$ dex at $z<0.5$ and $\simeq-0.55$ dex at $0.5<z<1$. At $1<z<2$, the deficit is confined to the most recent SFH bin ($\simeq-0.3$ dex at $\tlb\lesssim30$ Myr), and at $2<z<3.7$ the offset is close to zero (at most $+0.1$ dex, likely reflecting early similarity rather than reversal).

In summary, the largest group--field offsets occur at high mass, but they concern populations already quenched in both environments and are the least constrained. The best-constrained environmental signal is found at low to intermediate mass: the intermediate-mass bin shows a deficit of up to $\simeq-0.7$ dex at $z<1$, while the low-mass deficit is strongest at $0.5<z<1$, persists up to $2<z<3.7$, and is preceded by an earlier excess that grows with redshift.

\subsection{Migration direction \label{sec:results-phihalf}}
For the top-30 richest groups, we analyze the bounded $\Phi_{50}\in[-90^{\circ},90^{\circ}]$ distributions (see Sect.~\ref{sec:methods}) for group-member galaxies and matched field controls, using the common stellar-mass completeness floor defined in Sect.~\ref{sec:sample} so that the two populations are comparable over the same accessible mass range at all epochs.

Figure~\ref{fig:topN_theta_triptych}, upper panels (a), shows the normalized $\Phi_{50}$ distributions for group members (red) and field-matched samples (blue) across redshift bins in the low-mass regime, $\log(\Mstar/\Msol)<9.5$; the bottom panels (b) show the redshift evolution of $A_\rho$ (Eq.~\ref{eqn:A_rho}) for all three stellar-mass bins, with $A_\rho>0$ ($A_\rho<0$) indicating that more than 50\% of the population shows rising (declining) SFHs.

In the low-mass bin, the field remains rising-dominated at all redshifts ($A_\rho\simeq0.19$--$0.28$, i.e., $\sim60$--$64\%$ rising), while the group population is declining-dominated ($A_\rho\simeq-0.27$ to $-0.52$, i.e., $\sim64$--$76\%$ declining); the contrast is weakest at $1<z<2$, where the field stays stable but the group declining fraction drops to $\sim64\%$. In the intermediate-mass bin, both samples are already declining-dominated, but groups are systematically more negative than the field at $z<2$ (field $A_\rho\simeq-0.38$ to $-0.52$; groups $A_\rho\simeq-0.87$ to $-0.90$ at $z<1$), giving the clearest group--field separation at low redshift; this weakens at $1<z<2$ ($A_\rho\simeq-0.53$) and the two samples converge by $2<z<3.7$ (both $\sim70\%$ declining). The high-mass bin is the most declining-dominated of all: the field stays negative throughout ($A_\rho\simeq-0.37$ to $-0.54$), while the group value reaches near-saturation ($A_\rho\simeq-1$ at $z<1$, i.e., almost the entire population on the declining branch) before converging toward the field value at the highest redshift ($A_\rho\simeq-0.52$); given the small, broad low-redshift high-mass field sample, this contrast should be read cautiously.

In summary, group galaxies are systematically shifted toward the declining-SFH branch relative to the field, most clearly at $z<1$ and in the low- and intermediate-mass bins. The high-mass bin is already declining-dominated in both environments. This group--field contrast weakens toward the highest redshift bin at intermediate and high mass, whereas low-mass group galaxies remain declining-dominated against a rising-dominated field at all redshifts.

\begin{figure}[h]
  \centering
  \includegraphics[width=1.0\linewidth]{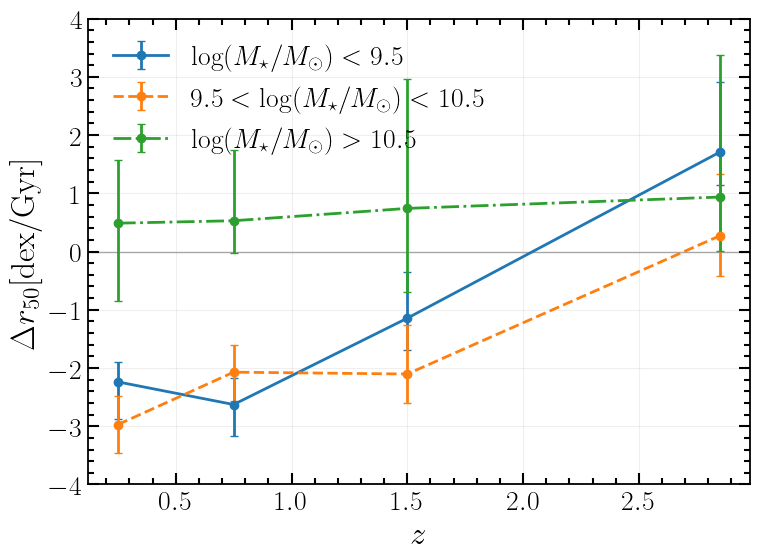}
  \caption{Redshift evolution of the migration-amplitude offset $\Delta r_{50}\equiv r_{50}^{\mathrm{group}}-r_{50}^{\mathrm{field}}$ for galaxies on the negative-$\Phi_{50}$ (declining) branch, restricted to the peak-centered 68\% angular interval around the negative $\Phi_{50}$ peak. Positive (negative) $\Delta r_{50}$ indicates faster (slower) declining SFHs in groups than in the matched field.}
  \label{fig:velocity}
\end{figure}

\subsection{Migration amplitude \label{sec:result-migration_amplitude}}
We next test whether environment affects not only the direction of recent SFH change ($\Phihalf$) but also its amplitude, $r_{50}$ (Eq.~\ref{eqn:amplitude}), measured within the 16th--84th percentile range of the $\Phi_{50}<0$ (declining) distribution. We quantify how much faster or slower groups decline than the field as $\Delta \rhalf\equiv \rhalf^{\mathrm{group}}-\rhalf^{\mathrm{field}}$, with positive (negative) values indicating faster (slower) declining SFHs in groups than in the field.

Figure~\ref{fig:velocity} shows the redshift evolution of median $\Delta r_{50}$ in three stellar-mass bins. In the low-mass bin, $\Delta r_{50}$ is negative at $z<2$, about $-2.3$ and $-2.6$ dex/Gyr at $z<0.5$ and $0.5<z<1$, weakening to $\sim-1.1$ dex/Gyr at $1<z<2$. It turns positive at $2<z<3.7$ ($\Delta r_{50}=+1.6^{+1.0}_{-0.5}$ dex/Gyr). The intermediate-mass bin follows the same pattern, but with a larger negative offset, $-3.0$ to $-2.0$ dex/Gyr at $z<2$, before becoming consistent with zero at $2<z<3.7$ ($\Delta r_{50}=+0.3^{+0.9}_{-0.4}$ dex/Gyr). In both bins, then, even where groups are declining-dominated (Sect.~\ref{sec:results-phihalf}), their decline is \emph{slower} than in the field at $z<2$. This contrast disappears, or reverses, at $2<z<3.7$. The high-mass bin shows only modest offsets at all redshifts ($|\Delta r_{50}|\lesssim1$ dex/Gyr), though broad, asymmetric percentile ranges (up to $\sim3$ dex/Gyr) leave room for a non-negligible faster-declining subpopulation; given the small high-mass samples, this is weak evidence at best.

Taken together, $\Phihalf$ and $\rhalf$ must be read jointly. At $z<2$, low- and intermediate-mass group galaxies are more often evolving toward a declining SFR (Sect.~\ref{sec:results-phihalf}), but they decline more slowly than their field counterparts. At $2<z<3.7$, this amplitude contrast becomes weaker and mass-dependent. The picture is therefore not a single quenching channel at all epochs, but a transition: from a low-redshift regime of frequent, slow decline, to a more heterogeneous, mass-dependent high-redshift regime.

\begin{figure*}[t]
  \centering
  \includegraphics[width=\textwidth]{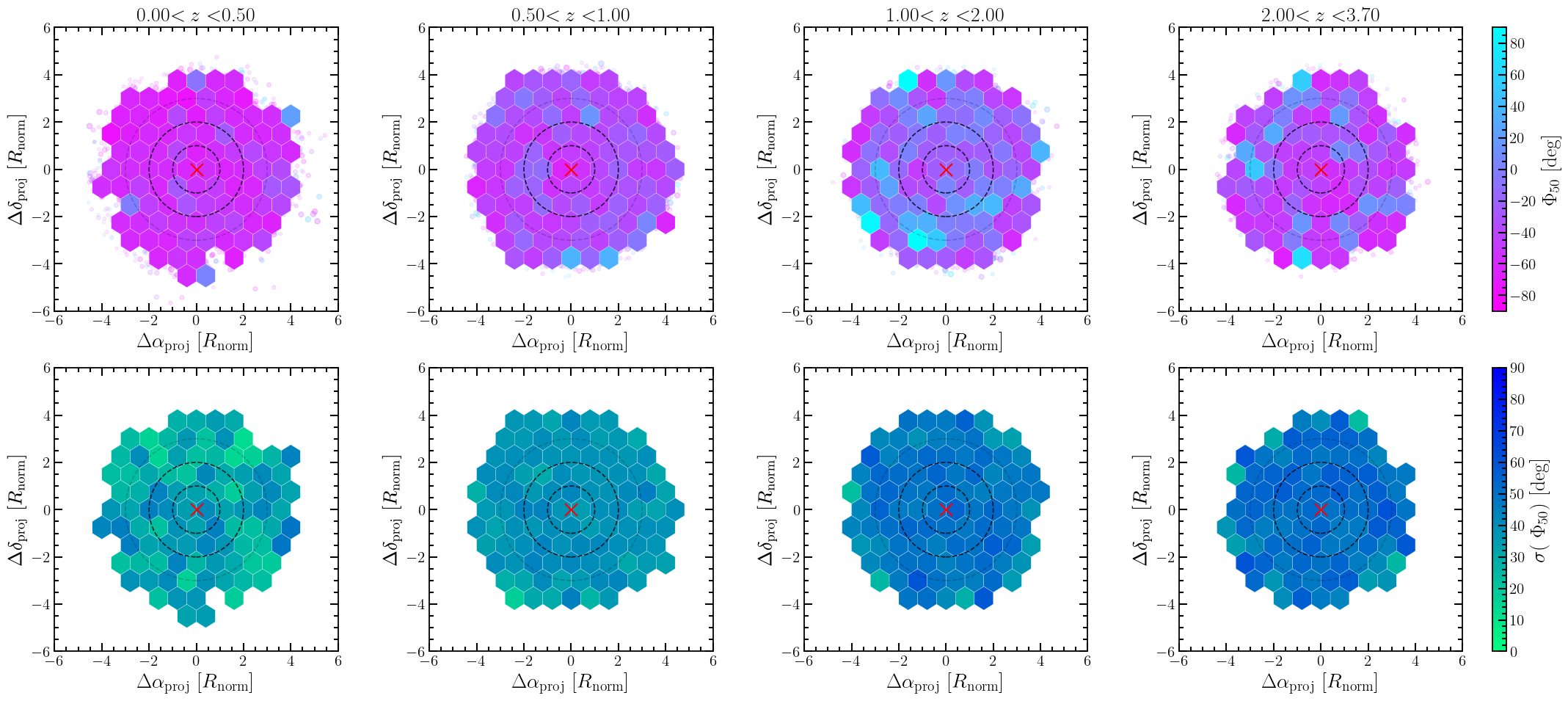}
  \caption{Spatial distribution of galaxies in the top-30 richest groups at different redshift bins, shown in normalized group-centric coordinates ($R_\mathrm{norm}$); $\Delta\alpha_\mathrm{proj}$ and $\Delta\delta_\mathrm{proj}$ are the projected offsets in right ascension and declination from the group barycenter, in units of $R_\mathrm{norm}$. Scatter points mark individual galaxies and are color-coded by $\Phihalf$, which traces the recent SFH evolution over the last half-mass assembly phase of each galaxy. Positive and negative $\Phihalf$ values indicate, respectively, rising and declining recent SFH. Hexagonal bins summarize the median $\Phihalf$ (upper panels) and local dispersion $\sigma(\Phihalf)$ (lower panels), requiring at least 10 galaxies per bin. Dashed circles indicate the inner ($R_\mathrm{norm}<1$), outer ($1<R_\mathrm{norm}<2$), and distant-outskirts ($R_\mathrm{norm}>2$) radial regimes.}
  \label{fig:top3_distri_radial}
\end{figure*}

\subsection{Group-centric radial dependence of SFH evolution}
\label{sec:results_radial}

We now examine how the SFH evolution of group members varies with distance from the group center, comparing both migration-vector components, $\Phihalf$ and $\rhalf$ (Sect.~\ref{sec:mig-vec}), as a function of normalized group-centric radius $\Rnorm$ (Sect.~\ref{sec:sample-radial}, Eq.~\ref{eqn:Rnorm}).

Figure \ref{fig:top3_distri_radial} gives a first visual summary, stacking the top-30 richest groups in normalized projected coordinates for all group members within the common mass-completeness cut (see the caption for the hexbin and radial-regime conventions). The pattern is redshift dependent. At $z<0.5$, the map is almost uniformly negative ($\Phihalf<0$) across the full group extent, indicating quenching-oriented SFHs with no clear inner--outer structure. At $z>0.5$, inner regions typically become more negative than the outskirts, which grow patchier and less regular. The dispersion $\sigma(\Phihalf)$ (lower panels) shows little radial dependence but a clear redshift trend, from low dispersion at $z<1$ to a more heterogeneous population at $z>1$.

\begin{figure}[h]
  \centering
  \includegraphics[width=\columnwidth]{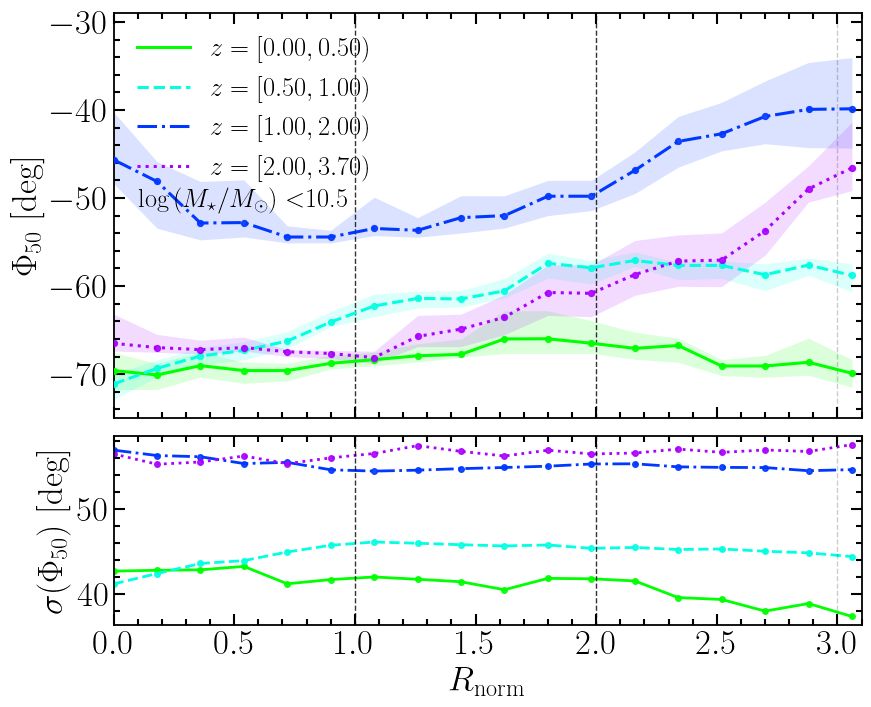}
  \caption{Radial profile of the recent SFH orientation metric $\Phihalf$ for galaxies with $\log(\Mstar/\Msol)<10.5$ in the top-30 richest groups, shown as a function of normalized group-centric radius $\Rnorm$ in redshift bins. Curves trace the median trend, and the shaded regions represent the uncertainties on the median estimated through bootstrap resampling. The three radial regimes are inner ($\Rnorm<1$), outer ($1<\Rnorm<2$), and distant outskirts ($\Rnorm>2$). The lower panel shows the dispersion $\sigma(\Phihalf)$.}
  \label{fig:full_phi_radial}
\end{figure}

Figure~\ref{fig:full_phi_radial} extracts this radial signal as median profiles for the low- to intermediate-mass population ($\log(\Mstar/\Msol)<10.5$). At $z<0.5$, the profile is flat at $\Phihalf\simeq-70^\circ$ to $-66^\circ$ across radius, consistent with no inner--outer gradient. At $z>0.5$, the median $\Phihalf$ is more negative in the inner regions and less negative toward larger radii: at $0.5<z<1$ it rises from $\simeq-71^\circ$ to $\simeq-57^\circ$ within $\Rnorm\lesssim1.5$ and then flattens, while at $1<z<2$ and $2<z<3.7$ the rise occurs mainly beyond $\Rnorm\simeq1$, reaching $\simeq-40^\circ$ and $\simeq-47^\circ$ near $\Rnorm\simeq3$, respectively. The dispersion is nearly independent of radius but increases with redshift, from $\sigma(\Phihalf)\simeq40^\circ$--$43^\circ$ at $z<0.5$ to $\simeq57^\circ$ at $2<z<3.7$, indicating a more heterogeneous population at early times.

\begin{figure}[h]
  \centering
  \includegraphics[width=\columnwidth]{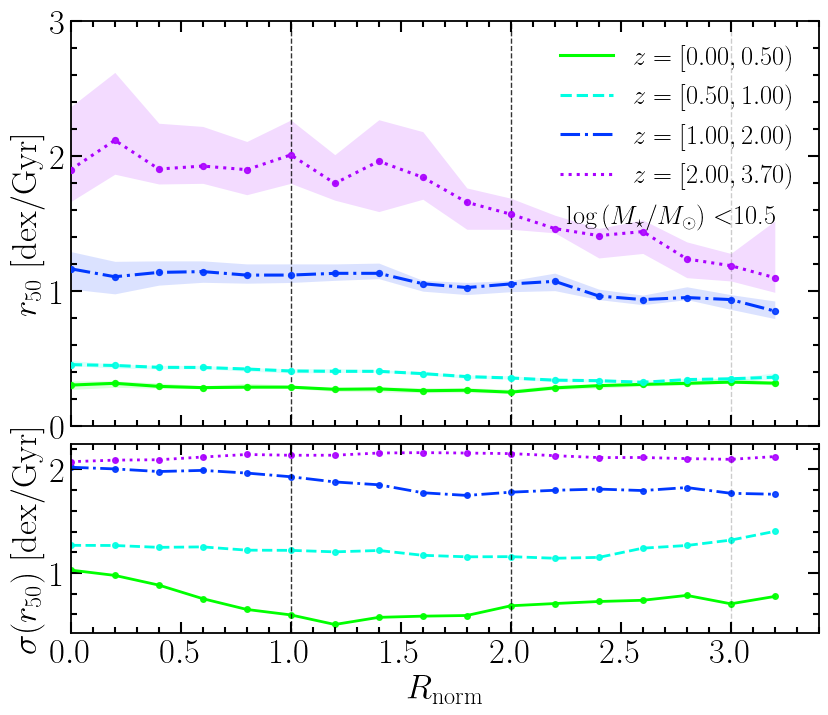}
  \caption{Same as Fig.~\ref{fig:full_phi_radial} but for the recent SFH amplitude metric $\rhalf$ in the low- to intermediate-mass population ($\log(\Mstar/\Msol)<10.5$). The upper panel shows the median radial profile, and the lower panel shows the dispersion $\sigma(\rhalf)$.}
  \label{fig:full_norm_radial}
\end{figure}

The complementary amplitude metric $\rhalf$ (Fig.~\ref{fig:full_norm_radial}) is instead driven mainly by epoch: typical values increase from $\simeq0.3$--$0.4$ dex/Gyr at $z<1$ to $\simeq1$ and $\simeq2$ dex/Gyr at $1<z<2$ and $2<z<3.7$, respectively, with broader dispersion at higher redshift. Its radial profiles are nearly flat at $z<2$; only at $2<z<3.7$ does $\rhalf$ decrease outward, from $\simeq2$ to $\simeq1.1$ dex/Gyr beyond $\Rnorm\simeq1.5$. Together, $\Phihalf$ and $\rhalf$ show that the direction of recent SFH evolution carries the clearest radial signal, while its amplitude is set mainly by cosmic epoch.

Finally, the same radial analysis for high-mass group members ($\log(\Mstar/\Msol)>10.5$) shows no clear monotonic variation with $\Rnorm$ in either $\Phihalf$ or $\rhalf$: massive group galaxies are predominantly declining, and the lack of a radial trend is consistent with their quenching being driven mainly by internal, mass-related processes rather than by the group environment.

\section{Discussion}
\label{sec:discussion}
The results reported in Sect.~\ref{sec:results} show that the nature of the environmental signal depends on redshift and galaxy mass, with the largest group--field offsets at high mass and the best-constrained response at low to intermediate mass (Sect.~\ref{sec:result-staked-SFHS}).

The migration vectors refine this picture: $\Phihalf$ shows that group galaxies more often have declining SFHs since $t_{50}$ than field galaxies, but among those already on this declining branch, low- and intermediate-mass group galaxies at $z<2$ show negative $\Delta r_{50}$ (i.e., $\rhalf^{\mathrm{field}}\gtrsim\rhalf^{\mathrm{group}}$), indicating slower recent evolution. Finally, the radial profiles (Figs.~\ref{fig:full_phi_radial} and~\ref{fig:full_norm_radial}) show that $\Phihalf$ is radially ordered at $z>0.5$, whereas $\rhalf$ depends mainly on redshift, with an outward decrease only at $2<z<3.7$.

We therefore interpret the results conservatively, emphasizing trends that are consistent across the SFH stacks, migration vectors, and radial profiles. In the high-mass bins and at high redshift, low statistics and large uncertainties can blur the signal, so we make no strong statements about individual detections. The \texttt{ASSOC\_PROB} weighting may under-represent backsplash galaxies, which could receive lower membership probabilities owing to their larger projected distances and less peaked photometric redshifts; this matters mainly for the outer radial bins ($\Rnorm>1$). Although catalog-identified AGN are excluded (Sect.~\ref{sec:sample}), unidentified AGN may remain and introduce residual uncertainty into the SED-based SFHs. Since our diagnostics rely only on the SFHs, mass changes not due to in situ star formation (mergers, winds, or tidal stripping) do not enter directly, although the SEDs of tidally stripped group galaxies could be misinterpreted by \cigale\ at second order. Passive galaxies, fainter at fixed mass, may be less complete near the mass limit; however, our common mass floor lies well above the local completeness limits except in the highest-redshift bin, and any residual loss of passive galaxies would affect groups more than the field, thus reducing rather than enhancing the group--field contrast. Finally, the reconstructed SFHs retain some dependence on the modeling assumptions \citep{boquien_cigale_2019,pacifici_art_2023,iyer_spectral_2025}: modeling-only uncertainties of order $\sim0.1$ dex in stellar mass \citep{pacifici_art_2023}, and shifts of up to $\sim0.2$ dex in the oldest-population mass fraction and by a factor of a few in inferred assembly timescales from the non-parametric SFH prior \citep{leja_deriving_2019,tacchella_stellar_2021,suessRecoveringStarFormation2022,pacifici_art_2023,wanStochasticPriorNonparametric2024}. Non-parametric codes relying on different stellar population models and dust treatments can also yield different SFH-based quantities as a function of stellar mass \citep{frequency_mahajan2011}. These systematics mainly affect absolute values, not population trends, and should affect groups and field alike, since both use the same code and priors.

\subsection{Early assembly followed by recent suppression\label{sec:disc-early}}

The stacked-SFH signal shown in Fig.~\ref{fig:topN_sfh} is best interpreted as a difference in when group and field galaxies assembled their stellar mass. When the group SFH lies above the field one at older lookback times, especially in the higher-redshift and lower-mass bins, it does not directly imply that group galaxies are more actively forming stars at that epoch. Instead, it suggests that part of the group population assembled its stellar mass earlier than comparable field galaxies \citep{dressler_galaxy_1980,peng_mass_2010,van_den_bosch_importance_2008}. The negative offsets at recent lookback times then indicate that this earlier assembly is often followed by suppression of star formation in groups \citep{wetzel_galaxy_2013,oman_homogeneous_2021,toni_cosmos-web_2026}.

This chronology is consistent with a regulatory view of environmental quenching. Rather than requiring an abrupt shutdown in every group galaxy, the measurements suggest that groups shift the population balance toward earlier growth, more often followed by an SFH decline. At high redshift this balance can still be mixed. Groups are overdense regions where galaxies may assemble earlier, but their low velocity dispersions can also make mergers, close passages, and tidal interactions efficient triggers of gas-rich star formation \citep{scudder_galaxy_2012,puskas_mergers_2025,rhee_origin_2026}. Together with continued filamentary accretion in the gas-rich conditions of Cosmic Noon, these processes can temporarily enhance star formation and accelerate stellar-mass build-up \citep{keresHowGalaxiesGet2005,tacconiHighMolecularGas2010,taamoli_cosmos2020_2024,paquereau_tracing_2025,jego_star-forming_2026,dovenEnvironmentalQuenchingHighRedshift2026}. This offers a natural explanation for high-redshift group SFHs lying above the matched field at older lookback times. This is compatible with the field-like SFHs of quiescent protocluster galaxies at $z\sim2.6$ \citep{pan_uncovermegascience_2025}. 

The same early growth can make later suppression easier to detect. Group galaxies may also consume gas earlier, a scenario known as overconsumption (or accelerated astration, i.e., the more rapid conversion of gas into stars): vigorous star formation and its associated outflows exhaust the internal gas supply on short timescales, once fresh accretion is cut off \citep{mcgee_overconsumption_2014, balogh_evidence_2016, pintos-castro_evolution_2019}. If infall then progressively reduces access to fresh accretion, their recent SFHs should decline more often than those of matched field galaxies \citep{vandevoortEnvironmentalDependenceGas2017}. This is the expected outcome of a starvation-like reduction of the gas supply \citep{reeves_gogreen_2021}, of stripping and heating of the circumgalactic gas \citep{bahe_star_2015,simpson_quenching_2018}, and of preprocessing before accretion into the present host \citep{hirschmann_Influence_Environmental_History2014,wetzel_satellite_2015,joshi_preprocessing_2017}. Moreover, this environmental suppression should be strongest for low-mass satellites, whose shallower potential wells make their remaining gas reservoirs easier to strip, heat, or exhaust \citep{boselli_environmental_2006,fillingham_under_2016,rhee_origin_2024,rhee_origin_2026}.

Massive galaxies require a different reading. Both group and field galaxies in the high-mass bin show early-declining SFHs, as expected from downsizing and from internal quenching processes that operate regardless of environment \citep{peng_mass_2010,shuntov_stellar_2026,ghaffari_star_2026}. Yet this bin also shows the largest group--field offsets (Fig.~\ref{fig:topN_sfh}). Part of this contrast may be technical, since the residual SFRs of quenched galaxies are weakly constrained, and part may reflect the inclusion of BGGs, whose central position favors early assembly and AGN-regulated quenching \citep{goubert_Env_comp2025,popesso_hot_2026,whitaker_quenching_2026}. If the offset is physical, it raises the question of why AGN feedback would be more effective in group galaxies at fixed stellar mass. One possibility is that group members are also affected by the AGN of the central and of nearby companions \citep{dashyan2019}. Another is the hot intragroup medium, which field galaxies, limited to their own halos, do not have: it can strip the hot gaseous halos of members through ram pressure and strangulation, even in low-mass groups \citep{mccarthy_ram_2008,kawata_strangulation_2008}, and cumulative heating by the central AGN can raise the quiescent fraction of satellites \citep{martin-navarro_quantifying_2019,eckert_feedback_2021}. These effects may be stronger in the most massive systems of our sample, at the boundary between massive groups and poor clusters (Sect.~\ref{sec:methods_staked-SFHs}), but our data carry no information on the hot gas, so we cannot test them here. If physical, the group environment would act here as an accelerator of an evolution already shaped by stellar mass, consistent with the absence of a radial trend for high-mass members (Sect.~\ref{sec:results_radial}).

\subsection{Migration vectors as environmental diagnostics}

The migration-vector diagnostics (Figs.~\ref{fig:topN_theta_triptych}, \ref{fig:velocity}; Sects.~\ref{sec:results-phihalf}, \ref{sec:result-migration_amplitude}) provide a complementary view of recent SFH evolution. While stacked SFHs identify when group and field histories diverge, $\Phihalf$ traces the direction of the recent evolution and $r_{50}$ measures its amplitude \citep{arango-toro_cosmos-web_2025}. The $\Phihalf$ distributions and their $A_\rho$ summaries show that group galaxies more often have a lower SFR at $t_\mathrm{now}$ than at $t_{50}$ than matched field galaxies, with the clearest contrast at low and intermediate stellar masses ($\log\,(\Mstar/\Msol)<10.5$) and at low-to-intermediate redshift \citep[see also][]{wetzel_satellite_2015,hatamnia_large-scale_2025}.

This combined signal, more negative $\Phihalf$ but slower declines at $z<2$, is therefore more naturally associated with slow regulatory quenching than with a single rapid shutdown channel. In the framework reviewed by \citet{whitaker_quenching_2026}, this corresponds to the same gas-regulation mechanisms discussed in Sect.~\ref{sec:disc-early}: reduced or unstable gas supply through inefficient accretion, stripping, and preprocessing \citep{boselli_environmental_2006,wetzel_galaxy_2013,malavasi_relative_2022,jego_star-forming_2026,oppenheimer_simulating_2021,popesso_hot_2026,bahe_star_2015,simpson_quenching_2018,oman_homogeneous_2021,rhee_origin_2026}.

The mass dependence supports this interpretation. Low- and intermediate-mass galaxies should respond more strongly to group processing because gas retention and recovery after tidal or hydrodynamic perturbations are less efficient in shallow potential wells \citep{fillingham_under_2016,rhee_origin_2024,rhee_origin_2026}. At high stellar mass ($\log\,(\Mstar/\Msol)>10.5$), the group--field amplitude differences are weaker and more uncertain, consistent with a regime where internal mass-related quenching dominates and environment acts mainly as a secondary modulator \citep{peng_mass_2010,shuntov_stellar_2026,toni_cosmos-web_2026}.

\subsection{Group radial trends in SFH evolution\label{sec:disc-radial}}

The radial profiles add a spatial dimension to this picture. The key point is not that every redshift bin follows the same radial sequence, but that this radial trend varies with epoch (Sect.~\ref{sec:results_radial}). At $z<0.5$, the group population is quenching-oriented (i.e., $\Phihalf<0$) across most of its extent, rather than showing an inner--outer gradient; given the bimodality of $\Phihalf$ (Fig.~\ref{fig:bimodal_phi}), this flat median profile may still be compatible with a radial decline of the quenched fraction \citep{wetzel_galaxy_2012}, as long as declining SFHs remain the majority at all radii. At $z>0.5$, inner regions are more quenching-oriented than the outskirts. The gradient is largest at $2<z<3.7$, where $\rhalf$ also decreases outward, but there the profiles are also the least reliable, because group centers and radii are most uncertain (Figs.~\ref{fig:top3_distri_radial}--\ref{fig:full_norm_radial}).

This is consistent with radius acting as a statistical proxy for residence time and orbital history: galaxies at $\Rnorm<1$ are more likely to have spent longer inside the group potential, experiencing repeated passages through dense intra-group regions, gas depletion, and orbit-coupled perturbations \citep{bahe_star_2015,simpson_quenching_2018}, whereas galaxies at $\Rnorm>1$ should include a broader mixture of first-infall systems, recently accreted satellites, preprocessed galaxies, backsplash objects, and photometric interlopers \citep{Mahajan+11,wetzel_satellite_2015,rhee_origin_2026}. The radial signal is therefore a superposition of local processing and inherited environmental history, not a simple inner-processed/outer-untouched split.

This interpretation should be read alongside the limitations of the tensor-based radial normalization (Sect.~\ref{sec:sample-radial}). Although model-independent, it remains sensitive to uncertain memberships, interlopers, and multiple density peaks, which can shift the barycenter and bias $\Rgroup$. Bernoulli resampling propagates membership uncertainty, but not systematics from unrecognized substructure or a barycenter--potential-minimum offset; these are expected to matter most for dynamically young, high-redshift systems \citep{toni_cosmos-web_2025,ghaffari_star_2026}. We therefore treat $\Rnorm$ as an empirical, population-level structural measure rather than a virial or orbital coordinate, and regard any residence-time interpretation at $z\gtrsim2$ as tentative.

Because the migration vector is anchored to $t_{50}$ (Sect.~\ref{sec:mig-vec}), the assembly-time maps provide a consistency check: if the positive group--field SFH excesses at high redshift and low mass (Fig.~\ref{fig:topN_sfh}) trace earlier assembly within the group, larger $t_{50}$ values should appear preferentially toward the inner group region, where early infall and long residence times are most likely.

\begin{figure}[h]
  \centering
  \includegraphics[width=\columnwidth]{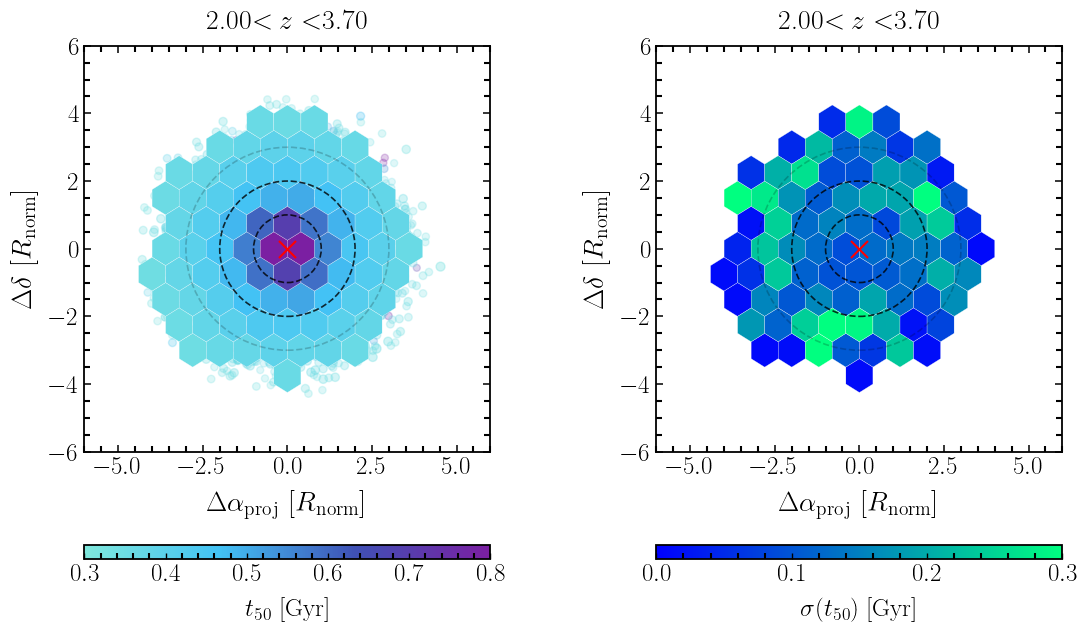}
\caption{Spatial distribution of the top-30 sample in normalized projected coordinates  for low- to intermediate-mass galaxies ($\log(\Mstar/\Msol)<10.5$) at $2<z<3.7$. The left panel shows the median $t_{50}$ in hexagonal bins containing at least ten galaxies, with individual galaxies overlaid; the right panel shows the standard deviation of $t_{50}$ among the galaxies in each bin, $\sigma(t_{50})$. Both statistics are membership weighted. Values are in Gyr, with larger $t_{50}$ indicating earlier assembly, in lookback time.}
  \label{fig:map_t_half}
\end{figure}

\begin{figure}[h]
  \centering
  \includegraphics[width=\columnwidth]{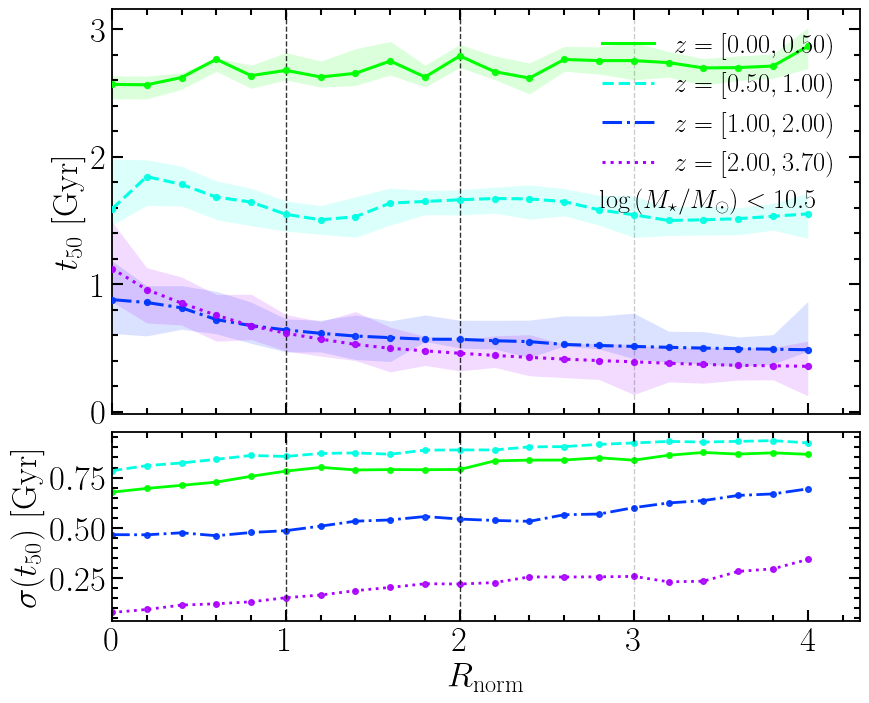}
  \caption{Median radial profile of $t_{50}$ (top) and its dispersion $\sigma(t_{50})$ (bottom) as a function of normalized group-centric radius $R_{\mathrm{norm}}$ for low- to intermediate-mass galaxies ($\log(\Mstar/\Msol)<10.5$) in the top-30 richest groups. Shaded bands show the 16th--84th percentile range of the median over 1000 bootstrap resamplings; $\sigma(t_{50})$ is the standard deviation of $t_{50}$ among the galaxies in each radial bin.}
  \label{fig:radial_t_half}
\end{figure}

Figure~\ref{fig:map_t_half} tests this for the highest-redshift, low- to intermediate-mass bin ($2<z<3.7$, $\log\,(\Mstar/\Msol)<10.5$), the same bin showing the clearest group--field SFH excess ($\delta\log\mathrm{SFR}_\mathrm{norm}\sim+0.4$ at $t_{\mathrm{lb}}\sim400$ Myr). The $t_{50}$ map is centrally peaked, with the largest values concentrated inside $R_{\mathrm{norm}}\lesssim1$, supporting an early-assembly component in the inner group population rather than a stochastic late fluctuation. The radial profile (Fig.~\ref{fig:radial_t_half}) confirms this trend. At low redshift, $t_{50}$ stays close to $2.6$ Gyr over $0\leq\Rnorm\leq4$, with no strong gradient. At $2<z<3.7$, the median decreases from $\sim1.0$ Gyr at $\Rnorm<0.5$ to $\sim0.3$ Gyr at $\Rnorm>3$: low-redshift groups show old assembly times across radius, while at high redshift the profile separates an earlier-assembled center from a more recent outer population.

This test is best viewed as a consistency check, not an independent orbital-history measurement: it relies on projected radii, probabilistic memberships, and the top-30 sample, and cannot separate residence time, preprocessing, backsplash, or interloper contamination galaxy-by-galaxy. Still, it adds useful time information: the same regions that appear more environmentally processed in their SFH evolution also show older assembly times, while the high-redshift outskirts remain younger and more mixed.

\section{Conclusions\label{sec:conclusions}}

We have presented a first SFH-based environmental analysis of \CWeb\ galaxy groups using the currently available catalog products. The analysis combines COSMOS2025/\CWeb\ stellar masses and non-parametric SFHs with \amico\ group detections, probabilistic memberships, richness estimates, and matched field controls. We used stacked SFHs and migration-vector statistics anchored to $t_{50}$ to test how galaxies in the top-30 sample differ from matched field galaxies as a function of stellar mass, redshift, and projected group-centric radius.

Our main conclusions are:

\begin{enumerate}

  \item \textbf{Environmental suppression of star formation is best constrained at low redshift and below $\log(\Mstar/\Msol)=10.5$}. In the top-30 sample, low- to intermediate-mass group galaxies show predominantly negative group--field SFH offsets at recent lookback times, with the clearest contrast at $z<1$ (recent deficits of up to $0.7$ dex) and weaker or more mixed signals at $z>1$. Massive galaxies show the largest group--field offsets, but they already have early-declining SFHs in both environments, consistent with downsizing; for them, the group environment acts as an additional modulation whose amplitude is the least constrained.

  \item \textbf{The high-redshift group--field contrast is more heterogeneous.} At $2<z<3.7$, the contrast becomes mass dependent: intermediate-mass group galaxies converge toward the field, whereas low-mass group galaxies show an earlier excess (up to $\sim+0.4$ dex) followed by a recent deficit. These cases are best interpreted as evidence for a mixed high-redshift regime, in which earlier assembly, continued gas accretion, interaction-driven activity, and emerging suppression can coexist, rather than as a uniform reversal of environmental quenching.

  \item \textbf{Migration vectors reveal an epoch- and mass-dependent population signal.} The angular component $\Phihalf$ shows that low- to intermediate-mass group galaxies are more often on quenching-oriented recent-SFH tracks than their matched field counterparts, whereas $r_{50}$ shows that, at $z<2$, galaxies on the declining branch evolve more slowly in groups than in the field. At $z<0.5$, the group population is broadly quenching-oriented across most radii, with no clear radial gradient. At $z>0.5$, the inner regions are generally more quenching-oriented than the outskirts, and the population becomes more heterogeneous with redshift. At $2<z<3.7$, low-mass galaxies in the inner group also assembled earlier (larger $t_{50}$) than those in the outskirts, consistent with an early-assembly component in group cores. Massive group galaxies show little systematic radial modulation, consistent with their evolution being dominated by mass and epoch.

  \item \textbf{The combined evidence supports an epoch-structured environmental sequence superimposed on mass-dependent galaxy evolution.} The results are consistent with a transition from mixed group--field SFH differences at early epochs to a cleaner, suppression-dominated environmental signal at later times, especially below $\log(\Mstar/\Msol)=10.5$. This interpretation is compatible with preprocessing, the greater environmental sensitivity of lower-mass galaxies, and residence-time-dependent satellite evolution. However, our photometric memberships, projected radii, and finite-time SFH summaries do not permit direct measurements of orbital histories or their associated quenching timescales. Possible residual contamination by unidentified AGN is an additional limitation.

\end{enumerate}

This work provides a reference layer for future \CWeb\ environmental studies, to be extended to second-order effects such as richness, large-scale connectivity, or position relative to the main sequence.

\section*{Data availability}
The COSMOS2025 photometric catalog, photometric redshifts, and non-parametric star-formation histories used in this work are publicly available at \url{https://cosmos2025.iap.fr} (see also the tutorial at \url{https://cosmos2025.iap.fr/tutorials/catalog_usage_tutorial.html}). The \amico\ group catalog and companion membership table for \CWeb, including \texttt{SN\_NOCL}, \texttt{MSKFRC}, \texttt{LAMBDA\_STAR}, \texttt{ASSOC\_PROB}, and \texttt{FIELD\_PROB}, are distributed at \url{https://cosmos2025.iap.fr/groups.html}. The derived quantities specific to this analysis, normalized group-centric radius ($\Rnorm$), formation times ($t_{10}$--$t_{90}$), and migration-vector components ($\Phihalf$, $r_{50}$), are available from the corresponding author upon reasonable request.

\begin{acknowledgements}
 We thank the anonymous referee for the thoughtful comments and constructive suggestions, which greatly helped to improve the clarity and depth of this manuscript. We appreciate the time and effort put into reviewing our work. We acknowledge funding from the French Agence Nationale de la Recherche for the project iMAGE (grant ANR-22-CE31-0007).
 This project has received financial support from the CNRS through the MITI interdisciplinary programs and the Initiative Physique des Infinis (IPI), a research training program of the Idex SUPER at Sorbonne Universit\'e.
  This work received support from the French government under the France 2030 investment plan, as part of the Initiative d’Excellence d’Aix-Marseille Université – A*MIDEX AMX-22-RE-AB-101. This project has received funding from the European Union’s Horizon 2020 research and innovation programme under the Marie Skłodowska-Curie grant agreement No 101148925.
  We warmly acknowledge the contributions of the COSMOS collaboration of more than 100 scientists all around the world. The HST-COSMOS program was supported through NASA grant HST-GO-09822. More information on the COSMOS survey is available at \url{https://cosmos.astro.caltech.edu}. This research is also partly supported by the Centre National d'Etudes Spatiales (CNES).
  This work was made possible by utilizing the CANDIDE cluster at the Institut d’Astrophysique de Paris. The cluster was funded through grants from the PNCG, CNES, DIM-ACAV, the Euclid Consortium, and the Danish National Research Foundation Cosmic Dawn Center (DNRF140). It is maintained by Stephane Rouberol.
  Generative artificial-intelligence tools were used to assist with language editing and to troubleshoot \LaTeX{} and analysis code. The authors reviewed all AI-assisted material and take full responsibility for the content of the manuscript.
\end{acknowledgements}

\bibliographystyle{bibtex/aa}
\bibliography{my_library}

@ARTICLE{puskas_mergers_2025,
   author = {{Pusk{\'a}s}, D{\'a}vid and {Tacchella}, Sandro and {Simmonds}, Charlotte and {Jones}, Gareth C. and {Juod{\v{z}}balis}, Ignas and {Scholtz}, Jan and {Baker}, William M. and {Bunker}, Andrew J. and {Carniani}, Stefano and {Curtis-Lake}, Emma and {Duan}, Qiao and {Eisenstein}, Daniel J. and {Hainline}, Kevin and {Johnson}, Benjamin D. and {Maiolino}, Roberto and {Rieke}, Marcia and {Robertson}, Brant and {Williams}, Christina C. and {Witstok}, Joris},
        title = {Mergers lighting the early Universe: modest star formation enhancement without evidence for AGN or Ly$\alpha$ triggering},
      journal = {\mnras},
         year = 2026,
        month = oct,
       volume = {552},
       number = {1},
          eid = {stag1698},
        pages = {stag1698},
          doi = {10.1093/mnras/stag1698},
archivePrefix = {arXiv},
       eprint = {2510.14743},
 primaryClass = {astro-ph.GA},
       adsurl = {https://ui.adsabs.harvard.edu/abs/2026MNRAS.552g1698P}
}

@ARTICLE{koekemoer_cosmos_2007,
   author = {{Koekemoer}, A.~M. and {Aussel}, H. and {Calzetti}, D. and {Capak}, P. and {Giavalisco}, M. and {Kneib}, J.-P. and {Leauthaud}, A. and {Le F{\`e}vre}, O. and {McCracken}, H.~J. and {Massey}, R. and {Mobasher}, B. and {Rhodes}, J. and {Scoville}, N. and {Shopbell}, P.~L.},
        title = {The COSMOS Survey: Hubble Space Telescope Advanced Camera for Surveys Observations and Data Processing},
      journal = {\apjs},
         year = 2007,
        month = sep,
       volume = {172},
       number = {1},
        pages = {196-202},
          doi = {10.1086/520086},
archivePrefix = {arXiv},
       eprint = {astro-ph/0703095},
 primaryClass = {astro-ph},
       adsurl = {https://ui.adsabs.harvard.edu/abs/2007ApJS..172..196K}
}

@ARTICLE{pan_uncovermegascience_2025,
       author = {{Pan}, Richard and {Suess}, Katherine A. and {Marchesini}, Danilo and {Wang}, Bingjie and {Leja}, Joel and {Cutler}, Sam E. and {Whitaker}, Katherine E. and {Bezanson}, Rachel and {Price}, Sedona H. and {Furtak}, Lukas J. and et al.},
        title = "{UNCOVER/MegaScience: No Evidence of Environmental Quenching in a z {\ensuremath{\sim}} 2.6 Protocluster}",
      journal = {\apjl},
         year = 2025,
        month = sep,
       volume = {990},
       number = {1},
          eid = {L24},
        pages = {L24},
          doi = {10.3847/2041-8213/adf7ab},
archivePrefix = {arXiv},
       eprint = {2504.06334},
 primaryClass = {astro-ph.GA},
       adsurl = {https://ui.adsabs.harvard.edu/abs/2025ApJ...990L..24P}
}

@ARTICLE{mccarthy_ram_2008,
       author = {{McCarthy}, I.~G. and {Frenk}, C.~S. and {Font}, A.~S. and {Lacey}, C.~G. and {Bower}, R.~G. and {Mitchell}, N.~L. and {Balogh}, M.~L. and {Theuns}, T.},
        title = {Ram pressure stripping the hot gaseous haloes of galaxies in groups and clusters},
      journal = {\mnras},
         year = 2008,
        month = jan,
       volume = {383},
       number = {2},
        pages = {593-605},
          doi = {10.1111/j.1365-2966.2007.12577.x},
archivePrefix = {arXiv},
       eprint = {0710.0964},
 primaryClass = {astro-ph},
       adsurl = {https://ui.adsabs.harvard.edu/abs/2008MNRAS.383..593M}
}

@ARTICLE{kawata_strangulation_2008,
       author = {{Kawata}, Daisuke and {Mulchaey}, John S.},
        title = {Strangulation in Galaxy Groups},
      journal = {\apjl},
         year = 2008,
        month = jan,
       volume = {672},
       number = {2},
        pages = {L103},
          doi = {10.1086/526544},
archivePrefix = {arXiv},
       eprint = {0707.3814},
 primaryClass = {astro-ph},
       adsurl = {https://ui.adsabs.harvard.edu/abs/2008ApJ...672L.103K}
}

@ARTICLE{martin-navarro_quantifying_2019,
       author = {{Mart{\'\i}n-Navarro}, I. and {Burchett}, Joseph N. and {Mezcua}, Mar},
        title = {Quantifying the Effect of Black Hole Feedback from the Central Galaxy on the Satellite Populations of Groups and Clusters},
      journal = {\apjl},
         year = 2019,
        month = oct,
       volume = {884},
       number = {2},
          eid = {L45},
        pages = {L45},
          doi = {10.3847/2041-8213/ab4885},
archivePrefix = {arXiv},
       eprint = {1909.12841},
 primaryClass = {astro-ph.GA},
       adsurl = {https://ui.adsabs.harvard.edu/abs/2019ApJ...884L..45M}
}

@ARTICLE{eckert_feedback_2021,
       author = {{Eckert}, Dominique and {Gaspari}, Massimo and {Gastaldello}, Fabio and {Le Brun}, Amandine M.~C. and {O'Sullivan}, Ewan},
        title = {Feedback from Active Galactic Nuclei in Galaxy Groups},
      journal = {Universe},
         year = 2021,
        month = may,
       volume = {7},
       number = {5},
          eid = {142},
        pages = {142},
          doi = {10.3390/universe7050142},
archivePrefix = {arXiv},
       eprint = {2106.13259},
 primaryClass = {astro-ph.GA},
       adsurl = {https://ui.adsabs.harvard.edu/abs/2021Univ....7..142E}
}

@ARTICLE{wetzel_galaxy_2012,
   author = {{Wetzel}, Andrew R. and {Tinker}, Jeremy L. and {Conroy}, Charlie},
        title = {Galaxy evolution in groups and clusters: star formation rates, red sequence fractions and the persistent bimodality},
      journal = {\mnras},
         year = 2012,
        month = jul,
       volume = {424},
       number = {1},
        pages = {232-243},
          doi = {10.1111/j.1365-2966.2012.21188.x},
archivePrefix = {arXiv},
       eprint = {1107.5311},
 primaryClass = {astro-ph.CO},
       adsurl = {https://ui.adsabs.harvard.edu/abs/2012MNRAS.424..232W}
}

@ARTICLE{Downsizing_Leung_2026,
   author = {{Leung}, Ho-Hin and {Carnall}, Adam C. and {Taylor}, Elizabeth and {Stevenson}, Struan D. and {Beverage}, Aliza G. and {Cullen}, Fergus and {Dunlop}, James S. and {McLeod}, Derek J. and {McLure}, Ross J. and {Begley}, Ryan and {Almaini}, Omar and {Antonogiannaki}, Stella and {Arellano-C{\'o}rdova}, Karla Z. and {Barrufet}, Laia and {Bondestam}, Cecilia and {Donnan}, Callum T. and {Holst}, Isaac J.~B. and {Liu}, Feng-Yuan F. and {Rowlands}, Kate and {Sanders}, Ryan L. and {Scholte}, Dirk and {Skarbinski}, Maya and {Stanton}, Thomas M. and {Wild}, Vivienne},
        title = {The JWST EXCELS survey: the ages and abundances of 3 < z < 5 massive quiescent galaxies show that downsizing was already in place by z $\simeq$ 4},
      journal = {\mnras},
         year = 2026,
        month = jun,
       volume = {549},
       number = {1},
          eid = {stag827},
        pages = {stag827},
          doi = {10.1093/mnras/stag827},
archivePrefix = {arXiv},
       eprint = {2602.05934},
 primaryClass = {astro-ph.GA},
       adsurl = {https://ui.adsabs.harvard.edu/abs/2026MNRAS.549ag827L}
}

@ARTICLE{Downsizing_Carnall_2019,
   author = {{Carnall}, A.~C. and {McLure}, R.~J. and {Dunlop}, J.~S. and {Cullen}, F. and {McLeod}, D.~J. and {Wild}, V. and {Johnson}, B.~D. and {Appleby}, S. and {Dav{\'e}}, R. and {Amorin}, R. and {Bolzonella}, M. and {Castellano}, M. and {Cimatti}, A. and {Cucciati}, O. and {Gargiulo}, A. and {Garilli}, B. and {Marchi}, F. and {Pentericci}, L. and {Pozzetti}, L. and {Schreiber}, C. and {Talia}, M. and {Zamorani}, G.},
        title = {The VANDELS survey: the star-formation histories of massive quiescent galaxies at 1.0 < z < 1.3},
      journal = {\mnras},
         year = 2019,
        month = nov,
       volume = {490},
       number = {1},
        pages = {417-439},
          doi = {10.1093/mnras/stz2544},
archivePrefix = {arXiv},
       eprint = {1903.11082},
 primaryClass = {astro-ph.GA},
       adsurl = {https://ui.adsabs.harvard.edu/abs/2019MNRAS.490..417C}
}

@ARTICLE{downsizing_Cowie_1996,
   author = {{Cowie}, Lennox L. and {Songaila}, Antoinette and {Hu}, Esther M. and {Cohen}, J.~G.},
        title = {New Insight on Galaxy Formation and Evolution From Keck Spectroscopy of the Hawaii Deep Fields},
      journal = {\aj},
         year = 1996,
        month = sep,
       volume = {112},
        pages = {839},
          doi = {10.1086/118058},
archivePrefix = {arXiv},
       eprint = {astro-ph/9606079},
 primaryClass = {astro-ph},
       adsurl = {https://ui.adsabs.harvard.edu/abs/1996AJ....112..839C}
}

@ARTICLE{frequency_mahajan2011,
   author = {{Mamon}, Gary A. and {Trevisan}, Marina and {Thuan}, Trinh X. and {Gallazzi}, Anna and {Dav{\'e}}, Romeel},
        title = {The frequency of very young galaxies in the local Universe - II. The view from SDSS spectra},
      journal = {\mnras},
         year = 2020,
        month = feb,
       volume = {492},
       number = {2},
        pages = {1791-1811},
          doi = {10.1093/mnras/stz3556},
archivePrefix = {arXiv},
       eprint = {1912.06522},
 primaryClass = {astro-ph.GA},
       adsurl = {https://ui.adsabs.harvard.edu/abs/2020MNRAS.492.1791M}
}

@ARTICLE{dashyan2019,
   author = {{Dashyan}, Gohar and {Choi}, Ena and {Somerville}, Rachel S. and {Naab}, Thorsten and {Quirk}, Amanda C.~N. and {Hirschmann}, Michaela and {Ostriker}, Jeremiah P.},
        title = {AGN-driven quenching of satellite galaxies},
      journal = {\mnras},
         year = 2019,
        month = aug,
       volume = {487},
       number = {4},
        pages = {5889-5901},
          doi = {10.1093/mnras/stz1697},
archivePrefix = {arXiv},
       eprint = {1906.07431},
 primaryClass = {astro-ph.GA},
       adsurl = {https://ui.adsabs.harvard.edu/abs/2019MNRAS.487.5889D}
}

@ARTICLE{Mahajan+11,
   author = {{Mahajan}, Smriti and {Mamon}, Gary A. and {Raychaudhury}, Somak},
        title = {The velocity modulation of galaxy properties in and near clusters: quantifying the decrease in star formation in backsplash galaxies},
      journal = {\mnras},
         year = 2011,
        month = oct,
       volume = {416},
       number = {4},
        pages = {2882-2902},
          doi = {10.1111/j.1365-2966.2011.19236.x},
archivePrefix = {arXiv},
       eprint = {1106.3062},
 primaryClass = {astro-ph.CO},
       adsurl = {https://ui.adsabs.harvard.edu/abs/2011MNRAS.416.2882M}
}

@ARTICLE{Song+21,
   author = {{Song}, Hyunmi and {Laigle}, Clotilde and {Hwang}, Ho Seong and {Devriendt}, Julien and {Dubois}, Yohan and {Kraljic}, Katarina and {Pichon}, Christophe and {Slyz}, Adrianne and {Smith}, Rory},
        title = {Beyond halo mass: quenching galaxy mass assembly at the edge of filaments},
      journal = {\mnras},
         year = 2021,
        month = mar,
       volume = {501},
       number = {3},
        pages = {4635-4656},
          doi = {10.1093/mnras/staa3981},
archivePrefix = {arXiv},
       eprint = {2009.00013},
 primaryClass = {astro-ph.GA},
       adsurl = {https://ui.adsabs.harvard.edu/abs/2021MNRAS.501.4635S}
}

@ARTICLE{mcgee_overconsumption_2014,
   author = {{McGee}, S.~L. and {Bower}, R.~G. and {Balogh}, M.~L.},
        title = {Overconsumption, outflows and the quenching of satellite galaxies.},
      journal = {\mnras},
         year = 2014,
        month = jul,
       volume = {442},
        pages = {L105-L109},
          doi = {10.1093/mnrasl/slu066},
archivePrefix = {arXiv},
       eprint = {1404.6251},
 primaryClass = {astro-ph.GA},
       adsurl = {https://ui.adsabs.harvard.edu/abs/2014MNRAS.442L.105M}
}

@ARTICLE{Foltz+18,
   author = {{Foltz}, R. and {Wilson}, G. and {Muzzin}, A. and {Cooper}, M.~C. and {Nantais}, J. and {van der Burg}, R.~F.~J. and {Cerulo}, P. and {Chan}, J. and {Fillingham}, S.~P. and {Surace}, J. and {Webb}, T. and {Noble}, A. and {Lacy}, M. and {McDonald}, M. and {Rudnick}, G. and {Lidman}, C. and {Demarco}, R. and {Hlavacek-Larrondo}, J. and {Yee}, H.~K.~C. and {Perlmutter}, S. and {Hayden}, B.},
        title = {The Evolution of Environmental Quenching Timescales to z {\ensuremath{\sim}} 1.6: Evidence for Dynamically Driven Quenching of the Cluster Galaxy Population},
      journal = {\apj},
         year = 2018,
        month = oct,
       volume = {866},
       number = {2},
          eid = {136},
        pages = {136},
          doi = {10.3847/1538-4357/aad80d},
archivePrefix = {arXiv},
       eprint = {1803.03305},
 primaryClass = {astro-ph.GA},
       adsurl = {https://ui.adsabs.harvard.edu/abs/2018ApJ...866..136F}
}

@ARTICLE{Jeon+22,
   author = {{Jeon}, Seyoung and {Yi}, Sukyoung K. and {Dubois}, Yohan and {Chung}, Aeree and {Devriendt}, Julien and {Han}, San and {Jackson}, Ryan A. and {Kimm}, Taysun and {Pichon}, Christophe and {Rhee}, Jinsu},
        title = {Star Formation History and Transition Epoch of Cluster Galaxies Based on the Horizon-AGN Simulation},
      journal = {\apj},
         year = 2022,
        month = dec,
       volume = {941},
       number = {1},
          eid = {5},
        pages = {5},
          doi = {10.3847/1538-4357/ac9d8c},
archivePrefix = {arXiv},
       eprint = {2210.05285},
 primaryClass = {astro-ph.GA},
       adsurl = {https://ui.adsabs.harvard.edu/abs/2022ApJ...941....5J}
}

@ARTICLE{tacchella_stellar_2021,
   author = {{Tacchella}, Sandro and {Finkelstein}, Steven L. and {Bagley}, Micaela and {Dickinson}, Mark and {Ferguson}, Henry C. and {Giavalisco}, Mauro and {Graziani}, Luca and {Grogin}, Norman A. and {Hathi}, Nimish and {Hutchison}, Taylor A. and {Jung}, Intae and {Koekemoer}, Anton M. and {Larson}, Rebecca L. and {Papovich}, Casey and {Pirzkal}, Norbert and {Rojas-Ruiz}, Sof{\'\i}a and {Song}, Mimi and {Schneider}, Raffaella and {Somerville}, Rachel S. and {Wilkins}, Stephen M. and {Yung}, L.~Y. Aaron},
        title = {On the Stellar Populations of Galaxies at z = 9-11: The Growth of Metals and Stellar Mass at Early Times},
      journal = {\apj},
         year = 2022,
        month = mar,
       volume = {927},
       number = {2},
          eid = {170},
        pages = {170},
          doi = {10.3847/1538-4357/ac4cad},
archivePrefix = {arXiv},
       eprint = {2111.05351},
 primaryClass = {astro-ph.GA},
       adsurl = {https://ui.adsabs.harvard.edu/abs/2022ApJ...927..170T}
}

@ARTICLE{navarro1997,
   author = {{Navarro}, Julio F. and {Frenk}, Carlos S. and {White}, Simon D.~M.},
        title = {A Universal Density Profile from Hierarchical Clustering},
      journal = {\apj},
         year = 1997,
        month = dec,
       volume = {490},
       number = {2},
        pages = {493-508},
          doi = {10.1086/304888},
archivePrefix = {arXiv},
       eprint = {astro-ph/9611107},
 primaryClass = {astro-ph},
       adsurl = {https://ui.adsabs.harvard.edu/abs/1997ApJ...490..493N}
}

@ARTICLE{king1962,
   author = {{King}, Ivan},
        title = {The structure of star clusters. I. an empirical density law},
      journal = {\aj},
         year = 1962,
        month = oct,
       volume = {67},
        pages = {471},
          doi = {10.1086/108756},
       adsurl = {https://ui.adsabs.harvard.edu/abs/1962AJ.....67..471K}
}

@ARTICLE{robotham2011,
   author = {{Robotham}, A.~S.~G. and {Norberg}, P. and {Driver}, S.~P. and {Baldry}, I.~K. and {Bamford}, S.~P. and {Hopkins}, A.~M. and {Liske}, J. and {Loveday}, J. and {Merson}, A. and {Peacock}, J.~A. and {Brough}, S. and {Cameron}, E. and {Conselice}, C.~J. and {Croom}, S.~M. and {Frenk}, C.~S. and {Gunawardhana}, M. and {Hill}, D.~T. and {Jones}, D.~H. and {Kelvin}, L.~S. and {Kuijken}, K. and {Nichol}, R.~C. and {Parkinson}, H.~R. and {Pimbblet}, K.~A. and {Phillipps}, S. and {Popescu}, C.~C. and {Prescott}, M. and {Sharp}, R.~G. and {Sutherland}, W.~J. and {Taylor}, E.~N. and {Thomas}, D. and {Tuffs}, R.~J. and {van Kampen}, E. and {Wijesinghe}, D.},
        title = {Galaxy and Mass Assembly (GAMA): the GAMA galaxy group catalogue (G$^{3}$Cv1)},
      journal = {\mnras},
         year = 2011,
        month = oct,
       volume = {416},
       number = {4},
        pages = {2640-2668},
          doi = {10.1111/j.1365-2966.2011.19217.x},
archivePrefix = {arXiv},
       eprint = {1106.1994},
 primaryClass = {astro-ph.CO},
       adsurl = {https://ui.adsabs.harvard.edu/abs/2011MNRAS.416.2640R}
}

@ARTICLE{zemp2011,
   author = {{Zemp}, Marcel and {Gnedin}, Oleg Y. and {Gnedin}, Nickolay Y. and {Kravtsov}, Andrey V.},
        title = {On Determining the Shape of Matter Distributions},
      journal = {\apjs},
         year = 2011,
        month = dec,
       volume = {197},
       number = {2},
          eid = {30},
        pages = {30},
          doi = {10.1088/0067-0049/197/2/30},
archivePrefix = {arXiv},
       eprint = {1107.5582},
 primaryClass = {astro-ph.IM},
       adsurl = {https://ui.adsabs.harvard.edu/abs/2011ApJS..197...30Z}
}

@ARTICLE{wojtak2013,
   author = {{Wojtak}, Rados{\l}aw},
        title = {Phase-space shapes of clusters and rich groups of galaxies},
      journal = {\aap},
         year = 2013,
        month = nov,
       volume = {559},
          eid = {A89},
        pages = {A89},
          doi = {10.1051/0004-6361/201322509},
archivePrefix = {arXiv},
       eprint = {1310.3624},
 primaryClass = {astro-ph.CO},
       adsurl = {https://ui.adsabs.harvard.edu/abs/2013A&A...559A..89W}
}

@ARTICLE{paz2006,
   author = {{Paz}, D.~J. and {Lambas}, D.~G. and {Padilla}, N. and {Merch{\'a}n}, M.},
        title = {Shapes of clusters and groups of galaxies: comparison of model predictions with observations},
      journal = {\mnras},
         year = 2006,
        month = mar,
       volume = {366},
       number = {4},
        pages = {1503-1510},
          doi = {10.1111/j.1365-2966.2005.09934.x},
archivePrefix = {arXiv},
       eprint = {astro-ph/0509062},
 primaryClass = {astro-ph},
       adsurl = {https://ui.adsabs.harvard.edu/abs/2006MNRAS.366.1503P}
}

@ARTICLE{plionis2004,
   author = {{Plionis}, M. and {Basilakos}, S. and {Tovmassian}, H.~M.},
        title = {The shape of poor groups of galaxies},
      journal = {\mnras},
         year = 2004,
        month = aug,
       volume = {352},
       number = {4},
        pages = {1323-1328},
          doi = {10.1111/j.1365-2966.2004.08023.x},
archivePrefix = {arXiv},
       eprint = {astro-ph/0312019},
 primaryClass = {astro-ph},
       adsurl = {https://ui.adsabs.harvard.edu/abs/2004MNRAS.352.1323P}
}

@ARTICLE{Basilakos2000,
   author = {{Basilakos}, S. and {Plionis}, M. and {Maddox}, S.~J.},
        title = {The apparent and intrinsic shape of the APM galaxy clusters},
      journal = {\mnras},
         year = 2000,
        month = aug,
       volume = {316},
       number = {4},
        pages = {779-785},
          doi = {10.1046/j.1365-8711.2000.03590.x},
archivePrefix = {arXiv},
       eprint = {astro-ph/0002459},
 primaryClass = {astro-ph},
       adsurl = {https://ui.adsabs.harvard.edu/abs/2000MNRAS.316..779B}
}

@ARTICLE{carter1980,
   author = {{Carter}, D. and {Metcalfe}, N.},
        title = {The morphology of clusters of galaxies.},
      journal = {\mnras},
         year = 1980,
        month = may,
       volume = {191},
        pages = {325-337},
          doi = {10.1093/mnras/191.2.325},
       adsurl = {https://ui.adsabs.harvard.edu/abs/1980MNRAS.191..325C}
}

@ARTICLE{wanStochasticPriorNonparametric2024,
   author = {{Wan}, Jenny T. and {Tacchella}, Sandro and {Johnson}, Benjamin D. and {Iyer}, Kartheik G. and {Speagle}, Joshua S. and {Maiolino}, Roberto},
        title = {Stochastic prior for non-parametric star-formation histories},
      journal = {\mnras},
         year = 2024,
        month = aug,
       volume = {532},
       number = {4},
        pages = {4002-4025},
          doi = {10.1093/mnras/stae1734},
archivePrefix = {arXiv},
       eprint = {2404.14494},
 primaryClass = {astro-ph.GA},
       adsurl = {https://ui.adsabs.harvard.edu/abs/2024MNRAS.532.4002W}
}

@ARTICLE{suessRecoveringStarFormation2022,
   author = {{Suess}, Katherine A. and {Leja}, Joel and {Johnson}, Benjamin D. and {Bezanson}, Rachel and {Greene}, Jenny E. and {Kriek}, Mariska and {Lower}, Sidney and {Narayanan}, Desika and {Setton}, David J. and {Spilker}, Justin S.},
        title = {Recovering the Star Formation Histories of Recently Quenched Galaxies: The Impact of Model and Prior Choices},
      journal = {\apj},
         year = 2022,
        month = aug,
       volume = {935},
       number = {2},
          eid = {146},
        pages = {146},
          doi = {10.3847/1538-4357/ac82b0},
archivePrefix = {arXiv},
       eprint = {2207.02883},
 primaryClass = {astro-ph.GA},
       adsurl = {https://ui.adsabs.harvard.edu/abs/2022ApJ...935..146S}
}

@ARTICLE{dovenEnvironmentalQuenchingHighRedshift2026,
   author = {{D{\"o}ven}, Aleyna and {Ayromlou}, Mohammadreza and {Porciani}, Cristiano},
        title = {Environmental quenching of high-redshift galaxies: Interpreting JWST observations with simulations},
      journal = {\aap},
         year = 2026,
        month = aug,
       volume = {712},
          eid = {A172},
        pages = {A172},
          doi = {10.1051/0004-6361/202660728},
archivePrefix = {arXiv},
       eprint = {2605.03008},
 primaryClass = {astro-ph.GA},
       adsurl = {https://ui.adsabs.harvard.edu/abs/2026A&A...712A.172D}
}

@ARTICLE{goubert_Env_comp2025,
   author = {{Goubert}, Paul H. and {Bluck}, Asa F.~L. and {Piotrowska}, Joanna M. and {Torrey}, Paul and {Maiolino}, Roberto and {Franco}, Thomas Pinto and {Casimiro}, Camilo and {Cea}, Nicolas},
        title = {Environmental versus intrinsic quenching at cosmic noon: predictions from cosmological hydrodynamical simulations for VLT-MOONRISE},
      journal = {\mnras},
         year = 2025,
        month = nov,
       volume = {543},
       number = {3},
        pages = {2006-2034},
          doi = {10.1093/mnras/staf1554},
archivePrefix = {arXiv},
       eprint = {2509.09626},
 primaryClass = {astro-ph.GA},
       adsurl = {https://ui.adsabs.harvard.edu/abs/2025MNRAS.543.2006G}
}

@ARTICLE{hirschmann_Influence_Environmental_History2014,
   author = {{Hirschmann}, Michaela and {De Lucia}, Gabriella and {Wilman}, Dave and {Weinmann}, Simone and {Iovino}, Angela and {Cucciati}, Olga and {Zibetti}, Stefano and {Villalobos}, {\'A}lvaro},
        title = {The influence of the environmental history on quenching star formation in a {\ensuremath{\Lambda}} cold dark matter universe},
      journal = {\mnras},
         year = 2014,
        month = nov,
       volume = {444},
       number = {3},
        pages = {2938-2959},
          doi = {10.1093/mnras/stu1609},
archivePrefix = {arXiv},
       eprint = {1407.5621},
 primaryClass = {astro-ph.GA},
       adsurl = {https://ui.adsabs.harvard.edu/abs/2014MNRAS.444.2938H}
}

@ARTICLE{leja_deriving_2019,
   author = {{Leja}, Joel and {Carnall}, Adam C. and {Johnson}, Benjamin D. and {Conroy}, Charlie and {Speagle}, Joshua S.},
        title = {How to Measure Galaxy Star Formation Histories. II. Nonparametric Models},
      journal = {\apj},
         year = 2019,
        month = may,
       volume = {876},
       number = {1},
          eid = {3},
        pages = {3},
          doi = {10.3847/1538-4357/ab133c},
archivePrefix = {arXiv},
       eprint = {1811.03637},
 primaryClass = {astro-ph.GA},
       adsurl = {https://ui.adsabs.harvard.edu/abs/2019ApJ...876....3L}
}

@ARTICLE{pacifici_art_2023,
   author = {{Pacifici}, Camilla and {Iyer}, Kartheik G. and {Mobasher}, Bahram and {da Cunha}, Elisabete and {Acquaviva}, Viviana and {Burgarella}, Denis and {Calistro Rivera}, Gabriela and {Carnall}, Adam C. and {Chang}, Yu-Yen and {Chartab}, Nima and {Cooke}, Kevin C. and {Fairhurst}, Ciaran and {Kartaltepe}, Jeyhan and {Leja}, Joel and {Ma{\l}ek}, Katarzyna and {Salmon}, Brett and {Torelli}, Marianna and {Vidal-Garc{\'\i}a}, Alba and {Boquien}, M{\'e}d{\'e}ric and {Brammer}, Gabriel G. and {Brown}, Michael J.~I. and {Capak}, Peter L. and {Chevallard}, Jacopo and {Circosta}, Chiara and {Croton}, Darren and {Davidzon}, Iary and {Dickinson}, Mark and {Duncan}, Kenneth J. and {Faber}, Sandra M. and {Ferguson}, Harry C. and {Fontana}, Adriano and {Guo}, Yicheng and {Haeussler}, Boris and {Hemmati}, Shoubaneh and {Jafariyazani}, Marziye and {Kassin}, Susan A. and {Larson}, Rebecca L. and {Lee}, Bomee and {Mantha}, Kameswara Bharadwaj and {Marchi}, Francesca and {Nayyeri}, Hooshang and {Newman}, Jeffrey A. and {Pandya}, Viraj and {Pforr}, Janine and {Reddy}, Naveen and {Sanders}, Ryan and {Shah}, Ekta and {Shahidi}, Abtin and {Stevans}, Matthew L. and {Triani}, Dian Puspita and {Tyler}, Krystal D. and {Vanderhoof}, Brittany N. and {de la Vega}, Alexander and {Wang}, Weichen and {Weston}, Madalyn E.},
        title = {The Art of Measuring Physical Parameters in Galaxies: A Critical Assessment of Spectral Energy Distribution Fitting Techniques},
      journal = {\apj},
         year = 2023,
        month = feb,
       volume = {944},
       number = {2},
          eid = {141},
        pages = {141},
          doi = {10.3847/1538-4357/acacff},
archivePrefix = {arXiv},
       eprint = {2212.01915},
 primaryClass = {astro-ph.GA},
       adsurl = {https://ui.adsabs.harvard.edu/abs/2023ApJ...944..141P}
}

@INPROCEEDINGS{iyer_spectral_2025,
   author = {{Iyer}, Kartheik G. and {Pacifici}, Camilla and {Calistro-Rivera}, Gabriela and {Lovell}, Christopher C.},
        title = {The spectral energy distributions of galaxies},
    booktitle = {Encyclopedia of Astrophysics},
         year = 2026,
       volume = {4},
        month = jan,
        pages = {236-281},
          doi = {10.1016/B978-0-443-21439-4.00127-9},
archivePrefix = {arXiv},
       eprint = {2502.17680},
 primaryClass = {astro-ph.GA},
       adsurl = {https://ui.adsabs.harvard.edu/abs/2026enap....4..236I}
}

@ARTICLE{vandevoortEnvironmentalDependenceGas2017,
   author = {{van de Voort}, Freeke and {Bah{\'e}}, Yannick M. and {Bower}, Richard G. and {Correa}, Camila A. and {Crain}, Robert A. and {Schaye}, Joop and {Theuns}, Tom},
        title = {The environmental dependence of gas accretion on to galaxies: quenching satellites through starvation},
      journal = {\mnras},
         year = 2017,
        month = apr,
       volume = {466},
       number = {3},
        pages = {3460-3471},
          doi = {10.1093/mnras/stw3356},
archivePrefix = {arXiv},
       eprint = {1611.03870},
 primaryClass = {astro-ph.GA},
       adsurl = {https://ui.adsabs.harvard.edu/abs/2017MNRAS.466.3460V}
}

@ARTICLE{keresHowGalaxiesGet2005,
   author = {{Kere{\v{s}}}, Du{\v{s}}an and {Katz}, Neal and {Weinberg}, David H. and {Dav{\'e}}, Romeel},
        title = {How do galaxies get their gas?},
      journal = {\mnras},
         year = 2005,
        month = oct,
       volume = {363},
       number = {1},
        pages = {2-28},
          doi = {10.1111/j.1365-2966.2005.09451.x},
archivePrefix = {arXiv},
       eprint = {astro-ph/0407095},
 primaryClass = {astro-ph},
       adsurl = {https://ui.adsabs.harvard.edu/abs/2005MNRAS.363....2K}
}

@ARTICLE{tacconiHighMolecularGas2010,
   author = {{Tacconi}, L.~J. and {Genzel}, R. and {Neri}, R. and {Cox}, P. and {Cooper}, M.~C. and {Shapiro}, K. and {Bolatto}, A. and {Bouch{\'e}}, N. and {Bournaud}, F. and {Burkert}, A. and {Combes}, F. and {Comerford}, J. and {Davis}, M. and {F{\"o}rster Schreiber}, N.~M. and {Garcia-Burillo}, S. and {Gracia-Carpio}, J. and {Lutz}, D. and {Naab}, T. and {Omont}, A. and {Shapley}, A. and {Sternberg}, A. and {Weiner}, B.},
        title = {High molecular gas fractions in normal massive star-forming galaxies in the young Universe},
      journal = {\nat},
         year = 2010,
        month = feb,
       volume = {463},
       number = {7282},
        pages = {781-784},
          doi = {10.1038/nature08773},
archivePrefix = {arXiv},
       eprint = {1002.2149},
 primaryClass = {astro-ph.CO},
       adsurl = {https://ui.adsabs.harvard.edu/abs/2010Natur.463..781T}
}

@ARTICLE{deluciaTracingQuenchingJourney2024,
   author = {{De Lucia}, Gabriella and {Fontanot}, Fabio and {Xie}, Lizhi and {Hirschmann}, Michaela},
        title = {Tracing the quenching journey across cosmic time},
      journal = {\aap},
         year = 2024,
        month = jul,
       volume = {687},
          eid = {A68},
        pages = {A68},
          doi = {10.1051/0004-6361/202349045},
archivePrefix = {arXiv},
       eprint = {2401.06211},
 primaryClass = {astro-ph.GA},
       adsurl = {https://ui.adsabs.harvard.edu/abs/2024A&A...687A..68D}
}

@ARTICLE{bluckGalaxyQuenchingHigh2024,
   author = {{Bluck}, Asa F.~L. and {Conselice}, Christopher J. and {Ormerod}, Katherine and {Piotrowska}, Joanna M. and {Adams}, Nathan and {Austin}, Duncan and {Caruana}, Joseph and {Duncan}, K.~J. and {Ferreira}, Leonardo and {Goubert}, Paul and {Harvey}, Thomas and {Trussler}, James and {Maiolino}, Roberto},
        title = {Galaxy Quenching at the High Redshift Frontier: A Fundamental Test of Cosmological Models in the Early Universe with JWST-CEERS},
      journal = {\apj},
         year = 2024,
        month = feb,
       volume = {961},
       number = {2},
          eid = {163},
        pages = {163},
          doi = {10.3847/1538-4357/ad0a98},
archivePrefix = {arXiv},
       eprint = {2311.02526},
 primaryClass = {astro-ph.GA},
       adsurl = {https://ui.adsabs.harvard.edu/abs/2024ApJ...961..163B}
}

@ARTICLE{balogh_evidence_2016,
   author = {{Balogh}, Michael L. and {McGee}, Sean L. and {Mok}, Angus and {Muzzin}, Adam and {van der Burg}, Remco F.~J. and {Bower}, Richard G. and {Finoguenov}, Alexis and {Hoekstra}, Henk and {Lidman}, Chris and {Mulchaey}, John S. and {Noble}, Allison and {Parker}, Laura C. and {Tanaka}, Masayuki and {Wilman}, David J. and {Webb}, Tracy and {Wilson}, Gillian and {Yee}, Howard K.~C.},
        title = {Evidence for a change in the dominant satellite galaxy quenching mechanism at z = 1},
      journal = {\mnras},
         year = 2016,
        month = mar,
       volume = {456},
       number = {4},
        pages = {4364-4376},
          doi = {10.1093/mnras/stv2949},
archivePrefix = {arXiv},
       eprint = {1511.07344},
 primaryClass = {astro-ph.GA},
       adsurl = {https://ui.adsabs.harvard.edu/abs/2016MNRAS.456.4364B}
}

@ARTICLE{whitaker_quenching_2026,
   author = {{Whitaker}, Katherine E. and {Bezanson}, Rachel},
        title = {Quenching of Star Formation in Massive Galaxies},
      journal = {\araa},
         year = 2026,
        month = aug,
       volume = {64},
       number = {1},
        pages = {179-222},
          doi = {10.1146/annurev-astro-043024-121502},
archivePrefix = {arXiv},
       eprint = {2606.12156},
 primaryClass = {astro-ph.GA},
       adsurl = {https://ui.adsabs.harvard.edu/abs/2026ARA&A..64..179W}
}

@ARTICLE{rhee_origin_2024,
   author = {{Rhee}, Jinsu and {Yi}, Sukyoung K. and {Ko}, Jongwan and {Contini}, Emanuele and {Jang}, J.~K. and {Jeon}, Seyoung and {Han}, San and {Pichon}, Christophe and {Dubois}, Yohan and {Kraljic}, Katarina and {Peirani}, S{\'e}bastien},
        title = {On the Origin of Star Formation Quenching of Galaxies in Group Environments Using the NewHorizon Simulation},
      journal = {\apj},
         year = 2024,
        month = aug,
       volume = {971},
       number = {1},
          eid = {111},
        pages = {111},
          doi = {10.3847/1538-4357/ad5a83},
archivePrefix = {arXiv},
       eprint = {2408.08353},
 primaryClass = {astro-ph.GA},
       adsurl = {https://ui.adsabs.harvard.edu/abs/2024ApJ...971..111R}
}

@ARTICLE{alberts_clusters_2022,
   author = {{Alberts}, Stacey and {Noble}, Allison},
        title = {From Clusters to Proto-Clusters: The Infrared Perspective on Environmental Galaxy Evolution},
      journal = {Universe},
         year = 2022,
        month = oct,
       volume = {8},
       number = {11},
          eid = {554},
        pages = {554},
          doi = {10.3390/universe8110554},
archivePrefix = {arXiv},
       eprint = {2209.02781},
 primaryClass = {astro-ph.GA},
       adsurl = {https://ui.adsabs.harvard.edu/abs/2022Univ....8..554A}
}

@ARTICLE{hwang_evolution_2019,
   author = {{Hwang}, Ho Seong and {Shin}, Jihye and {Song}, Hyunmi},
        title = {Evolution of star formation rate-density relation over cosmic time in a simulated universe: the observed reversal reproduced},
      journal = {\mnras},
         year = 2019,
        month = oct,
       volume = {489},
       number = {1},
        pages = {339-348},
          doi = {10.1093/mnras/stz2136},
archivePrefix = {arXiv},
       eprint = {1907.03895},
 primaryClass = {astro-ph.GA},
       adsurl = {https://ui.adsabs.harvard.edu/abs/2019MNRAS.489..339H}
}

@ARTICLE{scudder_galaxy_2012,
   author = {{Scudder}, Jillian M. and {Ellison}, Sara L. and {Torrey}, Paul and {Patton}, David R. and {Mendel}, J. Trevor},
        title = {Galaxy pairs in the Sloan Digital Sky Survey - V. Tracing changes in star formation rate and metallicity out to separations of 80 kpc},
      journal = {\mnras},
         year = 2012,
        month = oct,
       volume = {426},
       number = {1},
        pages = {549-565},
          doi = {10.1111/j.1365-2966.2012.21749.x},
archivePrefix = {arXiv},
       eprint = {1207.4791},
 primaryClass = {astro-ph.CO},
       adsurl = {https://ui.adsabs.harvard.edu/abs/2012MNRAS.426..549S}
}

@misc{bertin_sourcextractor_2022,
   author = {{Bertin}, E. and {Schefer}, M. and {Apostolakos}, N. and {{\'A}lvarez-Ayll{\'o}n}, A. and {Dubath}, P. and {K{\"u}mmel}, M.},
        title = {SourceXtractor++},
 howpublished = {ascl:2212.018},
         year = 2022,
        month = dec,
          eid = {ascl:2212.018},
archivePrefix = {ascl},
       eprint = {2212.018},
       adsurl = {https://ui.adsabs.harvard.edu/abs/2022ascl.soft12018B}
}

@ARTICLE{jego_star-forming_2026,
   author = {{Jego}, Baptiste and {Kraljic}, Katarina and {B{\'e}thermin}, Matthieu and {Dav{\'e}}, Romeel},
        title = {Star-forming galaxies in the cosmic web in the last 11 Gyr},
      journal = {\aap},
         year = 2026,
        month = feb,
       volume = {706},
          eid = {A150},
        pages = {A150},
          doi = {10.1051/0004-6361/202557368},
archivePrefix = {arXiv},
       eprint = {2509.18077},
 primaryClass = {astro-ph.CO},
       adsurl = {https://ui.adsabs.harvard.edu/abs/2026A&A...706A.150J}
}

@ARTICLE{reeves_gogreen_2021,
   author = {{Reeves}, Andrew M.~M. and {Balogh}, Michael L. and {van der Burg}, Remco F.~J. and {Finoguenov}, Alexis and {Kukstas}, Egidijus and {McCarthy}, Ian G. and {Webb}, Kristi and {Muzzin}, Adam and {McGee}, Sean and {Rudnick}, Gregory and {Biviano}, Andrea and {Cerulo}, Pierluigi and {Chan}, Jeffrey C.~C. and {Cooper}, M.~C. and {Demarco}, Ricardo and {Jablonka}, Pascale and {De Lucia}, Gabriella and {Vulcani}, Benedetta and {Wilson}, Gillian and {Yee}, Howard K.~C. and {Zaritsky}, Dennis},
        title = {The GOGREEN survey: dependence of galaxy properties on halo mass at z > 1 and implications for environmental quenching},
      journal = {\mnras},
         year = 2021,
        month = sep,
       volume = {506},
       number = {3},
        pages = {3364-3384},
          doi = {10.1093/mnras/stab1955},
archivePrefix = {arXiv},
       eprint = {2107.03425},
 primaryClass = {astro-ph.GA},
       adsurl = {https://ui.adsabs.harvard.edu/abs/2021MNRAS.506.3364R}
}

@ARTICLE{joshi_preprocessing_2017,
   author = {{Joshi}, Gandhali D. and {Wadsley}, James and {Parker}, Laura C.},
        title = {Preprocessing, mass-loss and mass segregation of galaxies in dark matter simulations},
      journal = {\mnras},
         year = 2017,
        month = jul,
       volume = {468},
       number = {4},
        pages = {4625-4634},
          doi = {10.1093/mnras/stx803},
archivePrefix = {arXiv},
       eprint = {1704.01959},
 primaryClass = {astro-ph.GA},
       adsurl = {https://ui.adsabs.harvard.edu/abs/2017MNRAS.468.4625J}
}

@ARTICLE{popesso_hot_2026,
   author = {{Popesso}, P. and {Biviano}, A. and {Marini}, I. and {Dolag}, K. and {Vladutescu-Zopp}, S. and {Csizi}, B. and {Biffi}, V. and {Lamer}, G. and {Robothan}, A. and {Bravo}, M. and {Lovisari}, L. and {Ettori}, S. and {Angelinelli}, M. and {Driver}, S. and {Toptun}, V. and {Dev}, A. and {Mazengo}, D. and {Merloni}, A. and {Comparat}, J. and {Ponti}, G. and {Mroczkowski}, T. and {Bulbul}, E. and {Grandis}, S. and {Bahar}, E.},
        title = {The hot gas mass fraction in halos: From Milky Way-like groups to massive clusters},
      journal = {\aap},
         year = 2026,
        month = mar,
       volume = {707},
          eid = {A362},
        pages = {A362},
          doi = {10.1051/0004-6361/202453256},
archivePrefix = {arXiv},
       eprint = {2411.16555},
 primaryClass = {astro-ph.GA},
       adsurl = {https://ui.adsabs.harvard.edu/abs/2026A&A...707A.362P}
}

@ARTICLE{oppenheimer_simulating_2021,
   author = {{Oppenheimer}, Benjamin D. and {Babul}, Arif and {Bah{\'e}}, Yannick and {Butsky}, Iryna S. and {McCarthy}, Ian G.},
        title = {Simulating Groups and the IntraGroup Medium: The Surprisingly Complex and Rich Middle Ground between Clusters and Galaxies},
      journal = {Universe},
         year = 2021,
        month = jun,
       volume = {7},
       number = {7},
          eid = {209},
        pages = {209},
          doi = {10.3390/universe7070209},
archivePrefix = {arXiv},
       eprint = {2106.13257},
 primaryClass = {astro-ph.GA},
       adsurl = {https://ui.adsabs.harvard.edu/abs/2021Univ....7..209O}
}

@ARTICLE{yang_galaxy_2007,
   author = {{Yang}, Xiaohu and {Mo}, H.~J. and {van den Bosch}, Frank C. and {Pasquali}, Anna and {Li}, Cheng and {Barden}, Marco},
        title = {Galaxy Groups in the SDSS DR4. I. The Catalog and Basic Properties},
      journal = {\apj},
         year = 2007,
        month = dec,
       volume = {671},
       number = {1},
        pages = {153-170},
          doi = {10.1086/522027},
archivePrefix = {arXiv},
       eprint = {0707.4640},
 primaryClass = {astro-ph},
       adsurl = {https://ui.adsabs.harvard.edu/abs/2007ApJ...671..153Y}
}

@ARTICLE{harish_cosmos-web_2025,
   author = {{Harish}, Santosh and {Kartaltepe}, Jeyhan S. and {Liu}, Daizhong and {Koekemoer}, Anton M. and {Casey}, Caitlin M. and {Franco}, Maximilien and {Akins}, Hollis B. and {Ilbert}, Olivier and {Shuntov}, Marko and {Drakos}, Nicole E. and {Engesser}, Mike and {Faisst}, Andreas L. and {Gozaliasl}, Ghassem and {Martin}, Crystal L. and {Hirschmann}, Michaela and {Kokorev}, Vasily and {Lambrides}, Erini and {McCracken}, Henry Joy and {McKinney}, Jed and {Paquereau}, Louise and {Rhodes}, Jason and {Robertson}, Brant E.},
        title = {COSMOS-Web: MIRI Data Reduction and Number Counts at 7.7 {\ensuremath{\mu}}m Using JWST},
      journal = {\apj},
         year = 2025,
        month = oct,
       volume = {992},
       number = {1},
          eid = {45},
        pages = {45},
          doi = {10.3847/1538-4357/adfa1e},
archivePrefix = {arXiv},
       eprint = {2506.03306},
 primaryClass = {astro-ph.GA},
       adsurl = {https://ui.adsabs.harvard.edu/abs/2025ApJ...992...45H}
}

@ARTICLE{casey_cosmos-web_2023,
   author = {{Casey}, Caitlin M. and {Kartaltepe}, Jeyhan S. and {Drakos}, Nicole E. and {Franco}, Maximilien and {Harish}, Santosh and {Paquereau}, Louise and {Ilbert}, Olivier and {Rose}, Caitlin and {Cox}, Isabella G. and {Nightingale}, James W. and {Robertson}, Brant E. and {Silverman}, John D. and {Koekemoer}, Anton M. and {Massey}, Richard and {McCracken}, Henry Joy and {Rhodes}, Jason and {Akins}, Hollis B. and {Allen}, Natalie and {Amvrosiadis}, Aristeidis and {Arango-Toro}, Rafael C. and {Bagley}, Micaela B. and {Bongiorno}, Angela and {Capak}, Peter L. and {Champagne}, Jaclyn B. and {Chartab}, Nima and {Ch{\'a}vez Ortiz}, {\'O}scar A. and {Chworowsky}, Katherine and {Cooke}, Kevin C. and {Cooper}, Olivia R. and {Darvish}, Behnam and {Ding}, Xuheng and {Faisst}, Andreas L. and {Finkelstein}, Steven L. and {Fujimoto}, Seiji and {Gentile}, Fabrizio and {Gillman}, Steven and {Gould}, Katriona M.~L. and {Gozaliasl}, Ghassem and {Hayward}, Christopher C. and {He}, Qiuhan and {Hemmati}, Shoubaneh and {Hirschmann}, Michaela and {Jahnke}, Knud and {Jin}, Shuowen and {Khostovan}, Ali Ahmad and {Kokorev}, Vasily and {Lambrides}, Erini and {Laigle}, Clotilde and {Larson}, Rebecca L. and {Leung}, Gene C.~K. and {Liu}, Daizhong and {Liaudat}, Tobias and {Long}, Arianna S. and {Magdis}, Georgios and {Mahler}, Guillaume and {Mainieri}, Vincenzo and {Manning}, Sinclaire M. and {Maraston}, Claudia and {Martin}, Crystal L. and {McCleary}, Jacqueline E. and {McKinney}, Jed and {McPartland}, Conor J.~R. and {Mobasher}, Bahram and {Pattnaik}, Rohan and {Renzini}, Alvio and {Rich}, R. Michael and {Sanders}, David B. and {Sattari}, Zahra and {Scognamiglio}, Diana and {Scoville}, Nick and {Sheth}, Kartik and {Shuntov}, Marko and {Sparre}, Martin and {Suzuki}, Tomoko L. and {Talia}, Margherita and {Toft}, Sune and {Trakhtenbrot}, Benny and {Urry}, C. Megan and {Valentino}, Francesco and {Vanderhoof}, Brittany N. and {Vardoulaki}, Eleni and {Weaver}, John R. and {Whitaker}, Katherine E. and {Wilkins}, Stephen M. and {Yang}, Lilan and {Zavala}, Jorge A.},
        title = {COSMOS-Web: An Overview of the JWST Cosmic Origins Survey},
      journal = {\apj},
         year = 2023,
        month = sep,
       volume = {954},
       number = {1},
          eid = {31},
        pages = {31},
          doi = {10.3847/1538-4357/acc2bc},
archivePrefix = {arXiv},
       eprint = {2211.07865},
 primaryClass = {astro-ph.GA},
       adsurl = {https://ui.adsabs.harvard.edu/abs/2023ApJ...954...31C}
}

@ARTICLE{malavasi_relative_2022,
   author = {{Malavasi}, Nicola and {Langer}, Mathieu and {Aghanim}, Nabila and {Gal{\'a}rraga-Espinosa}, Daniela and {Gouin}, C{\'e}line},
        title = {Relative effect of nodes and filaments of the cosmic web on the quenching of galaxies and the orientation of their spin},
      journal = {\aap},
         year = 2022,
        month = feb,
       volume = {658},
          eid = {A113},
        pages = {A113},
          doi = {10.1051/0004-6361/202141723},
archivePrefix = {arXiv},
       eprint = {2109.14623},
 primaryClass = {astro-ph.CO},
       adsurl = {https://ui.adsabs.harvard.edu/abs/2022A&A...658A.113M}
}

@ARTICLE{plummer_problem_1911,
   author = {{Plummer}, H.~C.},
        title = {On the problem of distribution in globular star clusters},
      journal = {\mnras},
         year = 1911,
        month = mar,
       volume = {71},
        pages = {460-470},
          doi = {10.1093/mnras/71.5.460},
       adsurl = {https://ui.adsabs.harvard.edu/abs/1911MNRAS..71..460P}
}

@ARTICLE{bellagamba_amico_2019,
   author = {{Bellagamba}, Fabio and {Sereno}, Mauro and {Roncarelli}, Mauro and {Maturi}, Matteo and {Radovich}, Mario and {Bardelli}, Sandro and {Puddu}, Emanuella and {Moscardini}, Lauro and {Getman}, Fedor and {Hildebrandt}, Hendrik and {Napolitano}, Nicola},
        title = {AMICO galaxy clusters in KiDS-DR3: weak lensing mass calibration},
      journal = {\mnras},
         year = 2019,
        month = apr,
       volume = {484},
       number = {2},
        pages = {1598-1615},
          doi = {10.1093/mnras/stz090},
archivePrefix = {arXiv},
       eprint = {1810.02827},
 primaryClass = {astro-ph.CO},
       adsurl = {https://ui.adsabs.harvard.edu/abs/2019MNRAS.484.1598B}
}

@ARTICLE{maturi_span_2019,
   author = {{Maturi}, Matteo and {Bellagamba}, Fabio and {Radovich}, Mario and {Roncarelli}, Mauro and {Sereno}, Mauro and {Moscardini}, Lauro and {Bardelli}, Sandro and {Puddu}, Emanuella},
        title = {AMICO galaxy clusters in KiDS-DR3: sample properties and selection function},
      journal = {\mnras},
         year = 2019,
        month = may,
       volume = {485},
       number = {1},
        pages = {498-512},
          doi = {10.1093/mnras/stz294},
archivePrefix = {arXiv},
       eprint = {1810.02811},
 primaryClass = {astro-ph.CO},
       adsurl = {https://ui.adsabs.harvard.edu/abs/2019MNRAS.485..498M}
}

@ARTICLE{shuntov_stellar_2026,
   author = {{Shuntov}, Marko and {Ilbert}, Olivier and {Lagos}, Claudia del P. and {Toft}, Sune and {Valentino}, Francesco and {Mercier}, Wilfried and {Akins}, Hollis B. and {Binh}, Nguyen and {Brinch}, Malte and {Casey}, Caitlin M. and {Franco}, Maximilien and {Gentile}, Fabrizio and {Gozaliasl}, Ghassem and {Haghjoo}, Aryana and {Harish}, Santosh and {Hirschmann}, Michaela and {Huertas-Company}, Marc and {Jin}, Shuowen and {Kartaltepe}, Jeyhan S. and {Koekemoer}, Anton M. and {Laigle}, Clotilde and {Lewis}, Joseph S.~W. and {Magdis}, Georgios E. and {Joy McCracken}, Henry and {Mobasher}, Bahram and {Moutard}, Thibaud and {Oesch}, Pascal A. and {Paquereau}, Louise and {Renzini}, Alvio and {Rich}, Michael R. and {Sanders}, David B. and {Toni}, Greta and {Tresse}, Laurence and {Weibel}, Andrea and {Weaver}, John R. and {Yang}, Lilan},
        title = {The stellar mass function of quiescent and star-forming galaxies and its dependence on morphology in COSMOS-Web},
      journal = {\aap},
         year = 2026,
        month = mar,
       volume = {707},
          eid = {A391},
        pages = {A391},
          doi = {10.1051/0004-6361/202558022},
archivePrefix = {arXiv},
       eprint = {2511.05259},
 primaryClass = {astro-ph.GA},
       adsurl = {https://ui.adsabs.harvard.edu/abs/2026A&A...707A.391S}
}

@ARTICLE{paquereau_tracing_2025,
   author = {{Paquereau}, L. and {Laigle}, C. and {McCracken}, H.~J. and {Shuntov}, M. and {Ilbert}, O. and {Akins}, H.~B. and {Allen}, N. and {Arango-Togo}, R. and {Berman}, E.~M. and {B{\'e}thermin}, M. and {Casey}, C.~M. and {McCleary}, J. and {Dubois}, Y. and {Drakos}, N.~E. and {Faisst}, A.~L. and {Franco}, M. and {Harish}, S. and {Jespersen}, C.~K. and {Kartaltepe}, J.~S. and {Koekemoer}, A.~M. and {Kokorev}, V. and {Lambrides}, E. and {Larson}, R. and {Liu}, D. and {Le Borgne}, D. and {Lewis}, J.~S.~W. and {McKinney}, J. and {Mercier}, W. and {Rhodes}, J.~D. and {Robertson}, B.~E. and {Toft}, S. and {Trebitsch}, M. and {Tresse}, L. and {Weaver}, J.~R.},
        title = {Tracing the galaxy-halo connection with galaxy clustering in COSMOS-Web from z = 0.1 to z {\ensuremath{\sim}} 12},
      journal = {\aap},
         year = 2025,
        month = oct,
       volume = {702},
          eid = {A163},
        pages = {A163},
          doi = {10.1051/0004-6361/202553828},
archivePrefix = {arXiv},
       eprint = {2501.11674},
 primaryClass = {astro-ph.GA},
       adsurl = {https://ui.adsabs.harvard.edu/abs/2025A&A...702A.163P}
}

@ARTICLE{taamoli_cosmos2020_2024,
   author = {{Taamoli}, Sina and {Nezhad}, Negin and {Mobasher}, Bahram and {Manesh}, Faezeh and {Chartab}, Nima and {Weaver}, John R. and {Capak}, Peter L. and {Casey}, Caitlin M. and {Gozaliasl}, Ghassem and {Heintz}, Kasper E. and {Ilbert}, Olivier and {Kartaltepe}, Jeyhan S. and {McCracken}, Henry J. and {Sanders}, David B. and {Scoville}, Nicholas and {Toft}, Sune and {Watson}, Darach},
        title = {COSMOS2020: Disentangling the Role of Mass and Environment in Star Formation Activity of Galaxies at 0.4 < z < 4},
      journal = {\apj},
         year = 2024,
        month = dec,
       volume = {977},
       number = {2},
          eid = {263},
        pages = {263},
          doi = {10.3847/1538-4357/ad94f3},
archivePrefix = {arXiv},
       eprint = {2411.09522},
 primaryClass = {astro-ph.GA},
       adsurl = {https://ui.adsabs.harvard.edu/abs/2024ApJ...977..263T}
}

@ARTICLE{lovisari_scaling_2021,
   author = {{Lovisari}, Lorenzo and {Ettori}, Stefano and {Gaspari}, Massimo and {Giles}, Paul A.},
        title = {Scaling Properties of Galaxy Groups},
      journal = {Universe},
         year = 2021,
        month = may,
       volume = {7},
       number = {5},
          eid = {139},
        pages = {139},
          doi = {10.3390/universe7050139},
archivePrefix = {arXiv},
       eprint = {2106.13256},
 primaryClass = {astro-ph.CO},
       adsurl = {https://ui.adsabs.harvard.edu/abs/2021Univ....7..139L}
}

@ARTICLE{paul_understanding_2017,
   author = {{Paul}, S. and {John}, R.~S. and {Gupta}, P. and {Kumar}, H.},
        title = {Understanding `galaxy groups' as a unique structure in the universe},
      journal = {\mnras},
         year = 2017,
        month = oct,
       volume = {471},
       number = {1},
        pages = {2-11},
          doi = {10.1093/mnras/stx1488},
archivePrefix = {arXiv},
       eprint = {1706.01916},
 primaryClass = {astro-ph.CO},
       adsurl = {https://ui.adsabs.harvard.edu/abs/2017MNRAS.471....2P}
}

@ARTICLE{hatamnia_large-scale_2025,
   author = {{Hatamnia}, Hossein and {Mobasher}, Bahram and {Taamoli}, Sina and {Kartaltepe}, Jeyhan S. and {Casey}, Caitlin M. and {Akins}, Hollis B. and {Brinch}, Malte and {Chartab}, Nima and {Drakos}, Nicole E. and {Faisst}, Andreas L. and {Finkelstein}, Steven L. and {Franco}, Maximilien and {Giddings}, Finn and {Gozaliasl}, Ghassem and {Hadi}, Ali and {Haghjoo}, Aryana and {Harish}, Santosh and {Ilbert}, Olivier and {Jablonka}, Pascale L. and {Jin}, Shuowen and {Khostovan}, Ali Ahmad and {Koekemoer}, Anton M. and {Laishram}, Ronaldo and {Liu}, Daizhong and {Maturi}, Matteo and {McCracken}, Henry Joy and {Martin}, Crystal L. and {Moscardini}, Lauro and {Scognamiglio}, Diana and {Shuntov}, Marko and {Toni}, Greta and {de la Vega}, Alexander and {Weaver}, John R. and {Yang}, Lilan},
        title = {Large-scale Structure in COSMOS-Web: Tracing Galaxy Evolution in the Cosmic Web up to z {\ensuremath{\sim}} 7 with the Largest JWST Survey},
      journal = {\apj},
         year = 2026,
        month = may,
       volume = {1002},
       number = {2},
          eid = {192},
        pages = {192},
          doi = {10.3847/1538-4357/ae5bac},
archivePrefix = {arXiv},
       eprint = {2511.10727},
 primaryClass = {astro-ph.GA},
       adsurl = {https://ui.adsabs.harvard.edu/abs/2026ApJ..1002..192H}
}

@ARTICLE{oman_homogeneous_2021,
   author = {{Oman}, Kyle A. and {Bah{\'e}}, Yannick M. and {Healy}, Julia and {Hess}, Kelley M. and {Hudson}, Michael J. and {Verheijen}, Marc A.~W.},
        title = {A homogeneous measurement of the delay between the onsets of gas stripping and star formation quenching in satellite galaxies of groups and clusters},
      journal = {\mnras},
         year = 2021,
        month = mar,
       volume = {501},
       number = {4},
        pages = {5073-5095},
          doi = {10.1093/mnras/staa3845},
archivePrefix = {arXiv},
       eprint = {2009.00667},
 primaryClass = {astro-ph.GA},
       adsurl = {https://ui.adsabs.harvard.edu/abs/2021MNRAS.501.5073O}
}

@ARTICLE{simpson_quenching_2018,
   author = {{Simpson}, Christine M. and {Grand}, Robert J.~J. and {G{\'o}mez}, Facundo A. and {Marinacci}, Federico and {Pakmor}, R{\"u}diger and {Springel}, Volker and {Campbell}, David J.~R. and {Frenk}, Carlos S.},
        title = {Quenching and ram pressure stripping of simulated Milky Way satellite galaxies},
      journal = {\mnras},
         year = 2018,
        month = jul,
       volume = {478},
       number = {1},
        pages = {548-567},
          doi = {10.1093/mnras/sty774},
archivePrefix = {arXiv},
       eprint = {1705.03018},
 primaryClass = {astro-ph.GA},
       adsurl = {https://ui.adsabs.harvard.edu/abs/2018MNRAS.478..548S}
}

@ARTICLE{bahe_star_2015,
   author = {{Bah{\'e}}, Yannick M. and {McCarthy}, Ian G.},
        title = {Star formation quenching in simulated group and cluster galaxies: when, how, and why?},
      journal = {\mnras},
         year = 2015,
        month = feb,
       volume = {447},
       number = {1},
        pages = {969-992},
          doi = {10.1093/mnras/stu2293},
archivePrefix = {arXiv},
       eprint = {1410.8161},
 primaryClass = {astro-ph.GA},
       adsurl = {https://ui.adsabs.harvard.edu/abs/2015MNRAS.447..969B}
}

@ARTICLE{pintos-castro_evolution_2019,
   author = {{Pintos-Castro}, I. and {Yee}, H.~K.~C. and {Muzzin}, A. and {Old}, L. and {Wilson}, G.},
        title = {The Evolution of the Quenching of Star Formation in Cluster Galaxies since z {\ensuremath{\sim}} 1},
      journal = {\apj},
         year = 2019,
        month = may,
       volume = {876},
       number = {1},
          eid = {40},
        pages = {40},
          doi = {10.3847/1538-4357/ab14ee},
archivePrefix = {arXiv},
       eprint = {1904.00023},
 primaryClass = {astro-ph.GA},
       adsurl = {https://ui.adsabs.harvard.edu/abs/2019ApJ...876...40P}
}

@ARTICLE{alberts_evolution_2014,
   author = {{Alberts}, Stacey and {Pope}, Alexandra and {Brodwin}, Mark and {Atlee}, David W. and {Lin}, Yen-Ting and {Dey}, Arjun and {Eisenhardt}, Peter R.~M. and {Gettings}, Daniel P. and {Gonzalez}, Anthony H. and {Jannuzi}, Buell T. and {Mancone}, Conor L. and {Moustakas}, John and {Snyder}, Gregory F. and {Stanford}, S. Adam and {Stern}, Daniel and {Weiner}, Benjamin J. and {Zeimann}, Gregory R.},
        title = {The evolution of dust-obscured star formation activity in galaxy clusters relative to the field over the last 9 billion years},
      journal = {\mnras},
         year = 2014,
        month = jan,
       volume = {437},
       number = {1},
        pages = {437-457},
          doi = {10.1093/mnras/stt1897},
archivePrefix = {arXiv},
       eprint = {1310.6040},
 primaryClass = {astro-ph.CO},
       adsurl = {https://ui.adsabs.harvard.edu/abs/2014MNRAS.437..437A}
}

@ARTICLE{brodwin_era_2013,
   author = {{Brodwin}, M. and {Stanford}, S.~A. and {Gonzalez}, Anthony H. and {Zeimann}, G.~R. and {Snyder}, G.~F. and {Mancone}, C.~L. and {Pope}, A. and {Eisenhardt}, P.~R. and {Stern}, D. and {Alberts}, S. and {Ashby}, M.~L.~N. and {Brown}, M.~J.~I. and {Chary}, R.-R. and {Dey}, Arjun and {Galametz}, A. and {Gettings}, D.~P. and {Jannuzi}, B.~T. and {Miller}, E.~D. and {Moustakas}, J. and {Moustakas}, L.~A.},
        title = {The Era of Star Formation in Galaxy Clusters},
      journal = {\apj},
         year = 2013,
        month = dec,
       volume = {779},
       number = {2},
          eid = {138},
        pages = {138},
          doi = {10.1088/0004-637X/779/2/138},
archivePrefix = {arXiv},
       eprint = {1310.6039},
 primaryClass = {astro-ph.CO},
       adsurl = {https://ui.adsabs.harvard.edu/abs/2013ApJ...779..138B}
}

@ARTICLE{williams_alma_2022,
   author = {{Williams}, Christina C. and {Alberts}, Stacey and {Spilker}, Justin S. and {Noble}, Allison G. and {Stefanon}, Mauro and {Willmer}, Christopher N.~A. and {Bezanson}, Rachel and {Narayanan}, Desika and {Whitaker}, Katherine E.},
        title = {ALMA Measures Molecular Gas Reservoirs Comparable to Field Galaxies in a Low-mass Galaxy Cluster at z = 1.3},
      journal = {\apj},
         year = 2022,
        month = apr,
       volume = {929},
       number = {1},
          eid = {35},
        pages = {35},
          doi = {10.3847/1538-4357/ac58fa},
archivePrefix = {arXiv},
       eprint = {2201.01304},
 primaryClass = {astro-ph.GA},
       adsurl = {https://ui.adsabs.harvard.edu/abs/2022ApJ...929...35W}
}

@ARTICLE{elbaz_reversal_2007,
   author = {{Elbaz}, D. and {Daddi}, E. and {Le Borgne}, D. and {Dickinson}, M. and {Alexander}, D.~M. and {Chary}, R.-R. and {Starck}, J.-L. and {Brandt}, W.~N. and {Kitzbichler}, M. and {MacDonald}, E. and {Nonino}, M. and {Popesso}, P. and {Stern}, D. and {Vanzella}, E.},
        title = {The reversal of the star formation-density relation in the distant universe},
      journal = {\aap},
         year = 2007,
        month = jun,
       volume = {468},
       number = {1},
        pages = {33-48},
          doi = {10.1051/0004-6361:20077525},
archivePrefix = {arXiv},
       eprint = {astro-ph/0703653},
 primaryClass = {astro-ph},
       adsurl = {https://ui.adsabs.harvard.edu/abs/2007A&A...468...33E}
}

@ARTICLE{boquien_cigale_2019,
   author = {{Boquien}, M. and {Burgarella}, D. and {Roehlly}, Y. and {Buat}, V. and {Ciesla}, L. and {Corre}, D. and {Inoue}, A.~K. and {Salas}, H.},
        title = {CIGALE: a python Code Investigating GALaxy Emission},
      journal = {\aap},
         year = 2019,
        month = feb,
       volume = {622},
          eid = {A103},
        pages = {A103},
          doi = {10.1051/0004-6361/201834156},
archivePrefix = {arXiv},
       eprint = {1811.03094},
 primaryClass = {astro-ph.GA},
       adsurl = {https://ui.adsabs.harvard.edu/abs/2019A&A...622A.103B}
}

@ARTICLE{gozaliasl_chandra_2019,
   author = {{Gozaliasl}, Ghassem and {Finoguenov}, Alexis and {Tanaka}, Masayuki and {Dolag}, Klaus and {Montanari}, Francesco and {Kirkpatrick}, Charles C. and {Vardoulaki}, Eleni and {Khosroshahi}, Habib G. and {Salvato}, Mara and {Laigle}, Clotilde and {McCracken}, Henry J. and {Ilbert}, Olivier and {Cappelluti}, Nico and {Daddi}, Emanuele and {Hasinger}, Guenther and {Capak}, Peter and {Scoville}, Nick Z. and {Toft}, Sune and {Civano}, Francesca and {Griffiths}, Richard E. and {Balogh}, Michael and {Li}, Yanxia and {Ahoranta}, Jussi and {Mei}, Simona and {Iovino}, Angela and {Henriques}, Bruno M.~B. and {Erfanianfar}, Ghazaleh},
        title = {Chandra centres for COSMOS X-ray galaxy groups: differences in stellar properties between central dominant and offset brightest group galaxies},
      journal = {\mnras},
         year = 2019,
        month = mar,
       volume = {483},
       number = {3},
        pages = {3545-3565},
          doi = {10.1093/mnras/sty3203},
archivePrefix = {arXiv},
       eprint = {1812.01604},
 primaryClass = {astro-ph.GA},
       adsurl = {https://ui.adsabs.harvard.edu/abs/2019MNRAS.483.3545G}
}

@ARTICLE{cucciati_progeny_2018,
   author = {{Cucciati}, O. and {Lemaux}, B.~C. and {Zamorani}, G. and {Le F{\`e}vre}, O. and {Tasca}, L.~A.~M. and {Hathi}, N.~P. and {Lee}, K.-G. and {Bardelli}, S. and {Cassata}, P. and {Garilli}, B. and {Le Brun}, V. and {Maccagni}, D. and {Pentericci}, L. and {Thomas}, R. and {Vanzella}, E. and {Zucca}, E. and {Lubin}, L.~M. and {Amorin}, R. and {Cassar{\`a}}, L.~P. and {Cimatti}, A. and {Talia}, M. and {Vergani}, D. and {Koekemoer}, A. and {Pforr}, J. and {Salvato}, M.},
        title = {The progeny of a cosmic titan: a massive multi-component proto-supercluster in formation at z = 2.45 in VUDS},
      journal = {\aap},
         year = 2018,
        month = nov,
       volume = {619},
          eid = {A49},
        pages = {A49},
          doi = {10.1051/0004-6361/201833655},
archivePrefix = {arXiv},
       eprint = {1806.06073},
 primaryClass = {astro-ph.CO},
       adsurl = {https://ui.adsabs.harvard.edu/abs/2018A&A...619A..49C}
}

@ARTICLE{darvish_cosmic_2017,
   author = {{Darvish}, Behnam and {Mobasher}, Bahram and {Martin}, D. Christopher and {Sobral}, David and {Scoville}, Nick and {Stroe}, Andra and {Hemmati}, Shoubaneh and {Kartaltepe}, Jeyhan},
        title = {Cosmic Web of Galaxies in the COSMOS Field: Public Catalog and Different Quenching for Centrals and Satellites},
      journal = {\apj},
         year = 2017,
        month = mar,
       volume = {837},
       number = {1},
          eid = {16},
        pages = {16},
          doi = {10.3847/1538-4357/837/1/16},
archivePrefix = {arXiv},
       eprint = {1611.05451},
 primaryClass = {astro-ph.GA},
       adsurl = {https://ui.adsabs.harvard.edu/abs/2017ApJ...837...16D}
}

@ARTICLE{darvish_comparative_2015,
   author = {{Darvish}, Behnam and {Mobasher}, Bahram and {Sobral}, David and {Scoville}, Nicholas and {Aragon-Calvo}, Miguel},
        title = {A Comparative Study of Density Field Estimation for Galaxies: New Insights into the Evolution of Galaxies with Environment in COSMOS out to z{\ensuremath{\sim}}3},
      journal = {\apj},
         year = 2015,
        month = jun,
       volume = {805},
       number = {2},
          eid = {121},
        pages = {121},
          doi = {10.1088/0004-637X/805/2/121},
archivePrefix = {arXiv},
       eprint = {1503.07879},
 primaryClass = {astro-ph.GA},
       adsurl = {https://ui.adsabs.harvard.edu/abs/2015ApJ...805..121D}
}

@ARTICLE{weaver_cosmos2020_2022,
   author = {{Weaver}, J.~R. and {Kauffmann}, O.~B. and {Ilbert}, O. and {McCracken}, H.~J. and {Moneti}, A. and {Toft}, S. and {Brammer}, G. and {Shuntov}, M. and {Davidzon}, I. and {Hsieh}, B.~C. and {Laigle}, C. and {Anastasiou}, A. and {Jespersen}, C.~K. and {Vinther}, J. and {Capak}, P. and {Casey}, C.~M. and {McPartland}, C.~J.~R. and {Milvang-Jensen}, B. and {Mobasher}, B. and {Sanders}, D.~B. and {Zalesky}, L. and {Arnouts}, S. and {Aussel}, H. and {Dunlop}, J.~S. and {Faisst}, A. and {Franx}, M. and {Furtak}, L.~J. and {Fynbo}, J.~P.~U. and {Gould}, K.~M.~L. and {Greve}, T.~R. and {Gwyn}, S. and {Kartaltepe}, J.~S. and {Kashino}, D. and {Koekemoer}, A.~M. and {Kokorev}, V. and {Le F{\`e}vre}, O. and {Lilly}, S. and {Masters}, D. and {Magdis}, G. and {Mehta}, V. and {Peng}, Y. and {Riechers}, D.~A. and {Salvato}, M. and {Sawicki}, M. and {Scarlata}, C. and {Scoville}, N. and {Shirley}, R. and {Silverman}, J.~D. and {Sneppen}, A. and {Smol{\v{c}}i{\'c}}, V. and {Steinhardt}, C. and {Stern}, D. and {Tanaka}, M. and {Taniguchi}, Y. and {Teplitz}, H.~I. and {Vaccari}, M. and {Wang}, W.-H. and {Zamorani}, G.},
        title = {COSMOS2020: A Panchromatic View of the Universe to z{\ensuremath{\sim}}10 from Two Complementary Catalogs},
      journal = {\apjs},
         year = 2022,
        month = jan,
       volume = {258},
       number = {1},
          eid = {11},
        pages = {11},
          doi = {10.3847/1538-4365/ac3078},
archivePrefix = {arXiv},
       eprint = {2110.13923},
 primaryClass = {astro-ph.GA},
       adsurl = {https://ui.adsabs.harvard.edu/abs/2022ApJS..258...11W}
}

@ARTICLE{laigle_cosmos2015_2016,
   author = {{Laigle}, C. and {McCracken}, H.~J. and {Ilbert}, O. and {Hsieh}, B.~C. and {Davidzon}, I. and {Capak}, P. and {Hasinger}, G. and {Silverman}, J.~D. and {Pichon}, C. and {Coupon}, J. and {Aussel}, H. and {Le Borgne}, D. and {Caputi}, K. and {Cassata}, P. and {Chang}, Y.-Y. and {Civano}, F. and {Dunlop}, J. and {Fynbo}, J. and {Kartaltepe}, J.~S. and {Koekemoer}, A. and {Le F{\`e}vre}, O. and {Le Floc'h}, E. and {Leauthaud}, A. and {Lilly}, S. and {Lin}, L. and {Marchesi}, S. and {Milvang-Jensen}, B. and {Salvato}, M. and {Sanders}, D.~B. and {Scoville}, N. and {Smolcic}, V. and {Stockmann}, M. and {Taniguchi}, Y. and {Tasca}, L. and {Toft}, S. and {Vaccari}, Mattia and {Zabl}, J.},
        title = {The COSMOS2015 Catalog: Exploring the 1 < z < 6 Universe with Half a Million Galaxies},
      journal = {\apjs},
         year = 2016,
        month = jun,
       volume = {224},
       number = {2},
          eid = {24},
        pages = {24},
          doi = {10.3847/0067-0049/224/2/24},
archivePrefix = {arXiv},
       eprint = {1604.02350},
 primaryClass = {astro-ph.GA},
       adsurl = {https://ui.adsabs.harvard.edu/abs/2016ApJS..224...24L}
}

@ARTICLE{scoville_cosmic_2007,
   author = {{Scoville}, N. and {Aussel}, H. and {Brusa}, M. and {Capak}, P. and {Carollo}, C.~M. and {Elvis}, M. and {Giavalisco}, M. and {Guzzo}, L. and {Hasinger}, G. and {Impey}, C. and {Kneib}, J.-P. and {LeFevre}, O. and {Lilly}, S.~J. and {Mobasher}, B. and {Renzini}, A. and {Rich}, R.~M. and {Sanders}, D.~B. and {Schinnerer}, E. and {Schminovich}, D. and {Shopbell}, P. and {Taniguchi}, Y. and {Tyson}, N.~D.},
        title = {The Cosmic Evolution Survey (COSMOS): Overview},
      journal = {\apjs},
         year = 2007,
        month = sep,
       volume = {172},
       number = {1},
        pages = {1-8},
          doi = {10.1086/516585},
archivePrefix = {arXiv},
       eprint = {astro-ph/0612305},
 primaryClass = {astro-ph},
       adsurl = {https://ui.adsabs.harvard.edu/abs/2007ApJS..172....1S}
}

@ARTICLE{rhee_origin_2026,
   author = {{Rhee}, Jinsu and {Pichon}, Christophe and {Dubois}, Yohan and {Yi}, Sukyoung K. and {Ko}, Jongwan and {Sheen}, Yun-Kyeong and {Han}, San and {Jeon}, Seyoung and {Jang}, J.~K. and {Lee}, Wonki and {Contini}, Emanuele and {Lee}, Bumhyun and {Lee}, Jaehyun and {Kraljic}, Katarina and {Peirani}, S{\'e}bastien},
        title = {The origin of gas stripping of galaxies in group environments},
      journal = {\aap},
         year = 2026,
        month = jan,
       volume = {705},
          eid = {A106},
        pages = {A106},
          doi = {10.1051/0004-6361/202556771},
archivePrefix = {arXiv},
       eprint = {2512.03145},
 primaryClass = {astro-ph.GA},
       adsurl = {https://ui.adsabs.harvard.edu/abs/2026A&A...705A.106R}
}

@ARTICLE{fillingham_under_2016,
   author = {{Fillingham}, Sean P. and {Cooper}, Michael C. and {Pace}, Andrew B. and {Boylan-Kolchin}, Michael and {Bullock}, James S. and {Garrison-Kimmel}, Shea and {Wheeler}, Coral},
        title = {Under pressure: quenching star formation in low-mass satellite galaxies via stripping},
      journal = {\mnras},
         year = 2016,
        month = dec,
       volume = {463},
       number = {2},
        pages = {1916-1928},
          doi = {10.1093/mnras/stw2131},
archivePrefix = {arXiv},
       eprint = {1606.07810},
 primaryClass = {astro-ph.GA},
       adsurl = {https://ui.adsabs.harvard.edu/abs/2016MNRAS.463.1916F}
}

@ARTICLE{wetzel_satellite_2015,
   author = {{Wetzel}, Andrew R. and {Deason}, Alis J. and {Garrison-Kimmel}, Shea},
        title = {Satellite Dwarf Galaxies in a Hierarchical Universe: Infall Histories, Group Preprocessing, and Reionization},
      journal = {\apj},
         year = 2015,
        month = jul,
       volume = {807},
       number = {1},
          eid = {49},
        pages = {49},
          doi = {10.1088/0004-637X/807/1/49},
archivePrefix = {arXiv},
       eprint = {1501.01972},
 primaryClass = {astro-ph.GA},
       adsurl = {https://ui.adsabs.harvard.edu/abs/2015ApJ...807...49W}
}

@ARTICLE{boselli_environmental_2006,
   author = {{Boselli}, Alessandro and {Gavazzi}, Giuseppe},
        title = {Environmental Effects on Late-Type Galaxies in Nearby Clusters},
      journal = {\pasp},
         year = 2006,
        month = apr,
       volume = {118},
       number = {842},
        pages = {517-559},
          doi = {10.1086/500691},
archivePrefix = {arXiv},
       eprint = {astro-ph/0601108},
 primaryClass = {astro-ph},
       adsurl = {https://ui.adsabs.harvard.edu/abs/2006PASP..118..517B}
}

@ARTICLE{van_den_bosch_importance_2008,
   author = {{van den Bosch}, Frank C. and {Aquino}, Daniel and {Yang}, Xiaohu and {Mo}, H.~J. and {Pasquali}, Anna and {McIntosh}, Daniel H. and {Weinmann}, Simone M. and {Kang}, Xi},
        title = {The importance of satellite quenching for the build-up of the red sequence of present-day galaxies},
      journal = {\mnras},
         year = 2008,
        month = jun,
       volume = {387},
       number = {1},
        pages = {79-91},
          doi = {10.1111/j.1365-2966.2008.13230.x},
archivePrefix = {arXiv},
       eprint = {0710.3164},
 primaryClass = {astro-ph},
       adsurl = {https://ui.adsabs.harvard.edu/abs/2008MNRAS.387...79V}
}

@ARTICLE{wetzel_galaxy_2013,
   author = {{Wetzel}, Andrew R. and {Tinker}, Jeremy L. and {Conroy}, Charlie and {van den Bosch}, Frank C.},
        title = {Galaxy evolution in groups and clusters: satellite star formation histories and quenching time-scales in a hierarchical Universe},
      journal = {\mnras},
         year = 2013,
        month = jun,
       volume = {432},
       number = {1},
        pages = {336-358},
          doi = {10.1093/mnras/stt469},
archivePrefix = {arXiv},
       eprint = {1206.3571},
 primaryClass = {astro-ph.CO},
       adsurl = {https://ui.adsabs.harvard.edu/abs/2013MNRAS.432..336W}
}

@ARTICLE{peng_mass_2010,
   author = {{Peng}, Ying-jie and {Lilly}, Simon J. and {Kova{\v{c}}}, Katarina and {Bolzonella}, Micol and {Pozzetti}, Lucia and {Renzini}, Alvio and {Zamorani}, Gianni and {Ilbert}, Olivier and {Knobel}, Christian and {Iovino}, Angela and {Maier}, Christian and {Cucciati}, Olga and {Tasca}, Lidia and {Carollo}, C. Marcella and {Silverman}, John and {Kampczyk}, Pawel and {de Ravel}, Loic and {Sanders}, David and {Scoville}, Nicholas and {Contini}, Thierry and {Mainieri}, Vincenzo and {Scodeggio}, Marco and {Kneib}, Jean-Paul and {Le F{\`e}vre}, Olivier and {Bardelli}, Sandro and {Bongiorno}, Angela and {Caputi}, Karina and {Coppa}, Graziano and {de la Torre}, Sylvain and {Franzetti}, Paolo and {Garilli}, Bianca and {Lamareille}, Fabrice and {Le Borgne}, Jean-Francois and {Le Brun}, Vincent and {Mignoli}, Marco and {Perez Montero}, Enrique and {Pello}, Roser and {Ricciardelli}, Elena and {Tanaka}, Masayuki and {Tresse}, Laurence and {Vergani}, Daniela and {Welikala}, Niraj and {Zucca}, Elena and {Oesch}, Pascal and {Abbas}, Ummi and {Barnes}, Luke and {Bordoloi}, Rongmon and {Bottini}, Dario and {Cappi}, Alberto and {Cassata}, Paolo and {Cimatti}, Andrea and {Fumana}, Marco and {Hasinger}, Gunther and {Koekemoer}, Anton and {Leauthaud}, Alexei and {Maccagni}, Dario and {Marinoni}, Christian and {McCracken}, Henry and {Memeo}, Pierdomenico and {Meneux}, Baptiste and {Nair}, Preethi and {Porciani}, Cristiano and {Presotto}, Valentina and {Scaramella}, Roberto},
        title = {Mass and Environment as Drivers of Galaxy Evolution in SDSS and zCOSMOS and the Origin of the Schechter Function},
      journal = {\apj},
         year = 2010,
        month = sep,
       volume = {721},
       number = {1},
        pages = {193-221},
          doi = {10.1088/0004-637X/721/1/193},
archivePrefix = {arXiv},
       eprint = {1003.4747},
 primaryClass = {astro-ph.CO},
       adsurl = {https://ui.adsabs.harvard.edu/abs/2010ApJ...721..193P}
}

@ARTICLE{dressler_galaxy_1980,
   author = {{Dressler}, A.},
        title = {Galaxy morphology in rich clusters: implications for the formation and evolution of galaxies.},
      journal = {\apj},
         year = 1980,
        month = mar,
       volume = {236},
        pages = {351-365},
          doi = {10.1086/157753},
       adsurl = {https://ui.adsabs.harvard.edu/abs/1980ApJ...236..351D}
}

@ARTICLE{toni_amico-cosmos_2024,
   author = {{Toni}, G. and {Maturi}, M. and {Finoguenov}, A. and {Moscardini}, L. and {Castignani}, G.},
        title = {AMICO-COSMOS galaxy cluster and group catalogue up to z = 2: Sample properties and X-ray counterparts},
      journal = {\aap},
         year = 2024,
        month = jul,
       volume = {687},
          eid = {A56},
        pages = {A56},
          doi = {10.1051/0004-6361/202348832},
archivePrefix = {arXiv},
       eprint = {2312.12501},
 primaryClass = {astro-ph.CO},
       adsurl = {https://ui.adsabs.harvard.edu/abs/2024A&A...687A..56T}
}

@ARTICLE{arango-toro_probing_2023,
   author = {{Arango-Toro}, R.~C. and {Ciesla}, L. and {Ilbert}, O. and {Magnelli}, B. and {Jim{\'e}nez-Andrade}, E.~F. and {Buat}, V.},
        title = {Probing the timescale of the 1.4 GHz radio emissions as a star formation tracer},
      journal = {\aap},
         year = 2023,
        month = jul,
       volume = {675},
          eid = {A126},
        pages = {A126},
          doi = {10.1051/0004-6361/202345848},
archivePrefix = {arXiv},
       eprint = {2304.13758},
 primaryClass = {astro-ph.GA},
       adsurl = {https://ui.adsabs.harvard.edu/abs/2023A&A...675A.126A}
}

@ARTICLE{ciesla_goods-alma_2023,
   author = {{Ciesla}, L. and {G{\'o}mez-Guijarro}, C. and {Buat}, V. and {Elbaz}, D. and {Jin}, S. and {B{\'e}thermin}, M. and {Daddi}, E. and {Franco}, M. and {Inami}, H. and {Magdis}, G. and {Magnelli}, B. and {Xiao}, M.},
        title = {GOODS-ALMA 2.0: Last gigayear star formation histories of the so-called starbursts within the main sequence},
      journal = {\aap},
         year = 2023,
        month = apr,
       volume = {672},
          eid = {A191},
        pages = {A191},
          doi = {10.1051/0004-6361/202245376},
archivePrefix = {arXiv},
       eprint = {2211.02510},
 primaryClass = {astro-ph.GA},
       adsurl = {https://ui.adsabs.harvard.edu/abs/2023A&A...672A.191C}
}

@ARTICLE{pozzetti_zcosmos_2010,
   author = {{Pozzetti}, L. and {Bolzonella}, M. and {Zucca}, E. and {Zamorani}, G. and {Lilly}, S. and {Renzini}, A. and {Moresco}, M. and {Mignoli}, M. and {Cassata}, P. and {Tasca}, L. and {Lamareille}, F. and {Maier}, C. and {Meneux}, B. and {Halliday}, C. and {Oesch}, P. and {Vergani}, D. and {Caputi}, K. and {Kova{\v{c}}}, K. and {Cimatti}, A. and {Cucciati}, O. and {Iovino}, A. and {Peng}, Y. and {Carollo}, M. and {Contini}, T. and {Kneib}, J.-P. and {Le F{\'e}vre}, O. and {Mainieri}, V. and {Scodeggio}, M. and {Bardelli}, S. and {Bongiorno}, A. and {Coppa}, G. and {de la Torre}, S. and {de Ravel}, L. and {Franzetti}, P. and {Garilli}, B. and {Kampczyk}, P. and {Knobel}, C. and {Le Borgne}, J.-F. and {Le Brun}, V. and {Pell{\`o}}, R. and {Perez Montero}, E. and {Ricciardelli}, E. and {Silverman}, J.~D. and {Tanaka}, M. and {Tresse}, L. and {Abbas}, U. and {Bottini}, D. and {Cappi}, A. and {Guzzo}, L. and {Koekemoer}, A.~M. and {Leauthaud}, A. and {Maccagni}, D. and {Marinoni}, C. and {McCracken}, H.~J. and {Memeo}, P. and {Porciani}, C. and {Scaramella}, R. and {Scarlata}, C. and {Scoville}, N.},
        title = {zCOSMOS - 10k-bright spectroscopic sample. The bimodality in the galaxy stellar mass function: exploring its evolution with redshift},
      journal = {\aap},
         year = 2010,
        month = nov,
       volume = {523},
          eid = {A13},
        pages = {A13},
          doi = {10.1051/0004-6361/200913020},
archivePrefix = {arXiv},
       eprint = {0907.5416},
 primaryClass = {astro-ph.CO},
       adsurl = {https://ui.adsabs.harvard.edu/abs/2010A&A...523A..13P}
}

@ARTICLE{toni_cosmos-web_2026,
   author = {{Toni}, Greta and {Maturi}, Matteo and {Castignani}, Gianluca and {Moscardini}, Lauro and {Gozaliasl}, Ghassem and {Finoguenov}, Alexis and {Taamoli}, Sina and {Gentile}, Fabrizio and {Akins}, Hollis B. and {Arango-Toro}, Rafael C. and {Casey}, Caitlin M. and {Drakos}, Nicole E. and {Faisst}, Andreas L. and {Flayhart}, Carter and {Franco}, Maximilien and {Hadi}, Ali and {Haghjoo}, Aryana and {Harish}, Santosh and {Hatamnia}, Hossein and {Ilbert}, Olivier and {Jin}, Shuowen and {Kartaltepe}, Jeyhan S. and {Khostovan}, Ali Ahmad and {Koekemoer}, Anton M. and {Leroy}, Gavin and {Magdis}, Georgios E. and {McCracken}, Henry Joy and {McKinney}, Jed and {Paquereau}, Louise and {Rhodes}, Jason and {Rich}, R. Michael and {Robertson}, Brant E. and {Samir}, Rasha M. and {Scognamiglio}, Diana and {Shamyati}, Samaneh and {Shuntov}, Marko and {Zavala}, Jorge A.},
        title = {COSMOS-Web galaxy groups: Evolution of red sequence and quiescent galaxy fraction},
      journal = {\aap},
         year = 2026,
        month = feb,
       volume = {707},
          eid = {A87},
        pages = {A87},
          doi = {10.1051/0004-6361/202557183},
archivePrefix = {arXiv},
       eprint = {2509.08040},
 primaryClass = {astro-ph.GA},
       adsurl = {https://ui.adsabs.harvard.edu/abs/2026A&A...707A..87T}
}

@ARTICLE{ghaffari_star_2026,
   author = {{Ghaffari}, Z. and {Gozaliasl}, G. and {Biviano}, A. and {Mamon}, G.~A. and {Toni}, G. and {Taamoli}, S. and {Maturi}, M. and {Moscardini}, L. and {Zacchei}, A. and {Gentile}, F. and {Haas}, M. and {Akins}, H. and {Arango-Toro}, R.~C. and {Cheng}, Y. and {Casey}, C. and {Franco}, M. and {Harish}, S. and {Hatamnia}, H. and {Ilbert}, O. and {Kartaltepe}, J. and {Khostovan}, A.~A. and {Koekemoer}, A.~M. and {Liu}, D. and {McCracken}, H.~J. and {McKinney}, J. and {Rhodes}, J. and {Robertson}, B. and {Shuntov}, M. and {Yang}, L.},
        title = {Star formation quenching precedes morphological transformation in COSMOS-Web's richest galaxy groups},
      journal = {\aap},
         year = 2026,
        month = apr,
       volume = {708},
          eid = {A129},
        pages = {A129},
          doi = {10.1051/0004-6361/202558283},
archivePrefix = {arXiv},
       eprint = {2601.06297},
 primaryClass = {astro-ph.GA},
       adsurl = {https://ui.adsabs.harvard.edu/abs/2026A&A...708A.129G}
}

@ARTICLE{toni_cosmos-web_2025,
   author = {{Toni}, Greta and {Gozaliasl}, Ghassem and {Maturi}, Matteo and {Moscardini}, Lauro and {Finoguenov}, Alexis and {Castignani}, Gianluca and {Gentile}, Fabrizio and {Virolainen}, Kaija and {Casey}, Caitlin M. and {Kartaltepe}, Jeyhan S. and {Akins}, Hollis B. and {Allen}, Natalie and {Arango-Toro}, Rafael C. and {Babul}, Arif and {Brinch}, Malte and {Drakos}, Nicole E. and {Faisst}, Andreas L. and {Franco}, Maximilien and {Griffiths}, Richard E. and {Harish}, Santosh and {Hasinger}, G{\"u}nther and {Ilbert}, Olivier and {Jin}, Shuowen and {Khostovan}, Ali Ahmad and {Koekemoer}, Anton M. and {Korpi-Lagg}, Maarit and {Larson}, Rebecca L. and {Lertprasertpong}, Jitrapon and {Liu}, Daizhong and {Magdis}, Georgios and {Massey}, Richard and {McCracken}, Henry Joy and {McKinney}, Jed and {Paquereau}, Louise and {Rhodes}, Jason and {Robertson}, Brant E. and {Sargent}, Mark and {Shuntov}, Marko and {Tanaka}, Masayuki and {Taamoli}, Sina and {Tempel}, Elmo and {Toft}, Sune and {Vardoulaki}, Eleni and {Yang}, Lilan},
        title = {The COSMOS-Web deep galaxy group catalog up to z = 3.7},
      journal = {\aap},
         year = 2025,
        month = may,
       volume = {697},
          eid = {A197},
        pages = {A197},
          doi = {10.1051/0004-6361/202553759},
archivePrefix = {arXiv},
       eprint = {2501.09060},
 primaryClass = {astro-ph.GA},
       adsurl = {https://ui.adsabs.harvard.edu/abs/2025A&A...697A.197T}
}

@ARTICLE{franco_cosmos-web_2026,
   author = {{Franco}, Maximilien and {Casey}, Caitlin M. and {Koekemoer}, Anton M. and {Liu}, Daizhong and {Bagley}, Micaela B. and {McCracken}, Henry Joy and {Kartaltepe}, Jeyhan S. and {Akins}, Hollis B. and {Ilbert}, Olivier and {Shuntov}, Marko and {Harish}, Santosh and {Robertson}, Brant E. and {Arango-Toro}, Rafael C. and {Battisti}, Andrew J. and {Chartab}, Nima and {Drakos}, Nicole E. and {Faisst}, Andreas L. and {Flayhart}, Carter and {Gozaliasl}, Ghassem and {Hirschmann}, Michaela and {Massey}, Richard and {Rhodes}, Jason and {Sattari}, Zahra and {Scognamiglio}, Diana and {Weaver}, John R. and {Yang}, Lilan and {Zavala}, Jorge A. and {Berman}, Edward M. and {Gentile}, Fabrizio and {Gillman}, Steven and {Long}, Arianna S. and {Magdis}, Georgios and {McCleary}, Jacqueline E. and {McKinney}, Jed and {Mobasher}, Bahram and {Paquereau}, Louise and {Rest}, Armin and {Sanders}, David B. and {Toft}, Sune and {Yu}, Si-Yue},
        title = {COSMOS-Web: Comprehensive Data Reduction for Wide-area JWST NIRCam Imaging},
      journal = {\apj},
         year = 2026,
        month = mar,
       volume = {999},
       number = {2},
          eid = {200},
        pages = {200},
          doi = {10.3847/1538-4357/ae3aa1},
archivePrefix = {arXiv},
       eprint = {2506.03256},
 primaryClass = {astro-ph.IM},
       adsurl = {https://ui.adsabs.harvard.edu/abs/2026ApJ...999..200F}
}

@ARTICLE{arango-toro_cosmos-web_2025,
   author = {{Arango-Toro}, R.~C. and {Ilbert}, O. and {Ciesla}, L. and {Shuntov}, M. and {Aufort}, G. and {Mercier}, W. and {Laigle}, C. and {Franco}, M. and {Bethermin}, M. and {Le Borgne}, D. and {Dubois}, Y. and {McCracken}, H.~J. and {Paquereau}, L. and {Huertas-Company}, M. and {Kartaltepe}, J. and {Casey}, C.~M. and {Akins}, H. and {Allen}, N. and {Andika}, I. and {Brinch}, M. and {Drakos}, N.~E. and {Faisst}, A. and {Gozaliasl}, G. and {Harish}, S. and {Kaminsky}, A. and {Koekemoer}, A. and {Kokorev}, V. and {Liu}, D. and {Magdis}, G. and {Martin}, C.~L. and {Moutard}, T. and {Rhodes}, J. and {Rich}, R.~M. and {Robertson}, B. and {Sanders}, D.~B. and {Sheth}, K. and {Talia}, M. and {Toft}, S. and {Tresse}, L. and {Valentino}, F. and {Vijayan}, A. and {Weaver}, J.},
        title = {COSMOS-Web: A history of galaxy migrations over the stellar mass--star formation rate plane},
      journal = {\aap},
         year = 2025,
        month = apr,
       volume = {696},
          eid = {A159},
        pages = {A159},
          doi = {10.1051/0004-6361/202452519},
archivePrefix = {arXiv},
       eprint = {2410.05375},
 primaryClass = {astro-ph.GA},
       adsurl = {https://ui.adsabs.harvard.edu/abs/2025A&A...696A.159A}
}

@ARTICLE{khostovan_cosmos_2026,
   author = {{Khostovan}, Ali Ahmad and {Kartaltepe}, Jeyhan S. and {Salvato}, Mara and {Ilbert}, Olivier and {Casey}, Caitlin M. and {Algera}, Hiddo and {Antwi-Danso}, Jacqueline and {Battisti}, Andrew and {Brinch}, Malte and {Brusa}, Marcella and {Calabr{\`o}}, Antonello and {Capak}, Peter L. and {Chartab}, Nima and {Cooper}, Olivia R. and {Cox}, Isa G. and {Darvish}, Behnam and {Drakos}, Nicole E. and {Faisst}, Andreas L. and {George}, Matthew R. and {Gozaliasl}, Ghassem and {Harish}, Santosh and {Hasinger}, G{\"u}nther and {Hatamnia}, Hossein and {Iovino}, Angela and {Jin}, Shuowen and {Kashino}, Daichi and {Koekemoer}, Anton M. and {Laishram}, Ronaldo and {Lee}, Khee-Gan and {Lertprasertpong}, Jitrapon and {Lilly}, Simon J. and {Liu}, Daizhong and {Masters}, Daniel C. and {Mobasher}, Bahram and {Nagao}, Tohru and {Onodera}, Masato and {Peng}, Yingjie and {Sanders}, David B. and {Sanders}, Ryan L. and {Sattari}, Zahra and {Scoville}, Nick and {Shah}, Ekta A. and {Silverman}, John D. and {Suzuki}, Nao and {Taamoli}, Sina and {Tanaka}, Masayuki and {Tasca}, Lidia A.~M. and {Toft}, Sune and {Toni}, Greta and {Trakhtenbrot}, Benny and {Trump}, Jonathan R. and {Vaccari}, Mattia and {Valentino}, Francesco and {Vanderhoof}, Brittany N. and {Weaver}, John R. and {Yun}, Min S. and {Zavala}, Jorge A.},
        title = {COSMOS Spectroscopic Redshift Compilation (First Data Release): 488,000 Redshifts Encompassing Two Decades of Spectroscopy},
      journal = {\apjs},
         year = 2026,
        month = jan,
       volume = {282},
       number = {1},
          eid = {6},
        pages = {6},
          doi = {10.3847/1538-4365/ae1cb9},
archivePrefix = {arXiv},
       eprint = {2503.00120},
 primaryClass = {astro-ph.GA},
       adsurl = {https://ui.adsabs.harvard.edu/abs/2026ApJS..282....6K}
}

@ARTICLE{ilbert_accurate_2006,
   author = {{Ilbert}, O. and {Arnouts}, S. and {McCracken}, H.~J. and {Bolzonella}, M. and {Bertin}, E. and {Le F{\`e}vre}, O. and {Mellier}, Y. and {Zamorani}, G. and {Pell{\`o}}, R. and {Iovino}, A. and {Tresse}, L. and {Le Brun}, V. and {Bottini}, D. and {Garilli}, B. and {Maccagni}, D. and {Picat}, J.~P. and {Scaramella}, R. and {Scodeggio}, M. and {Vettolani}, G. and {Zanichelli}, A. and {Adami}, C. and {Bardelli}, S. and {Cappi}, A. and {Charlot}, S. and {Ciliegi}, P. and {Contini}, T. and {Cucciati}, O. and {Foucaud}, S. and {Franzetti}, P. and {Gavignaud}, I. and {Guzzo}, L. and {Marano}, B. and {Marinoni}, C. and {Mazure}, A. and {Meneux}, B. and {Merighi}, R. and {Paltani}, S. and {Pollo}, A. and {Pozzetti}, L. and {Radovich}, M. and {Zucca}, E. and {Bondi}, M. and {Bongiorno}, A. and {Busarello}, G. and {de La Torre}, S. and {Gregorini}, L. and {Lamareille}, F. and {Mathez}, G. and {Merluzzi}, P. and {Ripepi}, V. and {Rizzo}, D. and {Vergani}, D.},
        title = {Accurate photometric redshifts for the CFHT legacy survey calibrated using the VIMOS VLT deep survey},
      journal = {\aap},
         year = 2006,
        month = oct,
       volume = {457},
       number = {3},
        pages = {841-856},
          doi = {10.1051/0004-6361:20065138},
archivePrefix = {arXiv},
       eprint = {astro-ph/0603217},
 primaryClass = {astro-ph},
       adsurl = {https://ui.adsabs.harvard.edu/abs/2006A&A...457..841I}
}

@ARTICLE{arnouts_measuring_2002,
   author = {{Arnouts}, S. and {Moscardini}, L. and {Vanzella}, E. and {Colombi}, S. and {Cristiani}, S. and {Fontana}, A. and {Giallongo}, E. and {Matarrese}, S. and {Saracco}, P.},
        title = {Measuring the redshift evolution of clustering: the Hubble Deep Field South},
      journal = {\mnras},
         year = 2002,
        month = jan,
       volume = {329},
       number = {2},
        pages = {355-366},
          doi = {10.1046/j.1365-8711.2002.04988.x},
archivePrefix = {arXiv},
       eprint = {astro-ph/0109453},
 primaryClass = {astro-ph},
       adsurl = {https://ui.adsabs.harvard.edu/abs/2002MNRAS.329..355A}
}

@ARTICLE{shuntov_cosmos2025_2025,
   author = {{Shuntov}, Marko and {Akins}, Hollis B. and {Paquereau}, Louise and {Casey}, Caitlin M. and {Ilbert}, Olivier and {Arango-Toro}, Rafael C. and {McCracken}, Henry Joy and {Franco}, Maximilien and {Harish}, Santosh and {Kartaltepe}, Jeyhan S. and {Koekemoer}, Anton M. and {Yang}, Lilan and {Huertas-Company}, Marc and {Berman}, Edward M. and {McCleary}, Jacqueline E. and {Toft}, Sune and {Gavazzi}, Rapha{\"e}l and {Achenbach}, Mark J. and {Bertin}, Emmanuel and {Brinch}, Malte and {Champagne}, Jackie and {Chartab}, Nima and {Drakos}, Nicole E. and {Egami}, Eiichi and {Endsley}, Ryan and {Faisst}, Andreas L. and {Fan}, Xiaohui and {Flayhart}, Carter and {Hartley}, William G. and {Hatamnia}, Hossein and {Gozaliasl}, Ghassem and {Gentile}, Fabrizio and {Jermann}, Iris and {Jin}, Shuowen and {Kakiichi}, Koki and {Khostovan}, Ali Ahmad and {K{\"u}mmel}, Martin and {Laigle}, Clotilde and {Laishram}, Ronaldo and {Lambrides}, Erini and {Liu}, Daizhong and {Lyu}, Jianwei and {Magdis}, Georgios and {Mobasher}, Bahram and {Moutard}, Thibaud and {Renzini}, Alvio and {Rich}, R. Michael and {Sanders}, David B. and {Sattari}, Zahra and {Robertson}, Brant E. and {Schefer}, Marc and {Scognamiglio}, Diana and {Scoville}, Nick and {Silverman}, John D. and {Taamoli}, Sina and {Trakhtenbrot}, Benny and {Valentino}, Francesco and {Wang}, Feige and {Weaver}, John R. and {Yang}, Jinyi},
        title = {COSMOS2025: The COSMOS-Web galaxy catalog of photometry, morphology, redshifts, and physical parameters from JWST, HST, and ground-based imaging},
      journal = {\aap},
         year = 2025,
        month = dec,
       volume = {704},
          eid = {A339},
        pages = {A339},
          doi = {10.1051/0004-6361/202555799},
archivePrefix = {arXiv},
       eprint = {2506.03243},
 primaryClass = {astro-ph.GA},
       adsurl = {https://ui.adsabs.harvard.edu/abs/2025A&A...704A.339S}
}

@ARTICLE{cooper_deep2_2007,
   author = {{Cooper}, Michael C. and {Newman}, Jeffrey A. and {Weiner}, Benjamin J. and {Yan}, Renbin and {Willmer}, Christopher N.~A. and {Bundy}, Kevin and {Coil}, Alison L. and {Conselice}, Christopher J. and {Davis}, Marc and {Faber}, S.~M. and {Gerke}, Brian F. and {Guhathakurta}, Puragra and {Koo}, David C. and {Noeske}, Kai G.},
        title = {The DEEP2 Galaxy Redshift Survey: the role of galaxy environment in the cosmic star formation history},
      journal = {\mnras},
         year = 2008,
        month = jan,
       volume = {383},
       number = {3},
        pages = {1058-1078},
          doi = {10.1111/j.1365-2966.2007.12613.x},
archivePrefix = {arXiv},
       eprint = {0706.4089},
 primaryClass = {astro-ph},
       adsurl = {https://ui.adsabs.harvard.edu/abs/2008MNRAS.383.1058C}
}

@INPROCEEDINGS{deluciaCosmicQuenching2025,
   author = {{De Lucia}, Gabriella and {Fontanot}, Fabio and {Hirschmann}, Michaela and {Xie}, Lizhi},
        title = {Cosmic quenching},
    booktitle = {Encyclopedia of Astrophysics},
         year = 2026,
       volume = {4},
        month = jan,
        pages = {282-311},
          doi = {10.1016/B978-0-443-21439-4.00077-8},
       adsurl = {https://ui.adsabs.harvard.edu/abs/2026enap....4..282D}
}

\end{document}